\pdfoutput=1
\documentclass[twocolumn,showpacs,prb,amsmath,amssymb,floatfix,eqsecnum]{revtex4-1}
\usepackage{amsmath,amssymb}
\usepackage{bm}
\usepackage{epsfig}
\usepackage{psfrag}
\usepackage{color}
\newcommand{\bd}{\bm}

\newcommand{\ktau}{\omega}
\newcommand{\barktau}{\bar{\omega}}
\newcommand{\barktaui}{\bar{\omega}_i}
\newcommand{\barktauone}{\bar{\omega}_1}
\newcommand{\barktautwo}{\bar{\omega}_2}
\newcommand{\barkprimetau}{\bar{\omega}^{\prime}}
\newcommand{\barqtau}{\bar{\epsilon}}

\begin{document}

\title{Functional renormalization group approach to the
Ising-nematic quantum critical point of 
two-dimensional metals}

\author{Casper Drukier, Lorenz Bartosch, Aldo Isidori, and Peter Kopietz}
  
\affiliation{Institut f\"{u}r Theoretische Physik, Universit\"{a}t
  Frankfurt,  Max-von-Laue Strasse 1, 60438 Frankfurt, Germany}

 \date{March 12, 2012}

 \begin{abstract}

Using functional renormalization group methods,
we study an effective low-energy model
describing the Ising-nematic quantum critical point in  two-dimensional metals.
We treat both
gapless fermionic and bosonic degrees of freedom on
equal footing and explicitly calculate the momentum and frequency dependent 
effective interaction between the fermions
mediated by the bosonic fluctuations.
Following earlier work by S.-S. Lee for a one-patch model, Metlitski and Sachdev 
[Phys. Rev. B {\bf{82}}, 075127] recently found within a
field-theoretical approach that certain three-loop diagrams strongly modify
the one-loop results, and that the  conventional
$1/N$ expansion breaks down in this problem.
We show that the singular three-loop diagrams considered by 
 Metlitski and Sachdev are included in a rather simple
truncation of the functional renormalization group flow
equations for this model involving only irreducible
vertices with two and three external legs.
Our approximate solution of these flow equations
explicitly yields the vertex corrections  
of this problem and allows us to calculate the 
anomalous dimension $\eta_{\psi}$
of the fermion field. 


\end{abstract}


\maketitle

\section{Introduction}

Inspired by the puzzling normal-state properties of the copper-oxide superconductors, 
the search for possible non-Fermi liquid states of metals continues to be a central topic in the theory of strongly correlated electrons.
An interesting new clue toward an understanding of the cuprates and other materials
comes from experiments on a variety of materials, indicating a nematic phase transition.\cite{Ando02,Borzi07,Kohsaka07,Hinkov08,Daou10,Nandi10,Fradkin10} 
This is a quantum phase transition which, while preserving translational symmetry,  breaks the lattice rotational symmetry from square to rectangular, i.e., the invariance under rotations of the system in the $x$-$y$ plane by $90^{\circ}$ is lost. At the quantum critical point, electrons couple strongly to order parameter fluctuations, leading to a destruction of the Fermi liquid state. The resulting distortion of the Fermi surface is also referred to as a Pomeranchuk transition.\cite{Pomeranchuk58,Halboth00}

The conventional theoretical approach to quantum critical phenomena is the so-called Hertz-Millis approach, where the interaction between electrons is decoupled via a (bosonic) Hubbard-Stratonovich transformation and the fermionic degrees of freedom are integrated out.\cite{Hertz76,Millis93,Belitz05,Loehneysen07} However, as in the metallic state the electrons are gapless, this approach usually leads to singular vertices, which especially in low dimensions need to be treated with care. It can therefore be advantageous not to integrate out the fermions at all.

As concerns the nematic phase transition, 
the most effective scattering processes of electrons take place when the 
momentum of the bosons is locally almost tangential to the Fermi surface.
The phase transition can therefore be modeled by coupling electrons in the vicinity of two patches of a Fermi surface to a gapless scalar Ising order parameter field. 
Indeed, the coupling of gapless fermions to gapless bosonic fluctuations is 
known to give rise to non-Fermi liquid behavior.\cite{Loehneysen07,Vojta03}
The corresponding field theory is very similar to a nonrelativistic gauge theory, which has been studied intensively,\cite{Holstein73,Reizer89,Lee92}  starting with the important work by Holstein, Norton, and Pincus.\cite{Holstein73} 
Such gauge theories have applications in a number of different problems, such as the description of the half-filled Landau level,\cite{Halperin93} the description of spin liquids in terms of spinons which can form a critical spinon Fermi surface,\cite{Motrunich05,SSLee05} or the description of the instability of a ferromagnetic quantum critical point.\cite{Rech06} Similar gauge theories have also been used to describe fermions on a honeycomb lattice interacting through an electromagnetic gauge field.\cite{Gonzales94,Giuliani10}
In all cases, the low-energy behavior is expected to be described by a scale-invariant scaling theory. The single-particle Green function $G (\omega, \bm{k} )$
was calculated\cite{Altshuler94,Polchinski94} for a spherical Fermi surface within the random phase approximation (RPA), resulting in 
 \begin{equation}
   \label{eq:GlowenergyRPA}
   G( \omega + i 0^+,\bm{k} )  \propto 
\frac{1}{ A_\omega|\omega|^{2/3} - \xi_{\bd{k}}} \;.
 \end{equation}
Here, $\omega$ and $\bm{k}$ 
are the frequency and momentum of the electron, $A_\omega=A^\prime\text{sgn}\left(\omega\right)+iA^{\prime\prime}$ is a complex constant depending on the sign of $\omega$, and
$\xi_{\bd{k}} = v_F ( | \bd{k} | - k_F )$, where $v_F$ is the Fermi velocity and $k_F$ is the Fermi momentum,
denotes the single-particle excitation energy.
While the static part of the self-energy remains unrenormalized, its dynamic part implies that both the renormalized energy and damping rate of the electron scale in exactly the same way. Consequently, there are no well-defined sharp quasiparticles and Landau's Fermi liquid theory breaks down.

\begin{figure}[tb]    
   \centering
  \includegraphics[width=0.45\textwidth]{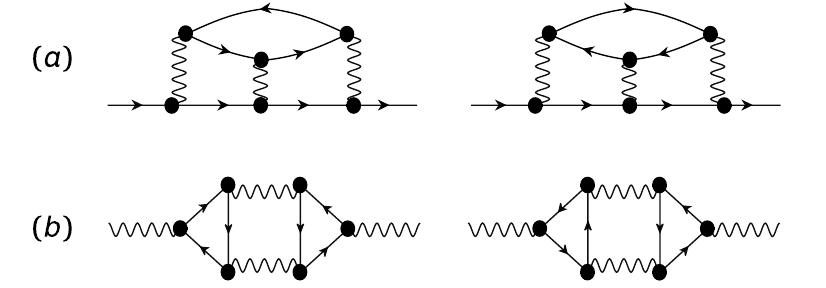}
  \caption{%
These three-loop diagrams have been identified by Metlitski and Sachdev \cite{Metlitski10a,Metlitski10b}
to give singular corrections to the one-loop results for the bosonic and fermionic self-energies.
Solid arrows denote the fermionic single-particle Green functions within the one-loop approximation,
while wavy lines represent the RPA propagator of the bosonic fluctuations.
The black dot is the bare interaction vertex between one boson and two fermion fields.
The diagrams in (a) are three-loop corrections to the
fermionic self-energy, while those in (b) represent corrections to the bosonic self-energy of the so-called
Aslamazov-Larkin type.
}
    \label{fig:MSdiagrams}
  \end{figure}

For a long time, it was thought that the above scenario holds true when going beyond the RPA.~\cite{Sedrakyan09} 
It was believed that this can be justified by considering the limit of large $N$, where $N$ is the number of fermion flavors. 
However, it was recently shown by Lee\cite{SSLee08,SSLee09} that, even 
if one considers only scattering processes in the vicinity of a single patch
of the Fermi surface, 
the coupling to a gapless gauge field results in very strong correlations such that, even in the large-$N$ limit, the theory remains strongly coupled.
Subsequently, \mbox{Metlitski} and Sachdev\cite{Metlitski10a,Metlitski10b} considered
a more realistic two-patch model where the two patches of the Fermi surface to which
a given bosonic momentum
is tangent are retained; they derived a scaling theory and explicitly calculated corrections to the bosonic and fermionic self-energies up to three loops,  
using the one-loop propagator in internal loop integrations. 
Metlitski and Sachdev identified certain three-loop contributions to the
bosonic and fermionic self-energies (see Fig.~\ref{fig:MSdiagrams}),  which give rise to logarithmically divergent corrections to the one-loop RPA results.
Exponentiating these logarithmic terms, they obtained the following expression for the
retarded propagator of the fermions, 
\begin{eqnarray}
  \label{eq:GlowenergyMetlitskiSachdev}
  G (\omega + i0^+,\bm{k}) & \propto & 
\frac{1}{\left[A_\omega|\omega|^{1/z} - \xi_{\bm{k}} \right]^{1-\eta_\psi/2} },
\end{eqnarray}
where $\eta_\psi$ is the anomalous dimension 
of the fermion field
and $z $ is the fermionic dynamic critical exponent.
Using scaling relations for the fermionic and bosonic Green functions,
Metlitski and Sachdev argued that the fermionic dynamic exponent is given by $z=z_b/2$, 
where $z_b$ is the corresponding bosonic dynamic exponent.
Their explicit calculations\cite{Metlitski10a} show that $z_b = 3$ is not renormalized by fluctuations
up to three loops,
implying that $z=3/2$ is correctly given by the one-loop approximation.
On the other hand, for the fermionic anomalous dimension,
Metlitski and Sachdev obtain the finite result  $\eta_\psi = 0.068$
at the Ising-nematic transition, whereas $\eta_{\psi}=0$ within the one-loop approximation.

On a technical level, the reason for the breakdown of the large-$N$ expansion can be traced back to the fact that the curvature of the fermion propagator comes with a factor of $N$. This in combination with a cancellation of the curvature in a set of planar diagrams eventually leads to the breakdown of the large-$N$ expansion. 
In the words of Chubukov,\cite{Chubukov10} 
there is {\em hidden one-dimensionality} in two-dimensional systems. \\
Albeit $1/N$ can not be used as a control parameter, it was suggested by Mross {\em et al.},\cite{Mross10} following earlier work at finite $N$ by Nayak and Wilczek,\cite{Nayak94a, *Nayak94b} to use $z_b$ as a tunable parameter. In this case it is possible to consider the limits $N \to \infty$ and $z_b - 2 \to 0$ while keeping the product $N(z_b - 2)$ finite to bring the calculation under control. 
Using a tunable $z_b$ is a sensible strategy 
because a non-local interaction is not expected to be 
renormalized. Extrapolating the results obtained by 
Mross {\em et al.} \cite{Mross10}
to the physically relevant case $z_b = 3$ and $N=2$, one obtains for 
the anomalous dimension of the fermion field $\eta_\psi \approx 0.6$ (using our 
definition  (\ref{eq:GlowenergyMetlitskiSachdev}) of $\eta_{\psi}$).
Obviously, this value is much larger than the estimate  $\eta_{\psi} \approx 0.068$  by 
Metlitski and Sachdev.~\cite{Metlitski10a}

Even though the calculations in Refs.~\onlinecite{Metlitski10a} and 
\onlinecite{Mross10} are based on the 
field-theoretical renormalization group, the fact that two independent calculations involving
different types of approximations
produce different values for $\eta_{\psi}$ 
shows that on a quantitative level there are still open questions. 
Due to the sign problem in quantum Monte Carlo calculations and the fact that dynamical mean field theory can essentially only predict mean field exponents, the number of alternative methods to verify the correctness of the anomalous scaling properties of the Ising-nematic transition is limited. 
In this work, we study this problem by means of
the one-particle irreducible 
implementation of the
functional renormalization group (FRG) method,\cite{Wetterich93,Berges02,Kopietz10,Metzner12}
which is a modern implementation of the Wilsonian renormalization group idea.
The flexibility of FRG methods to deal with systems involving
both fermionic and bosonic  fields has already been used by several 
authors.\cite{Baier04,Baier05,Schuetz05,Wetterich07,Strack08,Bartosch09a,Bartosch09b} 
In particular, in Refs. \onlinecite{Schuetz05,Bartosch09a, Bartosch09b} it has been shown
that it can be advantageous to  introduce a cutoff parameter $\Lambda$ which regularizes
the infrared divergences 
only in the momentum carried by the bosonic field (the momentum-transfer cutoff scheme).
We show in this work that this cutoff scheme is also convenient to
study the nematic quantum critical point.

The rest of this work is organized as follows.
After introducing the model system and defining our notation in Sec.~\ref{sec:model}, we give 
in  Sec.~\ref{sec:FRGfloweq} the FRG flow equations for the self-energies and vertex corrections
in general form. We also introduce the momentum-transfer cutoff scheme
and show that the singular three-loop diagrams
shown in Fig.~\ref{fig:MSdiagrams} are contained in a rather simple
truncation of the hierarchy of FRG flow equations involving
only irreducible two-point and three-point vertices.
In Sec.~\ref{sec:novertexcorrections}, we show how to recover the known one-loop results
for the momentum- and frequency-dependent fermionic and bosonic self-energies 
by integrating the FRG flow equations ignoring vertex corrections.
In the main part of this work, given in Sec.~\ref{sec:vertex}, we consider
the system of FRG flow equations including vertex corrections. 
We explicitly calculate the effect of vertex
corrections on the value of the fermionic anomalous dimension,
$\eta_{\psi}$, 
to leading order in the small parameter $z_b -2$.
In the concluding section~\ref{sec:conclusions}, we summarize our results and
discuss some open problems.
We have added two appendices with more technical details.
In Appendix~A we derive skeleton equations relating
the purely bosonic two-point and three-point functions to the 
fermionic propagators and irreducible vertices.  These skeleton equations
are useful to close the infinite hierarchy of FRG flow equations.
Finally, in Appendix B, we explicitly evaluate the symmetrized fermionic loop with three
external bosonic legs (the symmetrized three-loop vertex) for our model system.

\section{Definition of the model}
\label{sec:model}

\begin{figure}[tb]    
   \centering
\vspace{7mm}
  \includegraphics[width=0.25\textwidth]{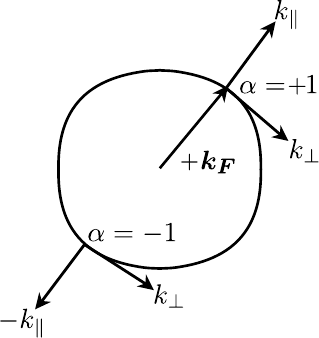}
  \caption{%
The two-patch model considered in this work
involves only two types of fermion fields with momenta close
to two opposite patches on the Fermi surface centered at
$\pm \bd{k}_F$.
The fermionic momenta are measured locally with respect to
$\pm \bd{k}_F$. We define
$k_\|$ as the  component of the momentum parallel to
the local Fermi surface normal, and $k_\bot$ as the component orthogonal to the surface normal. 
}
    \label{fig:patch}
  \end{figure}

{ 
We are interested in a minimal model describing the 
coupling of electrons 
in the proximity of a two-dimensional Fermi surface
to a gapless scalar Bose field. In particular, the 
bosonic field can describe the fluctuations of a scalar order parameter  
near the onset of a metallic Ising-nematic phase, such as 
a $d$-wave nematic state in a two-dimensional square lattice,
where the point-group symmetry of the lattice is reduced from square to rectangular.
However, the gapless scalar Bose field can also describe
the fluctuations of an emergent gauge field minimally coupled
to a two-dimensional Fermi surface. For example, this can be physically realized
in a system where a spin-liquid phase is described in terms of fermionic degrees
of freedom (spinons), thereby causing the emergence of a $U(1)$ gauge symmetry in the 
system: The critical fluctuations near the onset of the spin-liquid phase
are then described by the coupling of the spinon Fermi surface to the 
corresponding $U(1)$ gauge field.
The general form of the action for our model can be written as 
$S  =  S_{\psi} + S_{\phi} + S_{\rm int} $, with
 \begin{eqnarray}
 S_{\psi} & = & - \int_K \sum_{ \sigma}
 ( i \omega - \xi_{\bd{k}} ) \bar{\psi}_{K \sigma}
 \psi_{ K \sigma }, \label{eq:Spsi}
 \\
 S_{\phi} & = & \frac{1}{2} \int_{ \bar{K}} \left( \rho_0 + \tilde{\nu}_0 \bar{\bd{k}}^2 \right) 
 \phi_{ - \bar{K}} \phi_{\bar{K}},
 \\
 S_{\rm int} & = & 
 \int_{\bar{K}}  
 \hat{O}_{ \bar{K}}{[\bar{\psi},{\psi}]} \phi_{ - \bar{K} } , \label{eq:Sint}
 \end{eqnarray}
where $\psi$ and $\phi$ denote two-dimensional Fermi
and Bose fields, respectively, and $\hat{O}{[\bar{\psi},{\psi}]}$
is a bilinear operator in the fermion fields which has the same symmetry
as the order parameter field $\phi$.
In Eqs.~(\ref{eq:Spsi})--(\ref{eq:Sint}),
} 
$K = ( i \omega , \bd{k} )$ denotes fermionic Matsubara frequency and two-dimensional momentum,
while $\bar{K} = ( i \bar{\omega} , \bar{\bd{k}})$ denotes
the corresponding bosonic quantities. 
The integration symbols are defined by
$\int_K = ( \beta V )^{-1} \sum_{\omega} \sum_{\bd{k}}$,  and similarly
for the bosonic quantities, where $\beta$ is the inverse temperature and $V$ is the volume.
Throughout this work it is understood that we eventually take the
zero temperature limit ($\beta \rightarrow \infty$) and the
infinite volume limit ($V \rightarrow \infty$).
The index $\sigma = 1,  \ldots , N$ labels $N$ different flavors of the fermion field.
{
The fermionic energy dispersion $\xi_{\bd{k}}$ is defined relative to
the Fermi energy $\epsilon_{ F}$, i.e. $\xi_{\bd{k}} = 
\epsilon_{ \bd{k}} - \epsilon_{ F}$,
while, in the bosonic dispersion, $\rho_0$ plays the role 
of a mass (or gap) term which measures the distance to the quantum critical point:
At the quantum critical point, 
$\rho_0 = 0$, such that order parameter fluctuations become gapless.
The absence of higher-order terms in $\phi$ and gradients of $\phi$ in the 
action defining our model can be justified by a dimensional analysis, which shows 
that such higher-order terms become irrelevant at the critical point.

In the general form given in Eqs.~(\ref{eq:Spsi})--(\ref{eq:Sint}), 
the action of our model 
is still  too complicated to be treated
analytically with renormalization group or field-theoretical methods.
However, as pointed out by Metlitski and Sachdev\cite{Metlitski10a} 
(see also Ref.~\onlinecite{SachdevBook2nd}),
the relevant critical fluctuations can be described
by a simplified minimal action involving only fermion
fields with momenta close to two opposite patches on the Fermi surface.
The reason is that the most singular scattering processes 
mediated by a given bosonic mode with momentum $\bar{\bd{k}}$
involve only fermions lying on patches of the Fermi surface which are almost tangential
to the bosonic momentum $\bar{\bd{k}}$. The situation is shown 
graphically in Fig.~\ref{fig:patch}, where the label $\alpha = \pm 1$
denotes the two patches of the Fermi surface which are tangential to
a given bosonic mode with momentum parallel to $\bd{k}_\perp$ in the figure.
In order to describe the singular behavior of the fermionic and bosonic
Green functions at the critical point we can therefore restrict the general model 
involving fermions on the whole Fermi surface 
to a so-called two-patch model, characterized by the following Euclidean action:
}
\begin{equation}
 S_{\rm patches} [ \bar{\psi} , \psi , \phi ] =  S_0 [ \bar{\psi} , \psi ] + S_0 [ \phi ] +
 S_1 [  \bar{\psi} , \psi , \phi ], 
\end{equation}
 \begin{eqnarray}
 S_0 [ \bar{\psi} , \psi ] & = & - \int_K \sum_{\alpha, \sigma}
 ( i \omega - \xi^{\alpha}_{\bd{k}} ) \bar{\psi}^{\alpha}_{K \sigma}
 \psi^{\alpha}_{ K \sigma },
 \\
 S_0 [ \phi ] & = & \frac{1}{2} \int_{ \bar{K}} f^{-1}_{ \bar{\bd{k}} } 
 \phi_{ - \bar{K}} \phi_{\bar{K}},
 \\
 S_1 [  \bar{\psi} , \psi , \phi ] & = & 
 \int_{ K_1} \int_{ K_2} \int_{ \bar{K}_3} \sum_{ \alpha, \sigma }
 \delta_{ K_1 , K_2 + \bar{K}_3}
 \nonumber
 \\
 & & \times  \Gamma_0^{\bar{\psi}^{\alpha} \psi^{\alpha} \phi } ( K_1 ; K_2 ; \bar{K}_3 ) 
 \bar{\psi}_{ K_1 \sigma}^{\alpha} \psi^{\alpha}_{K_2 \sigma}
 \phi_{ \bar{K}_3 }.
 \hspace{7mm}
 \end{eqnarray} 
{
In the above expressions the fermion fields are now
characterized by an additional
upper index 
$\alpha = \pm 1$ labeling the two patches on the Fermi surface 
centered at the two opposite momenta 
$\bd{k}_F^{\alpha} = \alpha {\bd{k}}_F$, as shown in Fig.~\ref{fig:patch}. 
By construction, the momenta of the fields characterizing the two-patch model
are intended to lie in the vicinity of the  Fermi
momenta $\bd{k}_F^{\alpha}$, so that $|k_\parallel|$ and $|k_\perp|$
(the momenta relative to the Fermi momenta, as shown in Fig.~\ref{fig:patch}) 
should be much smaller than $|{\bd{k}}_F|$.
In particular, one should impose a cutoff, $\Lambda_\perp \sim k_F\Delta\theta$ 
on the momenta perpendicular to the Fermi surface normal, 
where $\Delta\theta$ is the angular extension of the patch. 
However, as long as integrals over such momenta turn out to be ultraviolet convergent, 
we can effectively send this cutoff to infinity without affecting the low-energy 
critical behavior 
of our theory.
} 
%
%
%
%
%
%

The energy dispersion relative to the true Fermi energy at patch $\alpha  $ is
assumed to be of the form
 \begin{equation}
  \xi^{\alpha}_{\bd{k}} = \epsilon_{ \bd{k}_F^{\alpha} +
 \bd{k}} - \epsilon_{ \bd{k}_F^{\alpha}} 
{ = v_F^\alpha k_{\parallel} + \frac{k_{\bot}^2}{2m} } 
 = \alpha v_F k_{\parallel} + \frac{ k_{\bot}^2}{2m}, \label{eq:ferm_dispers}
 \end{equation}
where $v_F$ is the Fermi velocity,
$k_{\parallel}$ is the component of $\bd{k}$ parallel to the local
normal to the Fermi surface, and $k_{\bot}$ is perpendicular
to the local Fermi surface normal. 
{
In the last equality of Eq.~(\ref{eq:ferm_dispers}) we have 
used the fact that the Fermi velocity has opposite sign at the two patches. 
}
We normalize the bosonic field $\phi_{\bar{K}}$ such 
that the bare 
{ 
fermion-boson interaction
}
vertex
$ \Gamma_0^{\bar{\psi}^{\alpha} \psi^{\alpha} \phi } ( K_1 ; K_2 ; \bar{K}_3 )$
is unity for the model describing the nematic quantum phase transition 
(which we discuss in detail below),
and assumes the values $\alpha = \pm 1$ for the gauge field model,
 i.e.,
 \begin{equation}
 \Gamma_0^{\bar{\psi}^{\alpha} \psi^{\alpha} \phi } ( K_1 ; K_2 ; \bar{K}_3 ) 
= \Gamma_0^{\alpha} = \left\{ \begin{array}{cl} 1 & \mbox{(nematic model),}
 \\
 \alpha & \mbox{(gauge model).}
 \end{array}
 \right.
 \label{eq:Gamma0}
 \end{equation}
Finally,  the 
{
coefficient of the quadratic term
in the bosonic part of the action 
}
is assumed to be
of the form
 \begin{equation}
  f^{-1}_{ \bar{\bd{k}} } = \rho_0  
  + \nu_0 \left( \frac{ | \bar{k}_{\bot} | }{2 mv_F} \right)^{ z_b -1 },
 \end{equation}
where $\rho_0$ and $\nu_0$ are dimensionful constants
with units of mass. (Recall that in two dimensions,
the density of states also has units of mass.)
We note that, 
for a nonspherical Fermi surface, $mv_F$ does not necessarily equal the 
Fermi momentum.
For simplicity, in this work, we do not keep track of the renormalization of
$\rho_0$ by fluctuations, so that we may set $\rho_0=0$  to describe
the quantum critical point.
We assume that the bosonic dynamic exponent 
$z_b$ is in the range $2 < z_b \leq 3$.
To construct a sensible limit of large $N$, the constants $\rho_0$ and $\nu_0$ should 
be proportional to $N$. However, 
according to Mross {\it{et al.}} \cite{Mross10} the limit of large-$N$ can be safely taken only when $z_b -2$ is sent to zero simultaneously, such that the product $N (z_b -2)$ remains finite.

{ Although the FRG, unlike the field-theoretical renormalization group, does not rely on the presence of a small expansion
parameter (which indeed is not present in the considered problem), it is convenient to express the above action in terms of rescaled 
dimensionless momenta, frequencies and fields, to carry out the renormalization group procedure. 
A general discussion of proper scaling in mixed Fermi-Bose systems can be found in Ref.~\onlinecite{Yamamoto10}.
}
Given an arbitrary momentum scale $\Lambda$,
we define dimensionless fermionic labels $Q = (i\epsilon, q_{\parallel}, q_{\bot})$
by setting
 \begin{subequations}
 \begin{eqnarray}
 k_{\parallel} & = & \frac{ \Lambda^2}{2 mv_F} q_{\parallel},
 \\
 k_{\bot} & = &  \Lambda q_{\bot},
\\
 \omega & = &  2 m v_F^2  \left( \frac{ \Lambda}{ 2 mv_F } \right)^{ z_b } \epsilon
 =  \frac{\Lambda^2}{2m} \left( \frac{ \Lambda}{ 2 mv_F } \right)^{z_b -2 }  \epsilon.
 \hspace{7mm}
 \end{eqnarray}
\end{subequations}
The corresponding bosonic labels
$\bar{Q} = (i\bar{\epsilon}, \bar{q}_{\parallel}, \bar{q}_{\bot})$ are defined in 
precisely the same way:
 \begin{subequations}
 \begin{eqnarray}
 \bar{k}_{\parallel} & = & \frac{\Lambda^2}{2 mv_F} \bar{q}_{\parallel},
 \\
 \bar{k}_{\bot} & = &  \Lambda \bar{q}_{\bot},
\\
 \bar{\omega} & = &   2 m v_F^2  \left( \frac{ \Lambda}{ 2 mv_F } \right)^{ z_b } 
 \bar{\epsilon} =    \frac{\Lambda^2}{2m} \left( \frac{ \Lambda}{ 2 mv_F } \right)^{z_b -2 }  \bar{\epsilon}. 
 \hspace{7mm}
 \end{eqnarray}
\end{subequations}
Introducing the rescaled dimensionless fields
 \begin{eqnarray}
 {\psi}^{\alpha}_{ Q \sigma} & = & 4 m^2 v_F^3 
 \left( \frac{ \Lambda}{2 mv_F } \right)^{ \frac{ z_b + 5}{2}} \psi^{\alpha}_{K \sigma},
 \\
 {\phi}_ { \bar{Q} }  & = & 
  4 m^2 v_F^2 \left( \frac{ \Lambda}{2 mv_F } \right)^{  z_b + 1} \phi_{ \bar{K} },
 \end{eqnarray}
the Euclidean action of our model can 
be written as
\begin{eqnarray}
 S_0 [ \bar{\psi} , \psi ] & = & - \int_Q \sum_{\alpha, \sigma}
 ( i  \zeta_\Lambda\epsilon - {\xi}^{\alpha}_{\bd{q}} ) \bar{\psi}^{\alpha}_{Q \sigma}
 \psi^{\alpha}_{ Q \sigma },
 \hspace{12mm}
 \\
 S_0 [ \phi ] & = & \frac{1}{2  } \int_{ \bar{Q}} 
 \left( {r}_{\Lambda} + c_{0} | \bar{q}_{\bot} |^{z_b-1} \right)
 \phi_{ - \bar{Q}} \phi_{\bar{Q}},
 \\
 S_1 [  \bar{\psi} , \psi , \phi ] & = &  
 \int_{ Q_1} \int_{ Q_2} \int_{ \bar{Q}_3} \sum_{ \alpha, \sigma }
 \delta_{ Q_1 , Q_2 + \bar{Q}_3}
 \nonumber
 \\
 &  \times  & {\Gamma}_0^{ \bar{\psi}^{\alpha} \psi^{\alpha} \phi } ( Q_1 ; Q_2 ; \bar{Q}_3 ) 
 \bar{\psi}_{ Q_1 \sigma}^{\alpha} \psi^{\alpha}_{Q_2 \sigma}
 \phi_{ \bar{Q}_3 },
 \hspace{7mm}
 \end{eqnarray}
where
 \begin{subequations}
 \begin{eqnarray}
 \zeta_{\Lambda} & = & 
\left( \frac{ \Lambda}{ 2 mv_F } \right)^{  z_b -2},
 \label{eq:gammadef}
 \\
  {\xi}^{\alpha}_{\bd{q} } & = & \alpha q_{\parallel} + q_{\bot}^2 ,
\\
{r}_{\Lambda} & = &  \frac{ \rho_0 }{2m} 
 \left( \frac{\Lambda}{2 mv_F} \right)^{ 
 1 - z_b  } ,
 \\
 c_{0} & = & \frac{ \nu_0}{2m},
 \end{eqnarray}
 \end{subequations}
and the mixed fermion-boson vertex is the same as before,
 \begin{equation}
{\Gamma}_0^{ \bar{\psi}^{\alpha} \psi^{\alpha} \phi }  ( Q_1 ; Q_2 ; \bar{Q}_3 ) 
 = 
\Gamma_0^{ \bar{\psi}^{\alpha} \psi^{\alpha} \phi }   ( K_1 ; K_2 ; \bar{K}_3 )  
=  \Gamma^{\alpha}_0.
\end{equation}
If we use the expression $\nu_0 = Nm /(2 \pi )$ 
for the density of states of free fermions
in two dimensions, we have $c_0 = N/(4 \pi )$.
Consequently, setting $r_\Lambda \to 0$ to describe the quantum critical point, 
our model does not depend on any free parameters. \\
\indent Our FRG procedure will  generate also higher-order 
purely bosonic contributions to the effective action, which describe
interactions between the boson fields, mediated by the fermions.
In an expansion in powers of the fields, the lowest-order interaction 
process is cubic in the bosonic fields,
 \begin{eqnarray}
   S_3 [ \phi ] & = & \frac{1}{3 !} \int_{ \bar{K}_1 } \int_{ \bar{K}_2 }\int_{ \bar{K}_3 }
  \delta_{ \bar{K}_1 + \bar{K}_2 + \bar{K}_3 , 0 } 
\nonumber
 \\
 & & \times
\Gamma^{\phi \phi \phi } ( 
 \bar{K}_1 , \bar{K}_2 , \bar{K}_3 ) 
 \phi_{ \bar{K}_1 } \phi_{ \bar{K}_2 } \phi_{ \bar{K}_3 }
 \nonumber
 \\
 & = & \frac{1}{3 !} \int_{ \bar{Q}_1 } \int_{ \bar{Q}_2 }\int_{ \bar{Q}_3 }
  \delta_{ \bar{Q}_1 + \bar{Q}_2 + \bar{Q}_3 , 0 } 
\nonumber
 \\
 & & \times
 \tilde{\Gamma}^{\phi \phi \phi } ( 
 \bar{Q}_1 , \bar{Q}_2 , \bar{Q}_3 ) 
 \phi_{ \bar{Q}_1 } \phi_{ \bar{Q}_2 } \phi_{ \bar{Q}_3 },
\end{eqnarray}
with
 \begin{eqnarray}
 \tilde{\Gamma}^{\phi \phi \phi }   ( 
 \bar{Q}_1 , \bar{Q}_2 , \bar{Q}_3 ) &= &v_F^2 
\left( \frac{\Lambda}{2 mv_F } \right)^{ 3-z_b } 
\Gamma^{\phi \phi \phi } ( 
 \bar{K}_1 , \bar{K}_2 , \bar{K}_3 ) .
 \nonumber
 \\
 & &
 \end{eqnarray}
Although for $ z_b < 3$ this vertex 
seems to be  irrelevant by power counting,
it turns out that it has a singular dependence on the
external momenta and frequencies and therefore cannot be neglected.
Because, within our bosonic momentum-transfer cutoff scheme, all 
vertices involving only bosonic external legs are finite
at the initial renormalization group scale,\cite{Schuetz05,Kopietz10} it is crucial
to keep track of the FRG flow of the vertex
$\Gamma^{\phi \phi \phi } ( 
 \bar{K}_1 , \bar{K}_2 , \bar{K}_3 ) $.
In this work, we do this by  means of 
a skeleton equation relating $\Gamma^{\phi \phi \phi } ( 
 \bar{K}_1 , \bar{K}_2 , \bar{K}_3 ) $
to the symmetrized fermion loop with three external bosonic legs and
renormalized fermionic propagators, as discussed in Appendix~A.  An explicit evaluation
of this vertex is given in Appendix~B.


If we identify $\Lambda$ with the renormalization group flow parameter
which is reduced under the renormalization group procedure, the
canonical dimensions of all quantities explicitly appear in the FRG flow equations
with the above rescaling.
To compare the FRG results with perturbation theory,
it is more convenient, however, not to include the canonical dimensions  
into the definition of the vertices.     
Therefore, we simply choose $\Lambda = 2 mv_F$
in the above expressions, so that $\zeta_{\Lambda} \rightarrow 1$ and
$r_{\Lambda} \rightarrow r_0 = \rho_0 / (2m)$.
Renaming again $Q \rightarrow K$, our bare action is then the sum
of the following three terms,
\begin{eqnarray}
 S_0 [ \bar{\psi} , \psi ] & = & - \int_K \sum_{\alpha, \sigma}
 ( i  \ktau - \xi^{\alpha}_{\bd{k}} ) \bar{\psi}^{\alpha}_{K \sigma}
 \psi^{\alpha}_{ K \sigma },
 \label{eq:S0new}
 \\
 S_0 [ \phi ] & = & \frac{1}{2} \int_{ \bar{K}} ( r_0 + c_0 | \bar{k}_{\bot} |^{z_b -1} )
 \phi_{ - \bar{K}} \phi_{\bar{K}},
 \\
 S_1 [  \bar{\psi} , \psi , \phi ] & = & 
 \int_{ K_1} \int_{ K_2} \int_{ \bar{K}_3} \sum_{ \alpha, \sigma }
 \delta_{ K_1 , K_2 + \bar{K}_3}
 \nonumber
 \\
 &  \times & \Gamma_0^{\bar{\psi}^{\alpha} \psi^{\alpha} \phi } ( K_1 ; K_2 ; \bar{K}_3 ) 
 \bar{\psi}_{ K_1 \sigma}^{\alpha} \psi^{\alpha}_{K_2 \sigma}
 \phi_{ \bar{K}_3 },
 \hspace{7mm}
 \end{eqnarray}
where now
 \begin{equation}
 \xi^{\alpha}_{\bd{k}} = \alpha k_{\parallel} + k_{\bot}^2 , \; \
 \; \; r_0 = \frac{\rho_0}{2m} , \; \; \; 
c_0 = \frac{\nu_0}{2m}.
 \end{equation}
The corresponding Gaussian propagators are
 \begin{eqnarray}
 G_0^{\alpha} ( K ) & = &  \frac{1}{ i \ktau - \xi^{\alpha}_{\bd{k}} }
 = \frac{1}{ i \ktau - \alpha k_{\parallel} - k_{\bot}^2 },
 \\
 F_0 ( \bar{K} ) & = & f_{ \bar{\bd{k}} } = \frac{1}{r_0 + c_0 | \bar{k}_{\bot} |^{z_b -1 }}.
 \label{eq:F0new}
 \end{eqnarray}
This dimensionless parametrization of our model
is what is used in the following sections.

\section{FRG flow equations}
\label{sec:FRGfloweq}

The starting point of our calculation is the Wetterich equation \cite{Wetterich93,Berges02}
for the coupled Fermi-Bose model defined above, 
which is an exact FRG flow equation
for the generating functional $ \Gamma_{\Lambda} [ \bar{\psi} , \psi , \phi ]$ 
of the one-line irreducible vertices of
our theory. This flow equation describes the exact evolution
of $\Gamma_\Lambda[\bar{\psi},\psi,\phi]$
as some (now dimensionless) cutoff parameter $\Lambda$ is reduced.
By expanding $ \Gamma_{\Lambda} [ \bar{\psi} , \psi , \phi ]$  in powers
of the fields, we obtain an infinite hierarchy of coupled integro-differential
equations for the one-line irreducible vertices of our model.
This hierarchy is formally exact, 
but, in practice, further approximations are usually necessary in order to
obtain explicit results for the vertex functions (see Refs. \onlinecite{Kopietz10,Metzner12} for recent reviews).
Moreover, the proper choice of the cutoff scheme is also very important.

For our effective low-energy model discussed above, 
it is, in principle, possible to introduce cutoffs in 
both the bosonic  and the fermionic sectors and regularize the inverse Gaussian 
propagators as follows: 
\begin{eqnarray}
 {[} F_{ 0 , \Lambda} ( \bar{K} ) ]^{-1} & = &  f^{-1}_{ \bar{\bd{k}} } + \bar{R}_{\Lambda}
 ( \bar{{K}} ),
\\
{ [} G^{\alpha}_{ 0, \Lambda} ( K ) ]^{-1} & = & i \omega - \xi^{\alpha}_{\bd{k}} 
 - R_{\Lambda} ( K ).
 \end{eqnarray}
For the calculations in the  present problem, we find it more 
convenient to introduce a sharp momentum-transfer cutoff only in the bosonic sector. 
Using a similar cutoff procedure, two of us were able to derive the exact scaling behavior of the Tomonaga-Luttinger model within an FRG approach.\cite{Schuetz05}
We therefore set  $R_{\Lambda} ( K ) = 0$ in the fermionic sector, and choose 
\begin{equation}
  \label{eq:cutoffsharp}
  \bar{R}_{\Lambda}
 ( \bar{{K}} ) = f^{-1}_{ \bar{\bd{k}} }  \left[ \Theta^{-1} ( | \bar{k}_{\bot} | - \Lambda ) - 1 \right] 
\end{equation}
for the boson cutoff.
This leads to the cutoff-dependent bare Gaussian propagator
\begin{equation}
  \label{eq:propgatorbarebosoncutoff}
   F_{ 0 , \Lambda} ( \bar{K} )  = \Theta  ( | \bar{k}_{\bot} | - \Lambda ) f_{ \bar{\bd{k}}} ,
\end{equation}
which vanishes for $| \bar{k}_{\bot} | < \Lambda$ and equals $f_{ \bar{\bd{k}}}$ for $| \bar{k}_{\bot} |  > \Lambda$.  As there is no cutoff function in the fermionic sector, all purely fermionic loops already have non-vanishing initial values at the beginning of the flow. We see explicitly below that these are highly singular and need to be treated with care.

The exact hierarchy of FRG flow equations for the one-line irreducible
vertices of our model can be obtained as a special case of the general 
hierarchy of FRG flow equations
for mixed Bose-Fermi theories written down in Refs. \onlinecite{Schuetz05,Kopietz10}.
For our purpose, it is sufficient to 
consider a  truncation of this hierarchy
which generates, after iteration (apart from  many other diagrams), 
the important three-loop diagrams identified by Metlitski and Sachdev,\cite{Metlitski10a} 
shown in Fig.~\ref{fig:MSdiagrams}. Our truncation is characterized 
by the following three points.

\begin{itemize}

\item
On the right-hand side of 
the flow equations for the fermionic and bosonic self-energies,
retain only contributions involving irreducible vertices with three external legs.

\item Renormalize all three-legged vertices by 
triangular diagrams involving all combinations of three-legged vertices.

\item On the right-hand side of the flow equations for all three-legged vertices,
approximate the vertex with one bosonic  and two fermionic external legs
by its bare value.

\end{itemize}

Let us now explicitly give the corresponding FRG flow equations.
The fermionic self-energy $\Sigma^{\alpha} ( K )$
and bosonic self-energy $\Pi ( \bar{K} )$  satisfy the flow equations
 \begin{widetext}
\begin{eqnarray}
  \partial_{\Lambda} {\Sigma}^{\alpha} ( K ) & = &  
  \int_{\bar{K}} 
  \left[   \dot{F} ( \bar{K} ) G^{\alpha}  ( K + \bar{K} ) 
    + {F} ( \bar{K} ) \dot{G}^{\alpha}  ( K + \bar{K} ) 
  \right]
  \Gamma^{\bar{\psi}^{\alpha} \psi^{\alpha} \phi} ( K + \bar{K} ;   K ;    \bar{K}  ) 
  \Gamma^{ \bar{\psi}^{\alpha} \psi^{\alpha} \phi  } ( K; K + \bar{K} ; -  \bar{K}  )  ,
 \hspace{7mm}
  \label{eq:flowSigmatrunc}
\end{eqnarray}
\begin{eqnarray}
  \partial_{\Lambda} {\Pi} ( \bar{K} ) & = & 
 \int_K \sum_{\alpha, \sigma }
  \left[ \dot{ G}^{\alpha} ( K  ) G^{\alpha} ( K + \bar{K} ) +    { G}^{\alpha} ( K  ) 
    \dot{G}^{\alpha} ( K + \bar{K} )  \right]
 \Gamma^{\bar{\psi}^{\alpha} \psi^{\alpha} \phi} ( K + \bar{K} ;   K ;    \bar{K}  ) 
  \Gamma^{ \bar{\psi}^{\alpha} \psi^{\alpha} \phi  } ( K; K + \bar{K} ; -  \bar{K}  ) 
 \nonumber
 \\
& - &  \int_{\bar{K}^{\prime}}
   \dot{ F} ( \bar{K}^{\prime}  ) F ( \bar{K}^{\prime} + \bar{K} ) 
 \Gamma^{\phi \phi \phi} (       \bar{K} ,  \bar{K}^{\prime}  , - \bar{K}   -  \bar{K}^{\prime}  ) 
  \Gamma^{\phi \phi \phi} ( - \bar{K}, -  \bar{K}^{\prime}  ,    \bar{K} +  \bar{K}^{\prime}  )  ,
  \label{eq:flowPitrunc}
\end{eqnarray}
 \end{widetext}
which  are shown graphically in Fig.~\ref{fig:flowsigma} for a general cutoff scheme.
Here  the scale-dependent bosonic and fermionic  propagators are
 \begin{eqnarray}
 F ( \bar{K} ) & = & \frac{1}{  [ F_{ 0, \Lambda} ( \bar{K} ) ]^{-1} + \Pi ( \bar{K} ) },
 \\
 G^{\alpha} ( K ) & = & \frac{1}{ 
 [ G^{\alpha}_{0 , \Lambda} ( K )  ]^{-1} - \Sigma^{\alpha} ( K ) },
 \end{eqnarray}
while the corresponding single-scale propagators are 
 \begin{eqnarray}
 \dot{F} ( \bar{K} ) & =  & - F^2 ( \bar{K} ) \partial_{\Lambda} 
 [ F_{ 0, \Lambda} ( \bar{K} ) ]^{-1}  ,
 \\
 \dot{G}^{\alpha} ( K ) & = & - [ G^{\alpha} ( K ) ]^2 
 \partial_{\Lambda} [ G^{\alpha}_{0 , \Lambda} ( K )  ]^{-1}.
 \end{eqnarray}
Note that in the momentum-transfer cutoff scheme $\dot{G}^{\alpha} ( K ) =0$ such 
that we should omit all diagrams involving fermionic single-scale propagators. 
With a sharp cutoff in the bosonic transverse momentum, the full bosonic 
propagator is
 \begin{equation}
 F_{\Lambda} ( \bar{K} ) = \frac{ \Theta ( | \bar{k}_{\bot} | - \Lambda )}{
 r_0 + c_0 | \bar{k}_{\bot} |^{z_b -1} +  \Theta ( | \bar{k}_{\bot} | - \Lambda )
 \Pi_{\Lambda} ( \bar{K} ) },
 \end{equation}
while the corresponding single-scale propagator is given by
 \begin{equation}
 \dot{F}_{\Lambda} ( \bar{K} ) = - \frac{ \delta ( | \bar{k}_{\bot} | - \Lambda )}{
 r_0 + c_0 \Lambda^{z_b -1} + \Pi_{\Lambda} ( \bar{K} ) }.
 \label{eq:Fdot}
 \end{equation}

The right-hand sides of Eqs.~(\ref{eq:flowSigmatrunc}) and (\ref{eq:flowPitrunc}) also 
depend on the one-line irreducible three-point vertex with two fermionic and one bosonic
external legs,
$\Gamma^{\bar{\psi}^{\alpha} \psi^{\alpha} \phi} ( K + \bar{K} ;   K ;    \bar{K}  ) $,
and on the  one-line irreducible three-point vertex with three bosonic external legs,
$ \Gamma^{\phi \phi \phi} (       \bar{K} ,  \bar{K}^{\prime}  , - \bar{K}   -  \bar{K}^{\prime}  ) $, where the superscripts indicate the fields associated with the energy-momentum labels.
Within our truncation, the flow of $\Gamma^{\bar{\psi}^\alpha\psi^\alpha\phi}$ is determined by the following flow equation,
 \begin{widetext}
\begin{eqnarray}
  \partial_{\Lambda} {\Gamma}^{\bar{\psi}^{\alpha} \psi^{\alpha} \phi } 
 ( K + \bar{K}    ; K  ; \bar{K}   ) & = &    \int_{\bar{K}^{\prime}} 
  \Bigl[ \dot{ F} ( \bar{K}^{\prime} ) G^{\alpha} ( K + \bar{K}^{\prime} )  
 G^{\alpha} ( K +  \bar{K} + \bar{K}^{\prime} ) 
+
  { F} ( \bar{K}^{\prime} ) \dot{G}^{\alpha} ( K + \bar{K}^{\prime} )  G^{\alpha} ( K +  \bar{K} + \bar{K}^{\prime} ) 
  \nonumber
  \\
  & &  
\hspace{5mm}  +
  { F} ( \bar{K}^{\prime} ) G^{\alpha} ( K + \bar{K}^{\prime} )  \dot{G}^{\alpha} ( K +  \bar{K} + \bar{K}^{\prime} ) 
  \Bigr] 
  \Gamma^{\bar{\psi}^{\alpha} \psi^{\alpha} \phi } ( K + \bar{K}^{\prime}  ; 
 K    ;  \bar{K}^{\prime}  ) 
  \nonumber 
  \\
  &  & 
 \hspace{5mm}
\times
\Gamma^{ \bar{\psi}^{\alpha} \psi^{\alpha} \phi } ( K + \bar{K}  ; 
 K + \bar{K} + \bar{K}^{\prime}  ; - \bar{K}^{\prime}  )
  \Gamma^{ \bar{\psi}^{\alpha} \psi^{\alpha} \phi } ( K + \bar{K} + \bar{K}^{\prime} 
  ; K + \bar{K}^{\prime}   ;  \bar{K}  )
 \nonumber
 \\
 & - &
 \int_{\bar{K}^{\prime}} 
  \Bigl[ \dot{ F} ( \bar{K}^{\prime} ) 
 F ( \bar{K} + \bar{K}^{\prime} ) 
G^{\alpha} ( K + \bar{K} + \bar{K}^{\prime} )  
+
 { F} ( \bar{K}^{\prime} ) 
 \dot{F} ( \bar{K} + \bar{K}^{\prime} ) 
G^{\alpha} ( K + \bar{K} + \bar{K}^{\prime} ) 
  \nonumber
  \\
  & &  
\hspace{5mm}  +
  { F} ( \bar{K}^{\prime} ) 
 {F} ( \bar{K} + \bar{K}^{\prime} ) 
\dot{G}^{\alpha} ( K + \bar{K} + \bar{K}^{\prime} ) 
\Bigr] 
\Gamma^{ \phi \phi \phi } (  - \bar{K}   - \bar{K}^{\prime} , \bar{K}, 
\bar{K}^{\prime}  )
  \nonumber 
  \\
  &  & 
 \hspace{5mm}
\times
  \Gamma^{ \bar{\psi}^{\alpha} \psi^{\alpha} \phi } ( K + \bar{K}
  ; K + \bar{K} + \bar{K}^{\prime}   ;  - \bar{K}^{\prime} )
  \Gamma^{\bar{\psi}^{\alpha} \psi^{\alpha} \phi } ( K +  \bar{K} +
\bar{K}^{\prime}  ; 
 K    ;  \bar{K} + \bar{K}^{\prime}  ) .
  \label{eq:flowGammatrunc}
\end{eqnarray}
  \begin{figure}[tb]    
   \centering
  \includegraphics[width=0.45\textwidth]{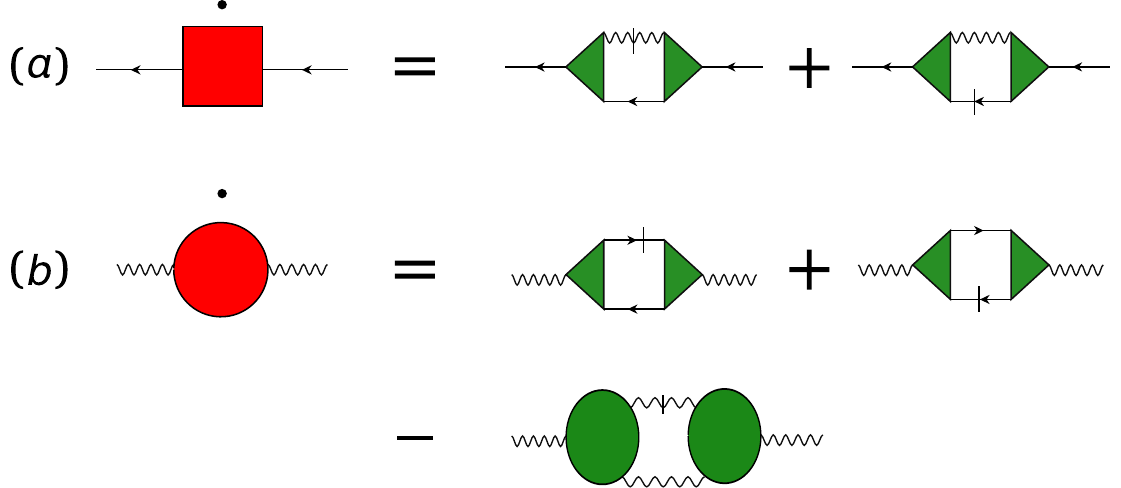}
  \caption{%
(Color online) 
(a) Graphical representation of the  FRG flow equation
(\ref{eq:flowSigmatrunc}) for the fermionic self-energy
$\Sigma^{\alpha} ( K )$, which is represented by a shaded rectangle with
one incoming arrow associated with $\psi^{\alpha}_{K \sigma}$, and
one outgoing arrow associated with $\bar{\psi}^{\alpha}_{ K \sigma}$.
The shaded triangles represent the three-legged vertex
$  \Gamma^{ \bar{\psi}^{\alpha} \psi^{\alpha} \phi  } ( K+\bar{K}; K ; \bar{K} )$
with two fermionic and one bosonic external legs.
The boson propagator is represented by a wavy line.
The black dot above the rectangle denotes a derivative with  respect to the
flow parameter, while the slashes attached to the propagators
on the right-hand side denote the corresponding single-scale propagators.
(b) Graphical representation of the FRG flow equation
(\ref{eq:flowPitrunc})
for the bosonic self-energy $\Pi ( \bar{K} )$.
The shaded circles on the right-hand side represent
the symmetrized bosonic three-point vertex. Note that, in the momentum-transfer cutoff scheme, all diagrams with a slash on internal fermionic propagators should be omitted.
}
    \label{fig:flowsigma}
  \end{figure}
A graphical representation of this flow equation is shown in Fig.~\ref{fig:flowGamma21}. 
In the momentum-transfer cutoff scheme, we should omit, again, all terms 
involving the fermionic single-scale propagator. Finally, the flow equation for the symmetrized bosonic three-legged vertex is

\begin{eqnarray}
 &&  \partial_{\Lambda} {\Gamma}^{\phi \phi \phi  } 
 ( \bar{K}_1    , \bar{K}_2  ,  - \bar{K}_1 - \bar{K}_2   )  
\nonumber
 \\
& = &    \int_{\bar{K}} 
 \bigl[
\dot{F} ( \bar{K} ) F ( \bar{K} - \bar{K}_1 ) F ( \bar{K} + \bar{K}_2 )
 +
{F} ( \bar{K} ) \dot{F} ( \bar{K} - \bar{K}_1 ) F ( \bar{K} + \bar{K}_2 )
+
{F} ( \bar{K} ) F ( \bar{K} - \bar{K}_1 ) \dot{F} ( \bar{K} + \bar{K}_2 )
\bigr] 
\nonumber
 \\
 & & \hspace{5mm}
  \times
\Gamma^{\phi \phi \phi} ( \bar{K}_1 , \bar{K} - \bar{K}_1 , - \bar{K} )
 \Gamma^{\phi \phi \phi } ( \bar{K}_2 , - \bar{K} - \bar{K}_2 ,  \bar{K} )
\Gamma^{\phi \phi \phi } ( - \bar{K}_1 - \bar{K}_2  ,  - \bar{K} + \bar{K}_1 ,  \bar{K} 
 +  \bar{K}_2  )
\nonumber
\\
 & + & \int_K \sum_{\alpha, \sigma}
 \Bigl\{ \bigl[
 \dot{G}^{\alpha} ( {K} ) G^{\alpha} ( {K} + \bar{K}_1 ) 
 G^{\alpha} ( {K} + \bar{K}_1 +   \bar{K}_2 ) 
 +
 {G}^{\alpha} ( {K} ) \dot{G}^{\alpha} ( {K} + \bar{K}_1 ) 
 G^{\alpha} ( {K} + \bar{K}_1 +   \bar{K}_2 ) 
 \nonumber
 \\
 & &
\hspace{13mm}
 +
 {G}^{\alpha} ( {K} ) G^{\alpha} ( {K} + \bar{K}_1 ) 
 \dot{G}^{\alpha} ( {K} + \bar{K}_1 +   \bar{K}_2 ) 
 \bigr]
 \Gamma^{\bar{\psi}^{\alpha} \psi^{\alpha} \phi}  ( K + \bar{K}_1 ; K; \bar{K}_1 )
 \Gamma^{\bar{\psi}^{\alpha} \psi^{\alpha} \phi}  ( K + \bar{K}_1 + \bar{K}_2  ; 
K + \bar{K}_1 ; \bar{K}_2 )
 \nonumber
 \\
 & &  \hspace{16mm} \times 
\Gamma^{\bar{\psi}^{\alpha} \psi^{\alpha} \phi}  ( K ; K + \bar{K}_1 + \bar{K}_2  ;
 -  \bar{K}_1 - \bar{K}_2 )
 + ( \bar{K}_1 \leftrightarrow \bar{K}_2 ) \Bigr\} ,
 \label{eq:flowGamma3}
 \end{eqnarray}
which is shown graphically in Fig.~\ref{fig:flowGamma3}.
  \begin{figure}[tb]    
   \centering
\vspace{7mm}
  \includegraphics[width=0.65\textwidth]{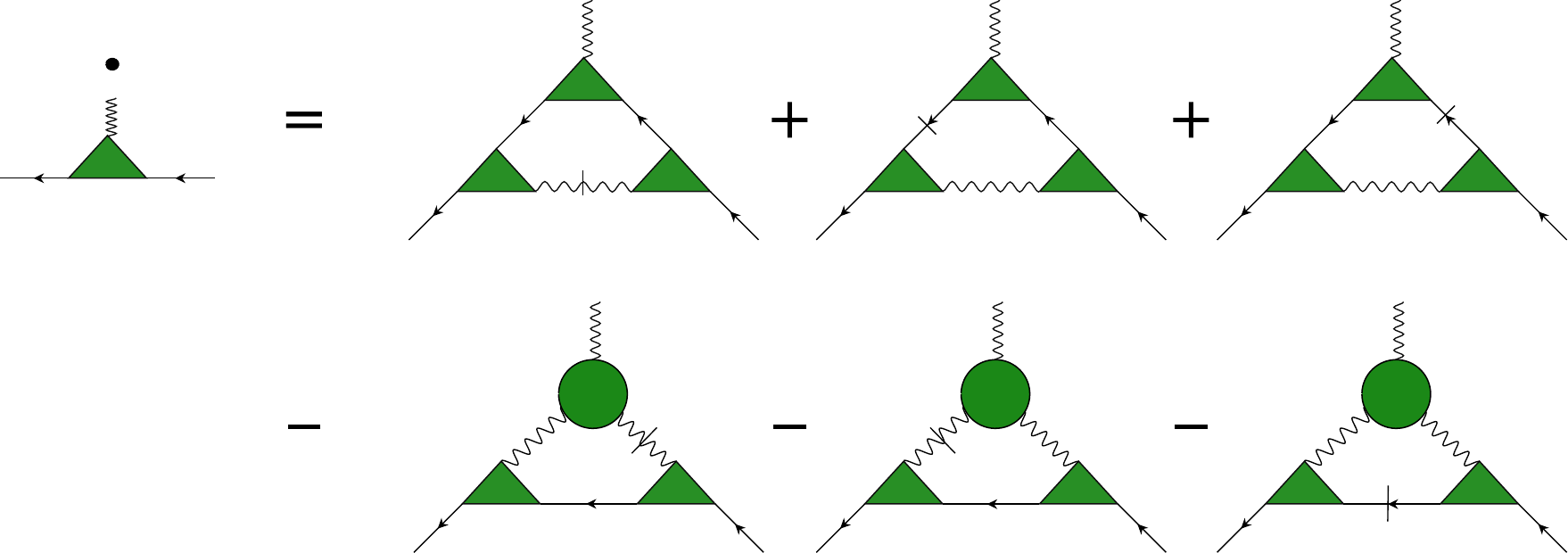}
  \caption{%
(Color online) 
Graphical representation of the  FRG flow equation
(\ref{eq:flowGammatrunc}) for the  three-legged vertex
$  \Gamma^{ \bar{\psi}^{\alpha} \psi^{\alpha} \phi  } ( K; K^{\prime} ; \bar{K} )$
with two fermionic and one bosonic external legs. 
}
    \label{fig:flowGamma21}
  \end{figure}
  \begin{figure}[tb]    
   \centering
\vspace{7mm}
  \includegraphics[width=0.65\textwidth]{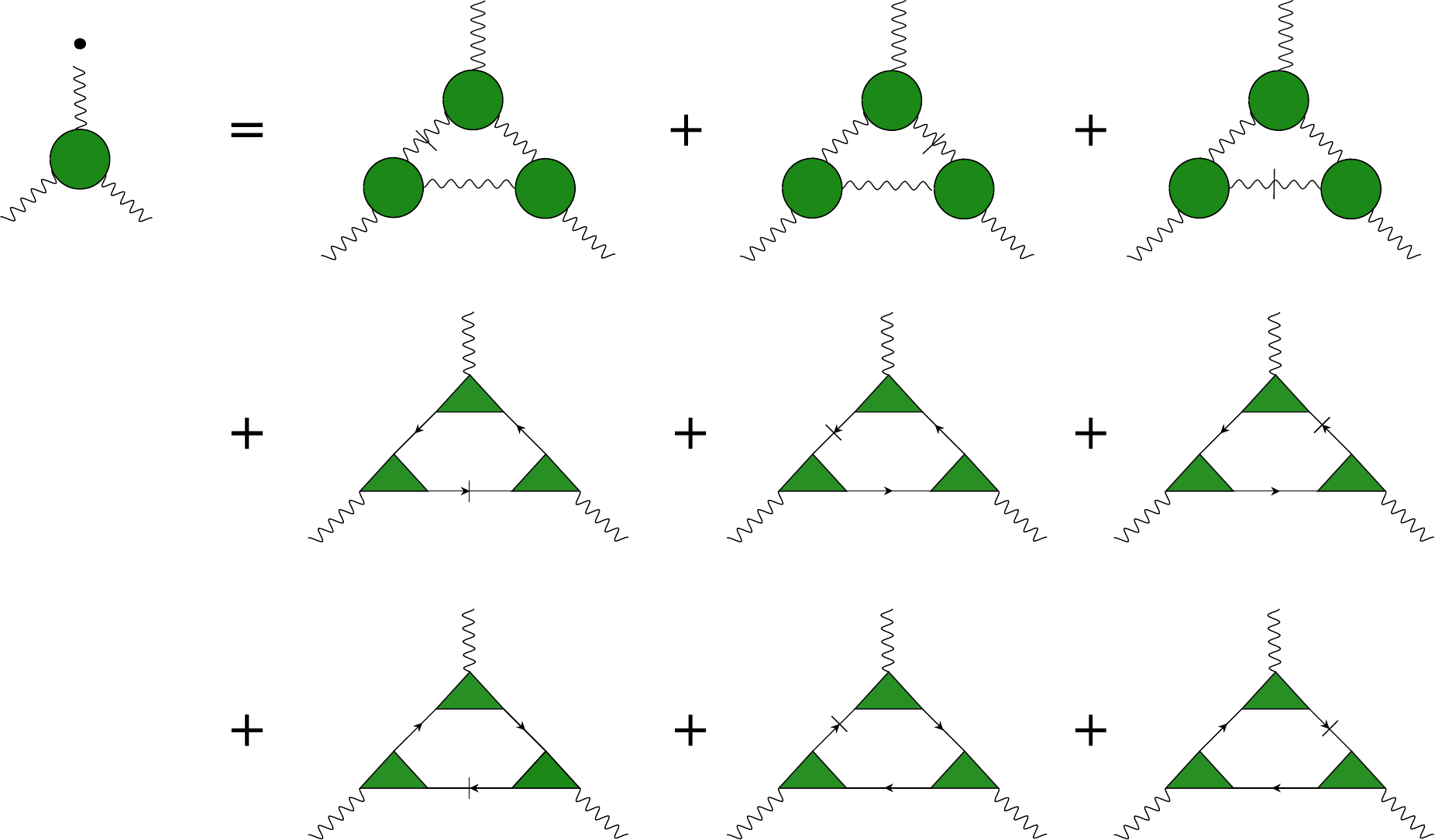}
  \caption{%
(Color online) 
Graphical representation of the  FRG flow equation
(\ref{eq:flowGamma3}) for the  three-legged boson vertex
$  \Gamma^{ \phi \phi  \phi  } ( \bar{K}_1 , \bar{K}_2 , \bar{K}_3 )$.
}
    \label{fig:flowGamma3}
  \end{figure}
\end{widetext}
If we ignore the vertex with three bosonic external legs, the 
above system of FRG flow equations
has  already been written down in Ref.~\onlinecite{Schuetz05} 
(see also Ref.~\onlinecite{Kopietz10}). 
Although the above flow equations involve only one-loop integrations,
the iterative solution of these equations generates higher-loop diagrams.
In particular, the  singular three-loop diagrams shown in Fig.~\ref{fig:MSdiagrams}
are generated as follows:
The Aslamazov-Larkin type of contribution to the
bosonic self-energy shown in 
Fig.~\ref{fig:MSdiagrams}\,(b) is generated 
by the last six diagrams on the right-hand side of the
flow equation for
$\Gamma^{ \phi \phi  \phi  } ( \bar{K}_1 , \bar{K}_2 , \bar{K}_3 )$
 shown in Fig.~\ref{fig:flowGamma3} after integrating this flow equation
over the flow parameter and substituting the result 
into the right-hand side of the flow equation for 
$\Pi ( \bar{K} )$ shown in Fig.~\ref{fig:flowsigma}\,(b).
The singular fermionic three-loop diagram in Fig.~\ref{fig:MSdiagrams}\,(a) 
is generated by substituting the same contribution to
 $\Gamma^{ \phi \phi  \phi  } ( \bar{K}_1 , \bar{K}_2 , \bar{K}_3 )$
into the flow equation for
$  \Gamma^{ \bar{\psi}^{\alpha} \psi^{\alpha} \phi  } ( K; K^{\prime} ; \bar{K} )$
shown in Fig.~\ref{fig:flowGamma21}. 
After integrating the resulting flow equation again over $\Lambda$
and substituting the resulting vertex correction
into the right-hand side of the flow equation for 
$\Sigma^{\alpha} ( K )$ shown in Fig.~\ref{fig:flowsigma}\,(a),
we generate the Metlitski-Sachdev diagrams shown in Fig.~\ref{fig:MSdiagrams}\,(a). In fact, there are even further contributions, e.g.\ using the renormalized vertex $\Gamma^{\bar{\psi}^\alpha\psi^\alpha\phi}\left(K;K^\prime;\bar{K}\right)$ containing the vertex $\Gamma^{\phi\phi\phi}( \bar{K}_1 , \bar{K}_2 , \bar{K}_3 )$ as depicted in the second line of Fig.~\ref{fig:flowGamma21} also gives rise to an Aslamazov-Larkin contribution when substituted into the diagram on the right-hand side of Fig.~\ref{fig:flowsigma}\,(b).

Although Eqs.~(\ref{eq:flowSigmatrunc})--(\ref{eq:flowGamma3}) form a
closed system of
integro-differential equations for the two-point and three-point functions of our model,
a direct numerical solution of these equations seems to be prohibitively
difficult, so that further approximations are necessary. 
As a first simplification, we 
use, below, truncated skeleton equations instead of the 
FRG flow equations (\ref{eq:flowPitrunc}) and (\ref{eq:flowGamma3})
to determine the bosonic self-energy $\Pi ( \bar{K})$ 
and three-point vertex
$ \Gamma^{ \phi \phi  \phi  } ( \bar{K}_1 , \bar{K}_2 , \bar{K}_3 )$.
As discussed in Appendix A, the Dyson-Schwinger equations of motion 
imply exact skeleton equations relating  $\Pi ( \bar{K})$  and
$ \Gamma^{ \phi \phi  \phi  } ( \bar{K}_1 , \bar{K}_2 , \bar{K}_3 )$
to the fermionic propagators, the three-point vertex 
$  \Gamma^{ \bar{\psi}^{\alpha} \psi^{\alpha} \phi  } ( K; K^{\prime} ; \bar{K} )$ with two fermionic
and one bosonic external legs
and the mixed four-point vertex $\Gamma^{\bar{\psi}\psi\phi\phi}( K; K^\prime ; \bar{K}_1; \bar{K}_2 )$ 
(see Eqs.~(\ref{eq:skeletonPi}) and (\ref{eq:SkeletonThreeLegged})).

\section{Truncation without vertex corrections}
\label{sec:novertexcorrections}

\begin{figure}[tb]    
   \centering
\vspace{7mm}
  \includegraphics[width=0.45\textwidth]{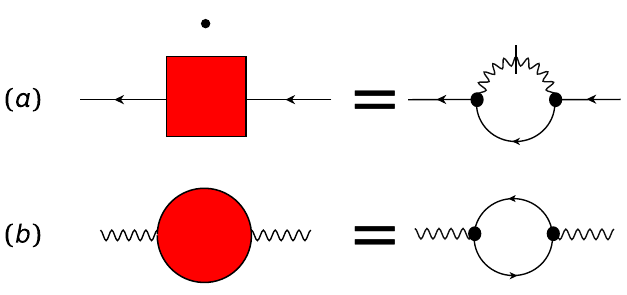}
  \caption{%
(Color online) 
(a) 
Graphical representation of the  FRG flow equation (\ref{eq:selfflow1loop})
for the fermionic self-energy 
in the momentum-transfer cutoff scheme in the simplest approximation where
the three-legged vertex is approximated by its  bare value (represented by a black dot).
(b) Truncated skeleton equation (\ref{eq:Pidef}) for the bosonic self-energy.
}
    \label{fig:flowoneloop}
  \end{figure}

In this section, we show how the known one-loop results for the self-energies
can be obtained within our FRG approach if we neglect vertex corrections.
Keeping in mind that, in the momentum-transfer cutoff scheme, we do not introduce any cutoff
in the fermionic sector, 
the scale-dependent fermionic propagator is 
 \begin{equation}
 G^{\alpha}_{\Lambda} ( K ) = 
\frac{1}{ i \ktau - \xi^{\alpha}_{\bd{k}} - [ \Sigma^{\alpha}_{\Lambda} ( K ) - 
\Sigma^{\alpha}_{\Lambda} (0 )]},
 \label{eq:Gdef}
 \end{equation}
where  $ \Sigma_{\Lambda}^{\alpha} (0) $ is the self-energy at 
the renormalized flowing Fermi surface. 
The subtraction of $ \Sigma_{\Lambda}^{\alpha} (0) $ is necessary 
because, by assumption,
we have expanded the wave vector at the renormalized Fermi surface 
of the underlying model.
Given the cutoff-dependent  
self-energy $\Sigma_{\Lambda} ( K )$, we may define the Fermi surface
for a given value of the cutoff parameter $\Lambda$ via
 \begin{equation}
 \epsilon_{\bd{k}_F} + \Sigma_{\Lambda} (  i0  , \bd{k}_F^{\alpha}   ) = \mu,
 \end{equation}
which, for $\Lambda \rightarrow 0$, reduces to the
definition of the true Fermi surface.
Hence, we may write
 \begin{equation}
 \epsilon_{\bd{k}_F^{\alpha}  + \bd{k}} - \mu
 = \epsilon_{\bd{k}_F^{\alpha}  + \bd{k}} -\epsilon_{\bd{k}_F^{\alpha}  }
 - \Sigma_{\Lambda} (  i0 , \bd{k}_F^{\alpha}   )
 = \xi^{\alpha}_{\bd{k}}  - \Sigma^{\alpha}_{\Lambda} ( 0    ).
 \end{equation}
Following 
Refs.~\onlinecite{Bartosch09a,Bartosch09b,Isidori10}, we now use the skeleton equation
(\ref{eq:skeletonPi})
to determine the flowing bosonic self-energy.
Using the fact that $ (\Gamma_0^{\alpha} )^2 =1$, the scale-dependent
bosonic self-energy is thus given by
 \begin{equation}
 \Pi_{\Lambda} ( \bar{K} ) 
= \frac{1}{2} \int_K \sum_{\alpha , \sigma} 
G^{\alpha}_{\Lambda} ( K ) \left[ G^{\alpha}_{\Lambda} ( K + \bar{K} )
 +  G^{\alpha}_{\Lambda} ( K - \bar{K} )   \right],
 \label{eq:Pidef}
 \end{equation}
while the fermionic self-energy satisfies the FRG flow equation
 \begin{equation} 
 \partial_{\Lambda} \Sigma^{\alpha}_{\Lambda} ( K ) = 
 \int_{ \bar{K}} \dot{F}_{\Lambda} ( \bar{K} ) G^{\alpha}_{\Lambda} ( \bar{K} + K ).
 \label{eq:selfflow1loop}
 \end{equation}
A graphical representation of Eqs.~(\ref{eq:Pidef}) and (\ref{eq:selfflow1loop})
is shown in Fig.~\ref{fig:flowoneloop}.
Actually, the integral on the right-hand side of Eq.~(\ref{eq:selfflow1loop})
is ultraviolet divergent
for our model, but since we do not keep track of the shape of the renormalized
Fermi surface, we can instead consider
the subtracted self-energy,
 \begin{equation}
 \Delta^{\alpha}_{\Lambda} ( K ) =  \Sigma^{\alpha}_{\Lambda} ( K ) 
- \Sigma^{\alpha}_{\Lambda} (0 ),
 \end{equation}
which appears in our cutoff-dependent fermionic 
propagator (\ref{eq:Gdef}).
The subtracted self-energy then satisfies the FRG flow equation
\begin{equation} 
 \partial_{\Lambda} {\Delta}^{\alpha}_{\Lambda} ( K ) = 
 \int_{ \bar{K}} \dot{F}_{\Lambda} ( \bar{K} ) 
 \left[ G^{\alpha}_{\Lambda} ( \bar{K} + K ) - G^{\alpha}_{\Lambda} ( \bar{K}  ) \right]  .
 \label{eq:sigmaflow}
 \end{equation}

To obtain an approximate solution of this flow equation, we expand the 
subtracted self-energy for small momenta and frequencies,
 \begin{equation}
 {\Delta}^{\alpha}_{\Lambda} ( K ) = - (Z_{\Lambda}^{-1} -1 ) i \ktau  + 
  (\tilde{Z}_{\Lambda}^{-1} -1 ) \xi^{\alpha}_{\bd{k}} + \ldots,
 \label{eq:selfexp}
 \end{equation}
 where we have used the fact that the $\bd{k}$-dependence of the self-energy
appears only in the combination $\xi^{\alpha}_{\bd{k}}$.
Because the cutoff, $\Lambda$, regularizes the singularity of the bare interaction, 
the self-energy is analytic for small momenta and frequencies
and can hence be expanded into a Taylor series.
The corresponding low-energy approximation for the fermion propagator is
 \begin{equation}
 G^{\alpha}_{\Lambda} ( K ) = \frac{1}{ Z_{\Lambda}^{-1} i \ktau - 
 \tilde{Z}_{\Lambda}^{-1} \xi^{\alpha}_{\bd{k}} } = \frac{Z_{\Lambda}}{ i \ktau - (Z_{\Lambda}/\tilde{Z}_{\Lambda}) 
 \xi^{\alpha}_{\bd{k}} } .
 \label{eq:Glow}
 \end{equation}
Substituting Eq.~(\ref{eq:Glow}) into Eq.~(\ref{eq:Pidef}), 
it is convenient to first perform the integration over $k_{\parallel}$,\cite{Metlitski10a} which can be done using the residue theorem,
 \begin{widetext}
 \begin{eqnarray}
 \Pi_{\Lambda} ( \bar{K} )
 & = &  \frac{N}{2} Z_{\Lambda} \tilde{Z}_{\Lambda} \int \frac{ d k_{\bot}}{2 \pi} 
 \sum_{\alpha} 
\frac{ i  \alpha }{ i \barktau - (Z_{\Lambda}/\tilde{Z}_{\Lambda}) ( \alpha \bar{k}_{\parallel} + 
 \bar{k}_{\bot}^2 + 2 
 \bar{k}_{\bot} k_{\bot} ) }
  \int \frac{ d \ktau}{2 \pi}  \left[
 \Theta ( \alpha ( \ktau  + \barktau)   ) -
\Theta ( \alpha  \ktau )  \right] + ( \bar{K} \rightarrow - \bar{K} ).
 \nonumber
 \\
 & &
 \end{eqnarray}
The  integration over $\ktau$ is now trivial,
 \begin{equation}
 \int \frac{ d \ktau}{2 \pi}  \left[
 \Theta ( \alpha ( \ktau  + \barktau)   ) -
\Theta ( \alpha  \ktau )  \right] = \frac{ \alpha \barktau}{2 \pi },
 \end{equation}
so that
\begin{eqnarray}
 \Pi_{\Lambda} ( \bar{K} )
 & = &  \frac{N}{4 \pi }  Z_{\Lambda} \tilde{Z}_{\Lambda} 
\int \frac{ d k_{\bot}}{2 \pi} 
 \sum_{\alpha} 
\Biggl[ 
\frac{  i  \barktau    }{ i \barktau - (Z_{\Lambda}/\tilde{Z}_{\Lambda}) ( \alpha \bar{k}_{\parallel} + 
 \bar{k}_{\bot}^2 + 2 
 \bar{k}_{\bot} k_{\bot} ) }
+ ( \bar{K} \rightarrow - \bar{K} ) \Biggr]  .
 \label{eq:Pi1}
 \end{eqnarray}
\end{widetext}
Note that the $k_{\bot}$ integral is still ultraviolet divergent.
Following Mross {\it{et al.}}, \cite{Mross10} we regularize
the divergence by symmetrizing
the integrand with respect to $k_\bot \leftrightarrow - k_\bot$,
so that the expression
in the square braces vanishes as $1/ k_{\bot}^2$ for large $k_{\bot}$.  
The $k_{\bot}$ integration can thus again be done using the method of residues, with the result
\begin{equation}
 \Pi_{\Lambda} ( \barktau , \bar{k}_{\bot} ) =
 b_{\Lambda} 
 \frac{ | \barktau |}{ | \bar{k}_{\bot} |},
 \label{eq:Pires1}
 \end{equation}
where
 \begin{equation}
 b_{\Lambda} = \frac{N}{4 \pi} \tilde{Z}_{\Lambda}^2.
 \end{equation}
Note that this is independent of $\bar{k}_{\parallel}$, the mathematical reason being
that the term $\alpha \bar{k}_{\parallel}$ 
in the denominator of Eq.~(\ref{eq:Pi1}) can be eliminated by means of a simple shift
of the integration variable $k_{\bot}$.
We show shortly that at one-loop level, the self-energy 
$\Sigma^{\alpha}_{\Lambda} ( K )$ is actually independent of $\bd{k}$, so that
$\tilde{Z}_{\Lambda} = 1$ and Eq.~(\ref{eq:Pires1}) reduces to
\begin{equation}
 \Pi_{\Lambda} ( \barktau , \bar{k}_{\bot} ) = b_0
 \frac{ | \barktau |}{ | \bar{k}_{\bot} |},
 \label{eq:Pires2}
 \end{equation}
where 
 \begin{equation}
 b_0 = \frac{N}{4 \pi} ,
 \end{equation}
in agreement with Metlitski and Sachdev.\cite{Metlitski10a}
It should be noted that the bosonic self-energy does not renormalize the
exponent $z_b$ which characterizes the
momentum dependence of the  bare boson propagator.\cite{Mross10}

Next, we substitute Eq.~(\ref{eq:Pires1}) into our
expression (\ref{eq:Fdot}) for the bosonic single-scale propagator and obtain
\begin{equation}
 \dot{F}_{\Lambda} ( \bar{K} ) \approx - \frac{ \delta ( | \bar{k}_{\bot} | - \Lambda )}{
 r_0 + c_0 \Lambda^{z_b -1} + b_{\Lambda} | \barktau | / \Lambda  }.
 \label{eq:Fdotoneloop}
 \end{equation}
Substituting this expression into the FRG flow equation 
(\ref{eq:sigmaflow}) for the subtracted self-energy, we may perform all
integrations on the right-hand side and obtain
 \begin{equation}
 \partial_{\Lambda} {\Delta}^{\alpha}_{\Lambda} ( K )  = 
 \frac{ i\,{\rm sgn}\left(\ktau\right) }{\pi b_\Lambda}  \frac{\Lambda}{2 \pi}
 \tilde{Z}_{\Lambda} \ln
 \left[ 1 + \frac{ b_{\Lambda} | \ktau | / \Lambda }{ r_0 + c_0 \Lambda^{ z_b -1 } }
 \right].
 \label{eq:sigmaflow1}
 \end{equation}
Relating both $Z_\Lambda$ and $\tilde{Z}_\Lambda$ to flowing anomalous dimensions 
via
\begin{equation}
  \eta_\Lambda =  \Lambda \partial_\Lambda \ln Z_\Lambda , 
 \qquad \tilde{\eta}_\Lambda =  \Lambda \partial_\Lambda \ln \tilde{Z}_\Lambda ,
 \label{eq:etaetadef}
\end{equation}
the corresponding flowing ``frequency'' anomalous dimension $\eta_\Lambda$ is given by\cite{Kopietz10}
 \begin{eqnarray}
 \eta_{\Lambda} & =  & \Lambda Z_{\Lambda}  
 \lim_{ \ktau  \rightarrow 0}
\frac{ \partial }{\partial ( i \ktau )}
 \partial_{\Lambda} {\Delta}^{\alpha}_{\Lambda} ( i \ktau , \bd{k} =0 )
 \nonumber
 \\
 & = & \frac{ \Lambda Z_{\Lambda} \tilde{Z}_{\Lambda}}{2 \pi^2 (    
 r_0 + c_0 \Lambda^{ z_b -1 } )    }.
 \label{eq:etaflow}
\end{eqnarray}
Keeping in mind that the right-hand side of the flow equation
(\ref{eq:sigmaflow1})  for the self-energy depends only on $\ktau$ and not on $\bm{k}$,
we conclude that $\tilde{Z}_{\Lambda} = 1$ in this approximation, so that
$b_{\Lambda} = b_0$.
At the quantum critical point, $r_0 =0$,
our expression (\ref{eq:etaflow}) for the flowing frequency anomalous dimension hence 
reduces to
 \begin{equation}
 \eta_{\Lambda} = \frac{ Z_{\Lambda}}{ 2 \pi^2 c_0 \Lambda^{ z_b -2}},
 \label{eq:etaoneloop} 
\end{equation}
so that $Z_{\Lambda}$ satisfies the flow equation
 \begin{equation}
 \Lambda \partial_{\Lambda} Z_{\Lambda} = \eta_{\Lambda} Z_{\Lambda}
 =  \frac{ Z^2_{\Lambda}}{ 2 \pi^2 c_0 \Lambda^{ z_b -2}}.
 \end{equation}
This differential equation can be easily integrated to give
 \begin{equation}
 Z_{\Lambda} = \frac{ Z_{\Lambda_0}}{
 1 + \frac{ Z_{\Lambda_0}}{ 2 \pi^2 c_0 ( z_b -2)}
( \Lambda^{2 - z_b } - \Lambda_0^{ 2 - z_b } )}.
 \label{eq:ZLambdaflow}
 \end{equation} 
Assuming $z_b > 2$, we see that, for $\Lambda \rightarrow 0$, 
the wave-function renormalization factor vanishes as
 \begin{equation}
 Z_{\Lambda} \sim 2 \pi^2 c_0 ( z_b -2 ) \Lambda^{ z_b -2 },
 \end{equation}
implying that, at the quantum critical point, the fermion field has the frequency anomalous dimension
 \begin{equation}
 \eta =  \lim_{\Lambda \rightarrow 0}  
 \frac{ \Lambda \partial \ln Z_{\Lambda} }{ \partial \Lambda} = z_b -2.
 \end{equation}
Because, for $\tilde{Z}_{\Lambda} = 1$,
the right-hand side of Eq.~(\ref{eq:sigmaflow1})
depends on the flow parameter $\Lambda$ only via the explicit 
$\Lambda$-dependence shown, 
we can obtain the flowing subtracted self-energy
$\Delta^{\alpha}_{\Lambda} ( K) $ for all frequencies by simply
integrating both sides of
Eq.~(\ref{eq:sigmaflow1}) over $\Lambda$,
 \begin{eqnarray}
 {\Delta}^{\alpha}_{\Lambda} ( K )   -  
 {\Delta}^{\alpha}_{\Lambda_0} ( K )  & = & 
-  \int_{ \Lambda}^{\Lambda_0} d \lambda\, 
\partial_{\lambda} {\Delta}^{\alpha}_{\lambda} ( K ) 
 \nonumber
 \\
 &  & \hspace{-32mm} =
- \frac{ i\, {\rm sgn} (\ktau) }{2 \pi^2 b_0}
 \int_{ \Lambda}^{\Lambda_0} d \lambda\,  \lambda
 \ln
 \left[ 1 + \frac{ b_{0} | \ktau | / \lambda }{ r_0 + c_0 {\lambda}^{ z_b -1 } }
 \right]. \hspace{7mm}
 \end{eqnarray}
For $z_b > 2$, the integral is ultraviolet convergent, so that we may
take the limit $\Lambda_0 \rightarrow  \infty$ where
$ {\Delta}^{\alpha}_{\Lambda_0} ( K )  \rightarrow 0$.
At the quantum critical point,
$r_0=0$, the dependence of the integral on $\ktau$ can be scaled out and we obtain
for $\Lambda \rightarrow 0$
 \begin{equation}
 \lim_{\Lambda \rightarrow 0}
 {\Delta}^{\alpha}_{\Lambda} ( K ) = - i\,  {\rm sgn} (\ktau) 
 \frac{a_0}{b_0}
 \left( \frac{ b_0 | \ktau | }{c_0 } \right)^{ 2/ z_b },
 \end{equation}
where
 \begin{equation}
 a_0 = \frac{1}{ 2 \pi^2 z_b} \int_0^{\infty} dx  \frac{\ln ( 1 + x )}{x^{ 1 + 2/z_b}}
 = \frac{1}{4 \pi \sin ( 2 \pi / z_b ) },
 \end{equation}
in agreement with previous work.\cite{Altshuler94,Polchinski94,Metlitski10a,Metlitski10b,Chubukov10,Mross10}
The corresponding one-loop corrected fermion propagator is thus
 \begin{eqnarray}
 G^{\alpha} ( K ) & = & \frac{1}{ i \omega - \xi^{\alpha}_{\bd{k}} +
  i\,  {\rm sgn} (\ktau) 
 \frac{a_0}{b_0}
 \left( \frac{ b_0 | \ktau | }{c_0 } \right)^{ 2/ z_b } }
 \nonumber
 \\
 & \approx & \frac{1}{   i\,  {\rm sgn} (\ktau) 
 \frac{a_0}{b_0}
 \left( \frac{ b_0 | \ktau | }{c_0 } \right)^{ 2/ z_b } - \alpha k_{\parallel} - k_{\bot}^2 },
 \label{eq:G1loopres}
 \end{eqnarray}
where we have used the fact that for $ z_b > 2$ the Matsubara frequency $ i \omega$
is small compared with the contribution from the self-energy at low energies.
Comparing Eq.~(\ref{eq:G1loopres}) with the 
general scaling form (\ref{eq:GlowenergyMetlitskiSachdev}), we conclude that
$z = z_b/2$ and $\eta_{\psi} =0$ within the one-loop approximation. { We note
that this is in agreement with the one-loop results of previous works based on the 
field theoretical renormalization group.\cite{Altshuler94,Polchinski94,Metlitski10a,Metlitski10b,Chubukov10,Mross10}}

\section{Vertex corrections}
\label{sec:vertex}

In this section, we use the hierarchy of FRG flow equations
given  in Sec.~\ref{sec:FRGfloweq} to estimate 
the effect of vertex corrections on the low-frequency behavior
of the
fermionic self-energy, $\Sigma^{\alpha}_{\Lambda} ( K )$.
We do not attempt to calculate the entire momentum and frequency dependence of
 $\Sigma^{\alpha}_{\Lambda} ( K )$, but focus on 
the flow of the two renormalization constants $Z_{\Lambda}$ and
$\tilde{Z}_{\Lambda}$ defined via the low-energy expansion
(\ref{eq:selfexp}) of the flowing self-energy, and on the associated anomalous dimensions
$\eta_{\Lambda}$ and $\tilde{\eta}_{\Lambda}$ defined
via the logarithmic derivatives of 
$Z_{\Lambda}$ and
$\tilde{Z}_{\Lambda}$ with respect to the flow parameter (see Eq.~(\ref{eq:etaetadef})).
Note that the definition (\ref{eq:Glow}) implies
 \begin{eqnarray}
 Z_\Lambda^{-1} & = & 1 - \lim_{ \ktau \rightarrow 0} 
 \frac{ \partial \Sigma^{\alpha}_{\Lambda} ( i \ktau , \bd{k}=0 )}{ \partial ( i \ktau ) },
 \\
 \tilde{Z}_\Lambda^{-1} & = & 1 + \lim_{ \bd{k}  \rightarrow 0} 
 \frac{ \partial \Sigma^{\alpha}_{\Lambda} ( i \ktau =0 , \bd{k} )}{ \partial  \xi^{\alpha}_{\bd{k}}  },
 \end{eqnarray}
while the definition (\ref{eq:etaetadef}) of the flowing anomalous dimensions, $\eta_{\Lambda}$ and $\tilde{\eta}_{\Lambda}$, allows us to
relate these quantities directly to the
derivative of the self-energy with respect to the flow parameter
 \begin{eqnarray}
 \eta_{\Lambda} & =  &    \Lambda \partial_{\Lambda} \ln Z_{\Lambda} =  Z_{\Lambda} \Lambda 
 \lim_{ \ktau  \rightarrow 0}
\frac{ \partial }{\partial ( i \ktau )}
 \partial_{\Lambda} {\Sigma}^{\alpha}_{\Lambda} ( i \ktau , \bd{k} =0 ),
 \nonumber
 \\
 & &
 \label{eq:etaflow2}
 \\
 \tilde{\eta}_{\Lambda} & = &   \Lambda \partial_{\Lambda} \ln \tilde{Z}_{\Lambda}  
 =   - \tilde{Z}_{\Lambda} \Lambda
 \lim_{ \bd{k}  \rightarrow 0}  
 \frac{ \partial}{\partial \xi^{\alpha}_{\bd{k}}} \partial_{\Lambda} 
 \Sigma^{\alpha}_{\Lambda} ( i \ktau =0, \bd{k} )
 \nonumber
 \\
 & = & 
  - \tilde{Z}_{\Lambda} \Lambda
 \lim_{ \bd{k}  \rightarrow 0}  
 \frac{ \partial}{\partial ( \alpha k_{\parallel})  } \partial_{\Lambda} 
 \Sigma^{\alpha}_{\Lambda} ( i \ktau =0, \bd{k} ),
 \label{eq:etatilderes}
 \end{eqnarray}
where, in the last line, we have used the fact that the momentum dependence
of the self-energy appears only in the combination $ \xi^{\alpha}_{\bd{k}}
 = \alpha k_{\parallel} + k_{\bot}^2$.
In the one-loop approximation, 
$ \lim_{\Lambda \rightarrow 0} \tilde{\eta}_{\Lambda} =0$, but
in this section we show that 
vertex corrections lead to a finite value of this limit.
In Sec.~\ref{subsec:anomalous} we further show 
that $\lim_{\Lambda \rightarrow 0 } \tilde{\eta}_{\Lambda}$
can be identified with the anomalous dimension
$\eta_{\psi}$ of the fermion field defined via
Eq.~(\ref{eq:GlowenergyMetlitskiSachdev}); moreover
we show how to express the fermionic dynamic exponent $z$
in terms of
$\eta_{\Lambda}$ and $\tilde{\eta}_{\Lambda}$.

\subsection{Truncated flow equations and skeleton equations}

\begin{figure}[tb]    
   \centering
  \includegraphics[width=0.45\textwidth]{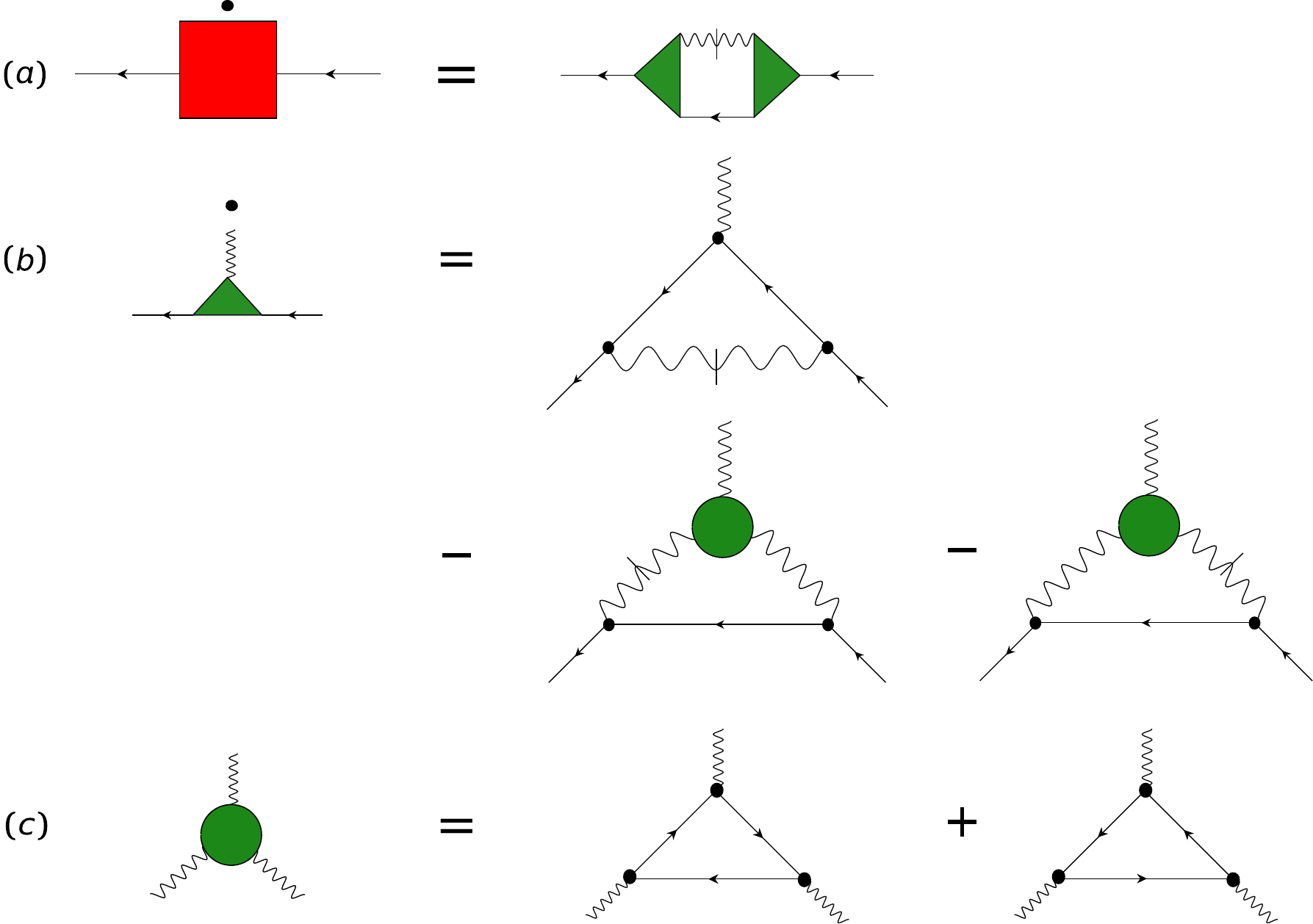}
  \caption{%
(Color online) 
(a) Graphical representation of the FRG flow equation 
(\ref{eq:flowSigma3}) for the fermionic self-energy 
in the momentum-transfer cutoff scheme. 
This flow equation can be obtained from the more general
flow equation (\ref{eq:flowSigmatrunc}) by omitting the contribution involving
the fermionic single-scale propagator.
(b)  Truncated flow equation for the three-legged vertex 
$\Gamma^{\bar{\psi}^\alpha\psi^\alpha\phi}(K+\bar{K};K;\bar{K})$ in the momentum-transfer cutoff scheme, which can be obtained from the more general flow equation
(\ref{eq:flowGammatrunc})  shown in Fig.~\ref{fig:flowGamma21} 
by neglecting the contribution from the fermionic single-scale propagator
and approximating the vertices
$\Gamma^{\bar{\psi}^\alpha\psi^\alpha\phi}(K+\bar{K};K;\bar{K})$
on the right-hand side by the bare vertices.
(c) Graphical representation of the approximate expression (\ref{eq:Skeleton3})
for  the  bosonic three-legged vertex used in our calculation.
Note that this expression can be obtained from the exact skeleton equation
(\ref{eq:SkeletonThreeLegged}) derived in Appendix~A by neglecting the
irreducible four-point vertex and then
making the same approximations
as in (b).
}
    \label{fig:fullselfenergy}
  \end{figure}

Using the momentum-transfer cutoff scheme in combination with the truncation
strategy discussed in the third paragraph of  Sec.~\ref{sec:FRGfloweq},
we obtain the fermionic self-energy from
 \begin{widetext}
\begin{eqnarray}
  \partial_{\Lambda} {\Sigma}^{\alpha} ( K ) & = &  
  \int_{\bar{K}} 
    \dot{F} ( \bar{K} ) G^{\alpha}  ( K + \bar{K} ) 
  \Gamma^{\bar{\psi}^{\alpha} \psi^{\alpha} \phi} ( K + \bar{K} ;   K ;    \bar{K}  ) 
  \Gamma^{ \bar{\psi}^{\alpha} \psi^{\alpha} \phi  } ( K; K + \bar{K} ; -  \bar{K}  )  ,
  \label{eq:flowSigma3}
  \end{eqnarray}
where the three-point vertex with one bosonic and two fermionic legs is determined by
\begin{eqnarray}
& &  \partial_{\Lambda} {\Gamma}^{\bar{\psi}^{\alpha} \psi^{\alpha} \phi } 
 ( K + \bar{K}    ; K  ; \bar{K}   )   =    ( \Gamma_0^{\alpha} )^3
  \int_{\bar{K}^{\prime}} 
  \dot{ F} ( \bar{K}^{\prime} ) 
G^{\alpha} ( K + \bar{K}^{\prime} )  
 G^{\alpha} ( K +  \bar{K} + \bar{K}^{\prime} ) 
 \nonumber
 \\
 &    &   -  ( \Gamma_0^{\alpha} )^2  \int_{\bar{K}^{\prime}} 
  \dot{ F} ( \bar{K}^{\prime} ) 
F ( \bar{K} + \bar{K}^{\prime} )  
 \left[
 G^{\alpha} ( K +  \bar{K} + \bar{K}^{\prime} ) +
 G^{\alpha} ( K - \bar{K}^{\prime} ) \right]
 \Gamma^{\phi \phi \phi } ( \bar{K} , \bar{K}^{\prime} , - \bar{K} - \bar{K}^{\prime} ).
  \label{eq:flowGammatrunc3}
\end{eqnarray}
\end{widetext}
Eq.~(\ref{eq:flowSigma3}) can be obtained from
the more general flow equation (\ref{eq:flowSigmatrunc})
by simply omitting the contribution involving the fermionic single-scale
propagator, while in Eq.~(\ref{eq:flowGammatrunc3}) we have, in addition,
replaced the flowing vertices  ${\Gamma}^{\bar{\psi}^{\alpha} \psi^{\alpha} \phi } 
 ( K + \bar{K}    ; K  ; \bar{K}   )$ on the right-hand side of the flow equation by their bare values.
The purely bosonic three-legged vertex 
in the second line of Eq.~(\ref{eq:flowGammatrunc3}) is approximated
by 
\begin{align}
    & \Gamma^{\phi\phi\phi}(\bar{K}_1, \bar{K}_2, - \bar{K}_1  - \bar{K}_2) = N\sum_\alpha \left(\Gamma_0^\alpha \right)^3  \nonumber \\
& \times \left[
   \int_K G^\alpha(K) G^\alpha(K + \bar{K}_1) G^\alpha(K + \bar{K}_1 + \bar{K}_2)  \right. \nonumber \\
 & 
\left. {} \qquad +  \int_K G^\alpha(K) G^\alpha(K + \bar{K}_1) G^\alpha(K - \bar{K}_2)  \right] ,
\label{eq:Skeleton3}
\end{align}
which is derived from the exact skeleton equation by making the
same approximations as in the derivation of
Eqs.~(\ref{eq:flowSigma3}) and (\ref{eq:flowGammatrunc3}).
A graphical representation of 
Eqs.~(\ref{eq:flowSigma3})--(\ref{eq:Skeleton3}) is shown in Fig.~\ref{fig:fullselfenergy}.
%
%
To obtain the corrections to the one-loop approximation,
it is sufficient to
approximate all propagators 
in
Eqs.~(\ref{eq:flowSigma3})--(\ref{eq:Skeleton3})
by the one-loop results given in
Sec.~\ref{sec:novertexcorrections},
i.e., we approximate on the right-hand sides 
  \begin{eqnarray}
G^{\alpha}_{\Lambda} ( K ) & =  & \frac{Z_{\Lambda}}{ i \ktau -  
 Z_{\Lambda} \xi^{\alpha}_{\bd{k}} },
 \label{eq:Glow2}
 \\
  {F}_{\Lambda} ( \bar{K} ) & = &   \frac{ \Theta ( | \bar{k}_{\bot} | - \Lambda )}{
 r_0 + c_0 | \bar{k}_{\bot}|^{z_b -1} +  \Theta ( | \bar{k}_{\bot} | - \Lambda )   
 b_{0}         | \barktau | /  | \bar{k}_{\bot} |   },
 \label{eq:Foneloop2}
 \hspace{7mm}
 \\
 \dot{F}_{\Lambda} ( \bar{K} ) & = &  -
\frac{ \delta ( | \bar{k}_{\bot} | - \Lambda )}{
 r_0 + c_0 \Lambda^{z_b -1} +  
 b_{0}      | \barktau | /  \Lambda   }.
 \label{eq:Fdotoneloop2}
 \end{eqnarray}
Here, $Z_{\Lambda}$ is the one-loop result for the flowing wave-function 
renormalization factor given in Eq.~(\ref{eq:ZLambdaflow}) and we have used the
fact that $\tilde{Z}_{\Lambda} =1$ within the one-loop approximation.
We thus arrive at a system of flow equations for
the momentum- and frequency-dependent two-point and three-point functions.
Having determined the right-hand side of the flow equation 
(\ref{eq:flowSigma3}) for the self-energy, we can 
substitute the result into Eqs.~(\ref{eq:etaflow2}) and (\ref{eq:etatilderes})
and obtain for the flowing anomalous dimension related to the frequency dependence
of the self-energy
\begin{eqnarray}
 {\eta}_{\Lambda}  & = &   
{Z}_{\Lambda} \Lambda
 \lim_{ K \rightarrow 0} \frac{\partial}{\partial ( i \ktau )}
 \int_{ \bar{K}} \dot{F}_{\Lambda} ( \bar{K} ) 
 [ G^{\alpha}_{\Lambda} ( \bar{K} + K ) - G^{\alpha}_{\Lambda} ( \bar{K}  ) ] 
 \nonumber
 \\
 & & \hspace{20mm} \times
 [ \Gamma^{\bar{\psi}^{\alpha} \psi^{\alpha} \phi}_{\Lambda} ( 
 \bar{K} ; 0 ; \bar{K} ) ]^2
 \nonumber
 \\
 &+ & 
 2 {Z}_{\Lambda} \Lambda
 \int_{ \bar{K}} \dot{F}_{\Lambda} ( \bar{K} ) 
    G^{\alpha}_{\Lambda} ( \bar{K}  )
\Gamma^{\bar{\psi}^{\alpha} \psi^{\alpha} \phi}_{\Lambda} ( 
 \bar{K} ; 0 ; \bar{K} )
 \nonumber
 \\
 & & \hspace{10mm}  \times 
 \lim_{ K \rightarrow 0}
\frac{ \partial \Gamma^{\bar{\psi}^{\alpha} \psi^{\alpha} \phi}_{\Lambda} ( 
 \bar{K} + K  ; K ; \bar{K} )}{ \partial ( i \ktau ) },
 \label{eq:etares2}
 \end{eqnarray}
and for the corresponding anomalous dimension associated with
the momentum dependence of the self-energy,
 \begin{eqnarray}
 \tilde{\eta}_{\Lambda}  & = &  
- \tilde{Z}_{\Lambda} \Lambda
 \lim_{ K \rightarrow 0} \frac{\partial}{\partial ( \alpha k_{\parallel} )}
 \int_{ \bar{K}} \dot{F}_{\Lambda} ( \bar{K} ) 
 [ G^{\alpha}_{\Lambda} ( \bar{K} + K ) - G^{\alpha}_{\Lambda} ( \bar{K}  ) ] 
 \nonumber
 \\
 & & \hspace{20mm} \times
 [ \Gamma^{\bar{\psi}^{\alpha} \psi^{\alpha} \phi}_{\Lambda} ( 
 \bar{K} ; 0 ; \bar{K} ) ]^2
 \nonumber
 \\
 &- & 
 2 \tilde{Z}_{\Lambda} \Lambda
 \int_{ \bar{K}} \dot{F}_{\Lambda} ( \bar{K} ) 
    G^{\alpha}_{\Lambda} ( \bar{K}  )
\Gamma^{\bar{\psi}^{\alpha} \psi^{\alpha} \phi}_{\Lambda} ( 
 \bar{K} ; 0 ; \bar{K} )
 \nonumber
 \\
 & & \hspace{10mm}  \times 
 \lim_{ K \rightarrow 0}
\frac{ \partial \Gamma^{\bar{\psi}^{\alpha} \psi^{\alpha} \phi}_{\Lambda} ( 
 \bar{K} + K  ; K ; \bar{K} )}{ \partial ( \alpha k_{\parallel} ) }.
 \label{eq:baretares2}
 \end{eqnarray}
In the first terms on the right-hand sides of these expressions we have
used the same regularization as in Eq.~(\ref{eq:sigmaflow}).

\subsection{Bosonic three-legged vertex}

In order to evaluate the  self-energy from Eq.~(\ref{eq:flowSigma3}),
we should calculate  the mixed fermion-boson  vertex
$ {\Gamma}^{\bar{\psi}^{\alpha} \psi^{\alpha} \phi } 
 ( K + \bar{K}    ; K  ; \bar{K}   ) $ by integrating the
flow equation (\ref{eq:flowGammatrunc3}), which in turn depends on the
purely bosonic three-legged vertex
$  \Gamma^{\phi\phi\phi}(\bar{K}_1, \bar{K}_2, - \bar{K}_1  - \bar{K}_2) $.
Fortunately, the integrations in the truncated skeleton equation 
(\ref{eq:Skeleton3}) can be carried out exactly in our model, as shown in Appendix B.
The result can be written as
\begin{eqnarray}
& & \Gamma^{\phi \phi \phi } ( \bar{K}_1 , \bar{K}_2 , - \bar{K}_1 - \bar{K}_2  )  
 \nonumber
 \\
& = &  2!N \sum_{\alpha}   (\Gamma_0^{\alpha} )^3 L^{\alpha}_{3} (  \bar{K}_1,  \bar{K}_2 , 
 - \bar{K}_1 - \bar{K}_2 ),
 \label{eq:Gamma3phires}
 \end{eqnarray} 
where the symmetrized fermion loop with three external bosonic
legs and one-loop renormalized fermionic propagators is given by
\begin{widetext}
\begin{eqnarray}
  L^{ \alpha}_{3} (  \bar{K}_1,  \bar{K}_2 ,  - \bar{K}_1 - \bar{K}_2 )  
 & = &
\frac{1}{4 \pi} 
 \left( \frac{1}{ \bar{k}_{\bot 1}} +    \frac{1}{ \bar{k}_{\bot 2}} \right)
 \frac{ \barktauone \Theta \left( \frac{ \barktauone}{\bar{k}_{\bot 1}} \right) + 
 \barktautwo \Theta \left(   \frac{ \barktautwo}{\bar{k}_{\bot 2}}   \right) - 
 ( \barktauone + \barktautwo ) 
 \Theta \left( \frac{\barktauone + \barktautwo}{
 \bar{k}_{\bot 1} + \bar{k}_{\bot 2} } \right)
       }{ 
 \left[  \frac{ \bar{k}_{\parallel 1}}{ \bar{k}_{\bot 1}} 
 -    \frac{ \bar{k}_{\parallel 2}}{ \bar{k}_{\bot 2}} -  
 \frac{i \alpha}{Z} (  \frac{ \barktauone}{ \bar{k}_{\bot 1}} -    
 \frac{ \barktautwo}{ \bar{k}_{\bot 2}}   )  \right]^2  
 -  ( \bar{k}_{\bot 1}  +  \bar{k}_{\bot 2}  )^2 } .
 \label{eq:L3resmod}
\end{eqnarray}
\end{widetext}
Note that this function represents a rather complicated 
momentum-  and frequency-dependent effective interaction between
the bosonic fluctuations, mediated by the fermions. 
Obviously, this function cannot simply be
approximated by a constant,
which is assumed to be possible in 
the Hertz-Millis approach to quantum critical phenomena.\cite{Hertz76,Millis93}

\subsection{Three-legged boson-fermion vertex}

We have now calculated all functions appearing on the right-hand side
of the FRG flow equation  (\ref{eq:flowGammatrunc3}) for the three-legged 
vertex with one bosonic 
and two fermionic external legs, so that we may next integrate this equation over the flow
parameter $\Lambda$.
Let us begin by evaluating the first term on the right-hand side
of  Eq.~(\ref{eq:flowGammatrunc3}), 
\begin{eqnarray}
& &  \partial_{\Lambda} {\Gamma}^{\bar{\psi}^{\alpha} \psi^{\alpha} \phi } 
 ( K + \bar{K}    ; K  ; \bar{K}   )^{(1)}  =   ( \Gamma_0^{\alpha} )^3
  \int_{\bar{K}^{\prime}} 
  \dot{ F} ( \bar{K}^{\prime} ) 
 \nonumber
 \\
 & & \times
G^{\alpha} ( K + \bar{K}^{\prime} )  
 G^{\alpha} ( K +  \bar{K} + \bar{K}^{\prime} ) ,
  \label{eq:flowGammatrunc2}
\end{eqnarray}
corresponding to the first diagram on the right-hand side of
Fig.~\ref{fig:fullselfenergy}\,(b).
The integrations in Eq.~(\ref{eq:flowGammatrunc2}) can be
explicitly carried out, with the result
 \begin{widetext}
 \begin{eqnarray}
  \Lambda  \partial_{\Lambda} {\Gamma}^{\bar{\psi}^{\alpha} \psi^{\alpha} \phi } 
 ( K + \bar{K}    ; K  ; \bar{K}   )^{(1)} &  = &
 - \frac{ ( \Gamma_0^{\alpha} )^3 }{ ( 2 \pi )^2}  \frac{ Z_{\Lambda}}{ 
 b_{0}} 
 \left[
 \frac{\Lambda^2}{ i \barktau - Z_{\Lambda} [ \alpha
 \bar{k}_{\parallel} + \bar{k}_{\bot}^2 + 2 \bar{k}_{\bot} ( k_{\bot} +
 \Lambda  ) ] } + ( \Lambda \rightarrow - \Lambda ) \right]
 \nonumber
 \\
 &\times & \left[
 i\, {\rm sgn} ( \ktau + \barktau ) 
 \ln \left( 1 + \frac{ b_{0} | \ktau + \barktau | }{r_0 \Lambda
 + c_0 \Lambda^{z_b} } 
 \right) 
- 
 i\, {\rm sgn}(\ktau) \ln \left( 1 + \frac{ b_{0} | \ktau | }{r_0 \Lambda
 + c_0 \Lambda^{z_b} } 
 \right) 
 \right].
 \label{eq:Gamma21flow1}
 \end{eqnarray}
 \end{widetext}
For the evaluation of the flowing anomalous dimensions in
Eqs.~(\ref{eq:etares2}) and (\ref{eq:baretares2}), we need only
the vertex ${\Gamma}^{\bar{\psi}^{\alpha} \psi^{\alpha} \phi } 
 (  \bar{K}    ; 0  ; \bar{K}   )$
at vanishing fermionic energy-momentum, as well as the derivatives of
 ${\Gamma}^{\bar{\psi}^{\alpha} \psi^{\alpha} \phi } 
 (  \bar{K} +K   ; K  ; \bar{K}   )$ with respect to the components of $K$  at $K=0$.
From Eq.~(\ref{eq:Gamma21flow1}), we see that the contribution
of the first diagram in Fig.~\ref{fig:fullselfenergy}\,(b) to the flow of these quantities 
can be written as
 \begin{eqnarray}
   & &  \Lambda \partial_{\Lambda} {\Gamma}^{\bar{\psi}^{\alpha} \psi^{\alpha} \phi } 
  (   \bar{K}    ; 0  ; \bar{K}   )^{(1) }   
 = 
 - I^{(1)}_{\Lambda} 
 \left( \frac{ \barktau}{\Lambda^{z_b}}, \frac{ \bar{k}_{\parallel}}{ \Lambda^2},
  \frac{ \bar{k}_{\bot}}{ \Lambda} \right),
 \hspace{7mm}
 \label{eq:I1}
 \\
 & &    \Lambda \partial_{\Lambda} 
\left. \frac{ \partial {\Gamma}^{\bar{\psi}^{\alpha} \psi^{\alpha} \phi } 
  (   \bar{K}    ; K  ; \bar{K}   )^{(1) }}{\partial   ( i \ktau )    }   \right|_{ K =0} 
 \nonumber
 \\
  & &  \hspace{20mm}  = -  \frac{1}{\Lambda^{ z_b}} I^{(1 \omega ) }_{\Lambda} 
 \left( \frac{ \barktau}{\Lambda^{z_b}}, \frac{ \bar{k}_{\parallel}}{ \Lambda^2},
  \frac{ \bar{k}_{\bot}}{ \Lambda} \right),
 \label{eq:I1omega}
 \\
  & &  \Lambda \partial_{\Lambda} 
\left. \frac{ \partial {\Gamma}^{\bar{\psi}^{\alpha} \psi^{\alpha} \phi } 
  (   \bar{K}    ; K  ; \bar{K}   )^{(1) }}{\partial   ( \alpha k_{\parallel} )    }   \right|_{ K =0} 
  =    0,
 \label{eq:I1k}
\end{eqnarray}
where we have defined the dimensionless scaling functions
\begin{eqnarray}
 & &  I^{(1)}_{\Lambda} 
 ( \barqtau ,  \bar{q}_{\parallel}, \bar{q}_{\bot} )  = 
 \frac{ \gamma_{\Lambda}}{     (2 \pi )^2    b_0 }
 i\, {\rm sgn} (\barqtau) 
 \ln \left( 1 + \frac{ b_{0} |  \barqtau | }{r_\Lambda + c_0  } \right) 
\nonumber
 \\
 & &
 \times
 \left[
 \frac{1}{ i \barqtau - \gamma_{\Lambda} [ \alpha
 \bar{q}_{\parallel} + \bar{q}_{\bot}^2 + 2 \bar{q}_{\bot}  ] } + 
 ( \bar{q}_{\bot} \rightarrow - \bar{q}_{\bot} ) \right],
 \label{eq:I1res}
 \end{eqnarray}
and
 \begin{eqnarray}
& &   I^{(1 \omega)}_{\Lambda} 
 ( \barqtau ,  \bar{q}_{\parallel}, \bar{q}_{\bot} )   = 
 - \frac{ \gamma_{\Lambda}  }{( 2 \pi )^2 ( r_{\Lambda} + c_0 )^2}
 \frac{   b_{0}    }{ 1 + \frac{ b_{0}}{ r_{\Lambda} + c_0 } 
 | \barqtau |  }
 \nonumber
 \\
 &  & \times  
 \left[
 \frac{  | \barqtau  | }{ i \barqtau - \gamma_{\Lambda} [ \alpha
 \bar{q}_{\parallel} + \bar{q}_{\bot}^2 + 2 \bar{q}_{\bot}  ] } + 
 ( \bar{q}_{\bot} \rightarrow - \bar{q}_{\bot} ) \right].
 \hspace{7mm}
 \label{eq:I1omegares}
 \end{eqnarray}
Here, $r_{\Lambda} = r_0 \Lambda^{1-z_b}$ vanishes at the quantum critical point,
and we have introduced the dimensionless parameter
  \begin{equation}
 \gamma_{\Lambda} =  \frac{Z_{\Lambda}}{ \Lambda^{  z_b- 2} }
 =   \frac{   2 \pi^2 c_0 ( z_b -2)     }{ 1 -  \Lambda^{z_b-2}
  \bigl[ \frac{2 \pi^2 c_0 ( z_b -2) }{Z_{\Lambda_0}} -  \Lambda_0^{  2 - z_b  }  \bigr] },
 \end{equation}
which approaches the limit 
$ 2 \pi^2 c_0 ( z_b -2 ) \propto N ( z_b -2 )$ for $\Lambda
 \rightarrow 0$.

Next, consider the contribution of the last two 
diagrams on the right-hand side of
Fig.~\ref{fig:fullselfenergy}\,(b) to the flow of the three-legged boson-fermion vertex,
corresponding to the second line in Eq.~(\ref{eq:flowGammatrunc3}),
\begin{widetext}
\begin{eqnarray}
  \partial_{\Lambda} {\Gamma}^{\bar{\psi}^{\alpha} \psi^{\alpha} \phi } 
 ( K + \bar{K}    ; K  ; \bar{K}   )^{(2)}   & = &     -  ( \Gamma_0^{\alpha} )^2  \int_{\bar{K}^{\prime}} 
  \dot{ F} ( \bar{K}^{\prime} ) 
F ( \bar{K} + \bar{K}^{\prime} )  
 \left[
 G^{\alpha} ( K +  \bar{K} + \bar{K}^{\prime} ) +
 G^{\alpha} ( K - \bar{K}^{\prime} ) \right]
 \nonumber
 \\
 & & \hspace{18mm} \times
 \Gamma^{\phi \phi \phi } ( \bar{K} , \bar{K}^{\prime} , - \bar{K} - \bar{K}^{\prime} ).
  \label{eq:gammamix2}
\end{eqnarray}
Due to the rather complicated form of
the vertex $\Gamma^{\phi \phi \phi } ( \bar{K} , \bar{K}^{\prime} , - \bar{K} - \bar{K}^{\prime} )$
given in Eqs.~(\ref{eq:Gamma3phires}) and (\ref{eq:L3resmod}), the
evaluation of the right-hand side of Eq.~(\ref{eq:gammamix2}) is
quite involved.
The 
$\bar{k}_{\parallel}^{\prime}$ integration
can still be performed by means of the residue theorem, while
the $\bar{k}_{\bot}^{\prime}$ integration is trivial 
due to the $\delta$-function in the single-scale propagator.
After these integrations, we obtain
 \begin{eqnarray}
& &  \partial_{\Lambda} {\Gamma}^{\bar{\psi}^{\alpha} \psi^{\alpha} \phi  } 
 ( K + \bar{K}    ; K  ; \bar{K}   )^{(2) }
 =   - i \alpha N \Lambda^2     \frac{ ( \Gamma_{0}^{\alpha} )^2}{ (2 \pi )^2 }
 \int_{ - \infty}^{\infty} \frac{ d \barkprimetau}{ 2 \pi }
 \frac{1}{ r_0 +c_0 \Lambda^{ z_b -1} + b_{0} 
 | \barkprimetau | / \Lambda }
 \nonumber
 \\
 & \times &  \sum_{ s = \pm } F ( \barktau + \barktau^{\prime} ,
 \bar{k}_{\bot} + s \Lambda )   
J (    \barktau ,    \bar{k}_{\bot},  \barktau^{\prime} , s \Lambda )
 \sum_{ \alpha^{\prime} = \pm } ( \Gamma_0^{\alpha^{\prime}} )^3
  \left[  \frac{ \Theta ( {\rm Im} (z_1) ) - \Theta ( {\rm Im} (z_3) ) }{ ( z_1 - z_3 )^2 - z_4^2 }
-  \frac{ \Theta ( {\rm Im} (z_2) ) - \Theta ( {\rm Im} (z_3) ) }{ ( z_2 - z_3 )^2 - z_4^2 }
\right] ,\hspace{5mm}
 \label{eq:flowGamma2int}
\end{eqnarray}
where
 \begin{equation}
J (    \barktau ,    \bar{k}_{\bot},  \barktau^{\prime} ,  \bar{k}_{\bot}^{\prime} )
 =  \left( \frac{1}{ \bar{k}_{\bot}} +  \frac{1}{ \bar{k}_{\bot}^{\prime}} \right)
 \left[ 
 \barktau \Theta \left( \frac{ \barktau}{\bar{k}_{\bot }} \right) + 
 \barktau^{\prime} \Theta \left(   \frac{ \barktau^{\prime}}{\bar{k}_{\bot }^{\prime}}   \right) - 
 ( \barktau + \barktau^{\prime} ) 
 \Theta \left( \frac{\barktau + \barktau^{\prime}}{
 \bar{k}_{\bot } + \bar{k}_{\bot }^{\prime} } \right)
 \right],
\end{equation}
\end{widetext}
and
 \begin{subequations}
 \begin{eqnarray}
 z_1 & = & - ( k_{\parallel} + \bar{k}_{\parallel} ) - \alpha [ \Lambda + s ( k_{\bot}
 + \bar{k}_{\bot}) ]^2 
 \nonumber
 \\
 &  & +    \frac{ i \barktau^{\prime} 
 + i \barktau + i \ktau }{Z_{\Lambda}},
 \\
 z_2 & = & k_{\parallel} + \alpha ( \Lambda - s k_{\bot} )^2 
 +  \alpha
 \frac{ i \barktau^{\prime} - i \ktau }{Z_{\Lambda}},
 \\
z_3  & = &  \Lambda \left[
   \frac{ \bar{k}_{\parallel} }{ s \bar{k}_{\bot}} - 
 \frac{\alpha^{\prime}}{Z_{\Lambda}} 
 \left(  \frac{ i \barktau }{ s \bar{k}_{\bot}} - \frac{ i \barktau^{\prime}}{\Lambda} 
 \right) \right],
 \label{eq:z3def}
 \\
 z_4 & = & \Lambda ( \Lambda + s \bar{k}_{\bot} ).
\end{eqnarray}
\end{subequations}
For the calculation of $\eta_{\Lambda}$ and $\tilde{\eta}_{\Lambda}$ in
Eqs.~(\ref{eq:etares2}) and (\ref{eq:baretares2}) we, again, need only
the vertex ${\Gamma}^{\bar{\psi}^{\alpha} \psi^{\alpha} \phi } 
 (  \bar{K}    ; 0  ; \bar{K}   )$ and 
the derivatives of
 ${\Gamma}^{\bar{\psi}^{\alpha} \psi^{\alpha} \phi } 
 (  \bar{K} +K   ; K  ; \bar{K}   )$ with respect to the components of the fermionic label
$K$  at $K=0$. In analogy with Eqs.~(\ref{eq:I1})--(\ref{eq:I1k}), we therefore define
 \begin{eqnarray}
   & &  \Lambda \partial_{\Lambda} {\Gamma}^{\bar{\psi}^{\alpha} \psi^{\alpha} \phi } 
  (   \bar{K}    ; 0  ; \bar{K}   )^{(2) }   
 = 
 - I^{(2)}_{\Lambda} 
 \left( \frac{ \barktau}{\Lambda^{z_b}}, \frac{ \bar{k}_{\parallel}}{ \Lambda^2},
  \frac{ \bar{k}_{\bot}}{ \Lambda} \right),
 \hspace{7mm}
 \label{eq:I2}
 \\
 & &    \Lambda \partial_{\Lambda} 
\left. \frac{ \partial {\Gamma}^{\bar{\psi}^{\alpha} \psi^{\alpha} \phi } 
  (   \bar{K}    ; K  ; \bar{K}   )^{(2) }}{\partial   ( i \ktau )    }   \right|_{ K =0} 
 \nonumber
 \\
  & &  \hspace{20mm}  = -  \frac{1}{\Lambda^{ z_b}} I^{(2 \omega ) }_{\Lambda} 
 \left( \frac{ \barktau}{\Lambda^{z_b}}, \frac{ \bar{k}_{\parallel}}{ \Lambda^2},
  \frac{ \bar{k}_{\bot}}{ \Lambda} \right),
 \label{eq:I2omega}
 \\
  & &  \Lambda \partial_{\Lambda} 
\left. \frac{ \partial {\Gamma}^{\bar{\psi}^{\alpha} \psi^{\alpha} \phi } 
  (   \bar{K}    ; K  ; \bar{K}   )^{(2) }}{\partial   ( \alpha k_{\parallel} )    }   \right|_{ K =0} 
  \nonumber
 \\
  & &  \hspace{20mm}  = -  \frac{1}{\Lambda^{ 2}} I^{(2 k ) }_{\Lambda} 
 \left( \frac{ \barktau}{\Lambda^{z_b}}, \frac{ \bar{k}_{\parallel}}{ \Lambda^2},
  \frac{ \bar{k}_{\bot}}{ \Lambda} \right).
 \label{eq:I2k}
\end{eqnarray}

From now on we focus on the Ising-nematic instability and explicitly set $\Gamma_0^{\alpha} =1$.
In principle,  the functions
 $I^{(2)}_{\Lambda}  ( \bar{\epsilon} , \bar{q}_{\parallel},
  \bar{q}_{\bot} )$,
 $I^{(2 \omega)}_{\Lambda}  ( \bar{\epsilon} , \bar{q}_{\parallel},
  \bar{q}_{\bot} )$,
and $I^{(2 k)}_{\Lambda}  ( \bar{\epsilon} , \bar{q}_{\parallel},
  \bar{q}_{\bot} )$
can be calculated analytically by performing the
remaining frequency integration in Eq. (\ref{eq:flowGamma2int}).
However, the result is rather cumbersome and not very transparent.
To simplify the calculation, let us assume that the  arguments
$\bar{\epsilon} $, $\bar{q}_{\parallel}$, and $\bar{q}_{\bot} $
of these functions are all small compared with unity. 
Then the resulting expressions
simplify and we obtain the approximate expressions
 \begin{eqnarray}
 I^{(2)}_{\Lambda} 
 ( \barqtau ,  \bar{q}_{\parallel}, \bar{q}_{\bot} ) & \approx &
  \frac{ \gamma_{\Lambda}    | \barqtau |N     }{   2  (2 \pi )^3    (r_{\Lambda} + c_0)^2 }
 \biggl[
\frac{1}{ | \bar{q}_{\bot}  | - w } 
  \nonumber
 \\ 
 & & 
  -   \frac{ | \bar{q}_{\bot} |}{ ( | \bar{q}_{\bot}  | - w )^2} 
 \ln \left( \frac{ | \bar{q}_{\bot}  | }{w} \right)
  \biggr],
 \label{eq:I2def}
 \\
  I^{(2 \omega) }_{\Lambda} 
 ( \barqtau ,  \bar{q}_{\parallel}, \bar{q}_{\bot} )  & \approx &
 \frac{ \gamma_{\Lambda}N}{( 2 \pi )^3 ( r_{\Lambda} + c_0 )^2}
 \Biggl[
 \frac{ \alpha \gamma_{\Lambda} \bar{q}_{\parallel} | \barqtau | }{
 \barqtau^2 + \gamma_{\Lambda}^2 \bar{q}_{\parallel}^2 }
 \nonumber
 \\
 & & -  i\,  {\rm sgn} (\barqtau)   | \bar{q}_{\bot} |
  \frac{  \barqtau^2 (  \barqtau^2 - \gamma_{\Lambda}^2 \bar{q}_{\parallel}^2  ) }{
 ( \barqtau^2 + \gamma_{\Lambda}^2 \bar{q}_{\parallel}^2)^2 }  
 \Biggr], \hspace{7mm}
 \label{eq:Itaudef}
 \\
  I^{(2k) }_{\Lambda} 
 ( \barqtau ,  \bar{q}_{\parallel}, \bar{q}_{\bot} )  & \approx &
 \frac{  i\,      {\rm sgn} (\barqtau)   \gamma_{\Lambda}^2   | \bar{q}_{\bot} | N
 }{( 2 \pi )^3 ( r_{\Lambda} + c_0 )^2}
  \frac{    \barqtau^2     ( \barqtau^2 
 - \gamma_{\Lambda}^2 \bar{q}_{\parallel}^2 )     }{
 ( \barqtau^2 + \gamma_{\Lambda}^2 \bar{q}_{\parallel}^2)^2 }  , \hspace{7mm}
 \label{eq:I2kdef}
 \end{eqnarray}
where, in Eq.~(\ref{eq:I2def}), the complex quantity $w$ is defined by
 \begin{equation}
 w = \frac{ b_0 }{2 ( r_{\Lambda} + c_0) }
 \left( | \barqtau | - i \alpha  \gamma_{\Lambda} 
 \bar{q}_{\parallel}    {\rm sgn} (\barqtau) \right).
 \end{equation} 

Let us now combine the contributions from all three diagrams
on the right-hand side of Fig.~\ref{fig:fullselfenergy}\,(b).
To be consistent with the approximations
made in the derivation of Eqs.~(\ref{eq:I2def})--(\ref{eq:I2kdef}),
we should also expand the right-hand sides of 
$ I^{(1)}_{\Lambda} 
 ( \barqtau ,  \bar{q}_{\parallel}, \bar{q}_{\bot} )$ and
$ I^{(1\omega)}_{\Lambda} 
 ( \barqtau ,  \bar{q}_{\parallel}, \bar{q}_{\bot} )$ 
in Eqs.~(\ref{eq:I1res}) and (\ref{eq:I1omegares})
for small $\bar{\epsilon}$.
We therefore define
 \begin{widetext} 
 \begin{subequations}
  \begin{eqnarray}
    I_{\Lambda} 
 ( \barqtau ,  \bar{q}_{\parallel}, \bar{q}_{\bot} )  &  = & 
 I^{(1)}_{\Lambda} 
 ( \barqtau ,  \bar{q}_{\parallel}, \bar{q}_{\bot} ) +  
 I^{(2)}_{\Lambda} 
 ( \barqtau ,  \bar{q}_{\parallel}, \bar{q}_{\bot} ) 
 \nonumber
 \\
 & = &
 \frac{ \gamma_{\Lambda}}{     (2 \pi )^2 ( r_{\Lambda} + c_0  )    }
 \Biggl\{
 \frac{  i  \barqtau  }{ i \barqtau - \gamma_{\Lambda} [ \alpha
 \bar{q}_{\parallel} + \bar{q}_{\bot}^2 + 2 \bar{q}_{\bot}  ] } + 
\frac{  i  \barqtau  }{ i \barqtau - \gamma_{\Lambda} [ \alpha
 \bar{q}_{\parallel} + \bar{q}_{\bot}^2 - 2 \bar{q}_{\bot}  ] }
 \nonumber
 \\
 & & \hspace{23mm} +  
  \frac{  N    }{   4 \pi   (r_{\Lambda} + c_0) }
 \Biggl[
\frac{  | \barqtau |    }{ | \bar{q}_{\bot}  | - w } 
  -   \frac{ | \bar{q}_{\bot} | | \barqtau |    }{ ( | \bar{q}_{\bot}  | - w )^2} 
 \ln \left( \frac{ | \bar{q}_{\bot} | } { w } \right) 
\Biggr]
 \Biggr\}, \hspace{7mm}
 \label{eq:Inulltot}
 \\
 I^{(\omega)}_{\Lambda} 
 ( \barqtau ,  \bar{q}_{\parallel}, \bar{q}_{\bot} ) &  = & 
 I^{(1\omega)}_{\Lambda} 
 ( \barqtau ,  \bar{q}_{\parallel}, \bar{q}_{\bot} ) +  
 I^{(2 \omega)}_{\Lambda} 
 ( \barqtau ,  \bar{q}_{\parallel}, \bar{q}_{\bot} ) 
 \nonumber
 \\
 & = &
  \frac{ \gamma_{\Lambda}  b_{\Lambda}   }{( 2 \pi )^2 ( r_{\Lambda} + c_0 )^2}
    \Biggl\{ -
 \frac{   | \barqtau | }{ i \barqtau - \gamma_{\Lambda} [ \alpha
 \bar{q}_{\parallel} + \bar{q}_{\bot}^2 + 2 \bar{q}_{\bot}  ] } 
 -
\frac{   | \barqtau | }{ i \barqtau - \gamma_{\Lambda} [ \alpha
 \bar{q}_{\parallel} + \bar{q}_{\bot}^2 - 2 \bar{q}_{\bot}  ] }
 \nonumber
 \\
 & &  \hspace{23mm} + \frac{N}{2 \pi b_{\Lambda}} 
\Biggl[
 \frac{ \alpha \gamma_{\Lambda} \bar{q}_{\parallel}  | \barqtau |   }{
 \barqtau^2 + \gamma_{\Lambda}^2 \bar{q}_{\parallel}^2 }
 -      | \bar{q}_{\bot} |
  \frac{  i \barqtau  | \barqtau |  (  \barqtau^2 - \gamma_{\Lambda}^2 \bar{q}_{\parallel}^2  ) }{
 ( \barqtau^2 + \gamma_{\Lambda}^2 \bar{q}_{\parallel}^2)^2 }  
 \Biggr]  \Biggr\},
 \label{eq:Itautot}
 \\
  I^{(k) }_{\Lambda} 
 ( \barqtau ,  \bar{q}_{\parallel}, \bar{q}_{\bot} )  & = &
 I^{(2k) }_{\Lambda} 
 ( \barqtau ,  \bar{q}_{\parallel}, \bar{q}_{\bot} )
=
 \frac{      \gamma_{\Lambda}^2   | \bar{q}_{\bot} |N    
 }{( 2 \pi )^3 ( r_{\Lambda} + c_0 )^2}
  \frac{   i \barqtau  |  \barqtau |    ( \barqtau^2 
 - \gamma_{\Lambda}^2 \bar{q}_{\parallel}^2 )     }{
 ( \barqtau^2 + \gamma_{\Lambda}^2 \bar{q}_{\parallel}^2)^2 } .
 \label{eq:Ipartot}
 \end{eqnarray}
 \end{subequations}
\end{widetext}
Finally, we integrate over the flow parameter $\Lambda$,
and obtain the following expressions for the 
three-legged boson-fermion vertex,
\begin{subequations}
  \begin{eqnarray}
   {\Gamma}^{\bar{\psi}^{\alpha} \psi^{\alpha} \phi }
  (   \bar{K}    ; 0  ; \bar{K}   )  & =  &   \tilde{\Gamma}_{\Lambda}
 ( \barqtau , \bar{q}_{\parallel} , \bar{q}_{\bot} ),
 \label{eq:Gammanull}
 \\
 \left. \frac{ \partial 
{\Gamma}^{\bar{\psi}^{\alpha} \psi^{\alpha} \phi } 
  (   \bar{K}    ; K  ; \bar{K}   )}{\partial ( i \ktau ) } \right|_{ K=0}  & =  &\frac{1}{\Lambda^{z_b}} 
 \tilde{\Gamma}^{\omega  }_{\Lambda}
 ( \barqtau , \bar{q}_{\parallel} , \bar{q}_{\bot} ),
 \label{eq:Gammaomega}
 \hspace{7mm}
 \\
 \left. \frac{ \partial 
{\Gamma}^{\bar{\psi}^{\alpha} \psi^{\alpha} \phi } 
  (   \bar{K}    ; K  ; \bar{K}   )}{\partial ( \alpha  k_{\parallel} ) } 
 \right|_{ K=0}  & = & \frac{1}{\Lambda^{2}} 
 \tilde{\Gamma}^{k  }_{\Lambda}
 ( \barqtau , \bar{q}_{\parallel} , \bar{q}_{\bot} ),
 \label{eq:Gammak} 
\end{eqnarray}
 \end{subequations}
where
 \begin{subequations}
 \begin{eqnarray}
 & & \tilde{\Gamma}_{\Lambda}
 ( \barqtau , \bar{q}_{\parallel} , \bar{q}_{\bot} )   = 1   +
 \int_{ \Lambda / \Lambda_0}^1 \frac{ ds}{s} I_{  \Lambda /s} 
 \left( s^{z_b}  \barqtau , s^2 \bar{q}_{\parallel},
 s \bar{q}_{\bot}  \right),
 \hspace{7mm}
 \nonumber
 \\
 & &
 \label{eq:xint1} 
 \\
 & & \tilde{\Gamma}^{\omega }_{\Lambda}
 ( \barqtau , \bar{q}_{\parallel} , \bar{q}_{\bot} )   = 
 \int_{ \Lambda / \Lambda_0}^1  ds\,s^{z_b-1} I^{\omega}_{  \Lambda /s} 
 \left( s^{z_b}  \barqtau , s^2 \bar{q}_{\parallel},
 s \bar{q}_{\bot}  \right),
 \nonumber
 \\
 & &
 \label{eq:xint2}
 \\
  &  & \tilde{\Gamma}^{k }_{\Lambda}
 ( \barqtau , \bar{q}_{\parallel} , \bar{q}_{\bot} ) 
 = \int_{ \Lambda / \Lambda_0}^1  ds\,s I^{k}_{  \Lambda /s} 
 \left( s^{z_b}  \barqtau , s^2 \bar{q}_{\parallel},
 s \bar{q}_{\bot}  \right).
 \label{eq:xint3}
 \end{eqnarray}
\end{subequations}
Recall that in deriving these expressions we have assumed that
$ | \barqtau | \lesssim 1$.

\subsection{Fermionic anomalous dimension and dynamic exponent}
\label{subsec:anomalous}

\begin{figure}[tb]    
   \centering
\vspace{7mm}
  \includegraphics[width=0.45\textwidth]{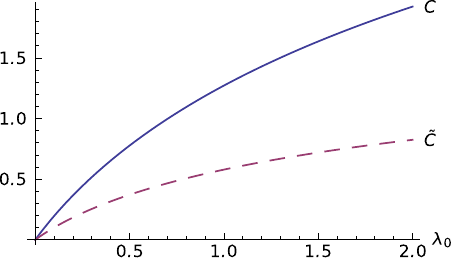}
  \caption{%
(Color online) 
Plot of the cutoff functions $C(\lambda_0)$ and $\tilde{C}(\lambda_0)$
defined in Eqs.~(\ref{eq:Cdef}) and (\ref{eq:Ctildedef}).
}
    \label{fig:Cplot}
  \end{figure}

We are now ready to calculate the anomalous dimensions 
$\eta = \lim_{\Lambda \rightarrow 0} \eta_{\Lambda}$ and
$\tilde{\eta} = \lim_{\Lambda \rightarrow 0} \tilde{\eta}_{\Lambda}$
at the quantum critical point.
We therefore substitute
Eqs.~(\ref{eq:Gammanull})--~(\ref{eq:Gammak})
into our general relations
(\ref{eq:etares2}) and (\ref{eq:baretares2}) for the flowing anomalous dimensions
${\eta}_{\Lambda}$ and $\tilde{\eta}_{\Lambda}$ and, introducing the
integration variables
$p = \gamma_{\Lambda} \alpha \bar{q}_{\parallel}$ and
$y = \barqtau$, we obtain for the flowing anomalous dimension associated with the
frequency dependence of the self-energy,
 \begin{widetext}
\begin{eqnarray}
\eta_{\Lambda} & = & 
 - 
\frac{\gamma_{\Lambda}}{2 \pi} \lim_{ \omega \rightarrow 0}
 \frac{\partial}{\partial  ( i \omega)  }
  \int_{ - \lambda_0 }^{\lambda_0} \frac{dy}{2 \pi}
 \int_{ - \infty}^{\infty} \frac{dp}{2 \pi}
 \left[
 \frac{1}{r_{\Lambda} + c_0 + b_{0}  | y | }
 \right]
 \left[ \frac{1}{ i y+ i \omega  - p - \gamma_{\Lambda} }
- 
\frac{1}{ i y - p - \gamma_{\Lambda} }
 \right]
\left[ \tilde{\Gamma}_{\Lambda} 
 \left( y, \frac{p}{\alpha \gamma_{\Lambda}} , 1 \right) \right]^2
 \nonumber
 \\
 &  & -
\frac{\gamma_{\Lambda}}{\pi}
  \int_{ - \lambda_0}^{\lambda_0} \frac{dy}{2 \pi}
 \int_{ - \infty}^{\infty} \frac{dp}{2 \pi}
 \left[
 \frac{1}{r_{\Lambda} + c_0 + b_{0}  | y | }
 \right]
 \left[
 \frac{1}{ i y - p - \gamma_{\Lambda} }
 \right]
  \tilde{\Gamma}_{\Lambda} \left( y, \frac{p}{\alpha \gamma_{\Lambda}} , 1 \right)
 \tilde{\Gamma}^\omega_{\Lambda} \left( y, \frac{p}{\alpha \gamma_{\Lambda}} , 1 \right),
 \label{eq:etaint1}
 \end{eqnarray}
and for the corresponding anomalous dimension associated with
the momentum dependence of the self-energy,
 \begin{eqnarray}
\tilde{\eta}_{\Lambda} & = &  
\frac{\gamma_{\Lambda}}{2 \pi} \lim_{ k \rightarrow 0}
 \frac{\partial}{\partial k }
  \int_{ - \lambda_0}^{\lambda_0} \frac{dy}{2 \pi}
 \int_{ - \infty}^{\infty} \frac{dp}{2 \pi}
 \left[
 \frac{1}{r_{\Lambda} + c_0 + b_{0}  | y | }
 \right]
 \left[ \frac{1}{ i y  - p - k - \gamma_{\Lambda} }
- 
\frac{1}{ i y - p - \gamma_{\Lambda} }
 \right]
\left[ \tilde{\Gamma}_{\Lambda} 
 \left( y, \frac{p}{\alpha \gamma_{\Lambda}} , 1 \right) \right]^2
 \nonumber
 \\
 & & +
\frac{1}{\pi}
  \int_{ - \lambda_0}^{\lambda_0} \frac{dy}{2 \pi}
 \int_{ - \infty}^{\infty} \frac{dp}{2 \pi}
 \left[
 \frac{1}{r_{\Lambda} + c_0 + b_{0}  | y | }
 \right]
 \left[
 \frac{1}{ i y - p - \gamma_{\Lambda} }
 \right]
  \tilde{\Gamma}_{\Lambda} \left( y, \frac{p}{\alpha \gamma_{\Lambda}} , 1 \right)
 \tilde{\Gamma}^k_{\Lambda} \left( y, \frac{p}{\alpha \gamma_{\Lambda}} , 1 \right).
 \label{eq:etabarint1}
\end{eqnarray}
Here, $\lambda_0$ is an ultraviolet cutoff of the order of unity which takes into account
that our expressions (\ref{eq:Inulltot})--(\ref{eq:Ipartot}) which we use to calculate the
vertices in Eqs.~(\ref{eq:xint1})--(\ref{eq:xint3})
are only valid for small frequencies.

Consider now the  
limit $\Lambda \rightarrow 0$. At the quantum critical point
we may then set $r_{\Lambda} \rightarrow 0$.
For simplicity, we also set $c_0 = b_{0} = N / (4 \pi )$. 
To make progress analytically, let us further assume that 
the parameter
$\gamma =\lim_{\Lambda\rightarrow0}\gamma_\Lambda=2 \pi^2 c_0 (z_b -2 ) = \frac{\pi}{2} N ( z_b -2 )$ is small compared with unity.
To leading order in $z_b -2$ the $s$-integrations in Eqs.~(\ref{eq:xint1})--~(\ref{eq:xint3})
can then be carried out analytically, with the result
 \begin{eqnarray}
  \lim_{\Lambda \rightarrow 0} \tilde{\Gamma}_{\Lambda} 
 \left( y , \frac{p}{\alpha \gamma_{\Lambda}} , 1 \right) 
 & = &  1 
+ \frac{\gamma  }{ (2 \pi)^2 c_0} 
 \biggl\{
\frac{ iy }{iy-p - \gamma} \ln
 \Bigl[ 1 - \Bigl( \frac{ iy - p - \gamma }{2 \gamma } \Bigr)^2 \Bigr] 
  \nonumber
 \\
 & &  \hspace{20mm}
 - \frac{iyN}{2 \pi c_0}
 \biggl[
 \frac{ \ln \bigl[ \frac{ 2 i\, {\rm sgn} (y) }{ iy + p }  \bigr]}{ 2 i\, {\rm sgn} (y) - (iy + p ) }
   + 2 \frac{\ln \bigl[ 1 - \frac{iy+p}{ 2 i\, {\rm sgn} (y) } \bigr]}{iy+p}
 \biggr] 
\biggr\},
 \hspace{7mm}
 \label{eq:vertexres1}
 \\
  \lim_{\Lambda \rightarrow 0}
 \tilde{\Gamma}^\omega_{\Lambda} 
 \left( y , \frac{p}{\alpha \gamma_{\Lambda}} , 1 \right) 
& = &
\frac{\gamma}{ (2 \pi)^2 c_0}
  \biggl\{ - \frac{ | y |}{iy - p }
+
 \frac{N}{2 \pi c_0} 
 \biggl[ \frac{  | y |p }{ y^2 + p^2}  -  \frac{ iy | y | ( y^2 - p^2 )}{ 3 (y^2 + p^2)^2 }
 \biggr] \biggr\},
 \label{eq:vertexres2}
 \\
  \lim_{\Lambda \rightarrow 0}
 \tilde{\Gamma}^k_{\Lambda} 
 \left( y , \frac{p}{\alpha \gamma_{\Lambda}} , 1 \right) 
& = &
 \frac{\gamma^2N}{(2 \pi )^3 c_0^2}
  \frac{ iy | y | ( y^2 - p^2 )}{ 3 (y^2 + p^2)^2 }.
 \label{eq:vertexres3}
 \end{eqnarray}
 \end{widetext}
Note that these vertex functions describe the renormalized effective interaction between
two fermions and one boson; clearly, this interaction has a rather complicated dependence
on momenta and frequencies and cannot be approximated by a constant.
%
%
If we now substitute Eqs.~(\ref{eq:vertexres1})--(\ref{eq:vertexres3})
into Eqs.~(\ref{eq:etaint1}) and (\ref{eq:etabarint1}), we may
perform the $p$-integration analytically  
by means of the method of residues.
Note that the first 
term in the curly braces of Eq.~(\ref{eq:vertexres1}) and the first term in 
the curly braces of Eq.~(\ref{eq:vertexres2})
do not contribute to the integrals, because we may close the
integration contour in a half plane where the integrand is analytic.
These terms arise from the first diagram on the right-hand side of
Fig.~\ref{fig:fullselfenergy}\,(b), so that the
three-boson vertex is crucial to obtain the leading corrections to the anomalous dimensions. 
We finally arrive at the following estimates for the
anomalous dimensions at the nematic quantum critical point,
\begin{eqnarray}
 \eta & = & z_b -2 +   \frac{( z_b-2)^2}{2} C ( \lambda_0 ) + {\cal{O}} ( (z_b -2 )^3 ) ,
 \label{eq:etafix}
 \\
 \tilde{\eta} & = & \frac{ ( z_b-2)^2}{2} \tilde{C} ( \lambda_0 )  + {\cal{O}} ( (z_b -2 )^3 )   ,
 \label{eq:etabarfix}
 \end{eqnarray}
where the cutoff functions are given by
 \begin{eqnarray}
 C ( \lambda_0 ) & = & \int_0^{\lambda_0} \frac{dy}{1+y} \left[
 \frac{7}{6} + \frac{1}{1-y} + \frac{y \ln y }{(1-y)^2} \right],
 \label{eq:Cdef}
 \\
 \tilde{C} ( \lambda_0 ) & = & \int_0^{\lambda_0} \frac{dy}{1+y} \left[
 \frac{1}{6} + \frac{1}{1-y} + \frac{y \ln y }{(1-y)^2} \right]
 \nonumber
 \\
  & = & C ( \lambda_0 ) - \ln ( 1 + \lambda_0 ).
 \label{eq:Ctildedef}
 \end{eqnarray}
A plot of the cutoff functions 
is shown in Fig.~\ref{fig:Cplot}.
Note that these functions depend logarithmically on the
ultraviolet cutoff, $\lambda_0$. This is due to the fact that we have
assumed that the rescaled frequencies and momenta in
 Eqs.~(\ref{eq:I2def})--(\ref{eq:I2kdef})
are small compared with unity in the evaluation of 
the vertex-correction diagrams shown in Fig.~\ref{fig:fullselfenergy}\,(b).
By retaining the full momentum- and frequency dependence
on the right-hand side of the flow equation (\ref{eq:flowGamma2int}),
one obtains ultraviolet convergent results from our FRG approach. However, as the cutoff dependence in our final result, Eqs.~(\ref{eq:etafix}) and (\ref{eq:etabarfix}), is only logarithmic, we have not attempted to carry out this calculation which requires substantial  
numerical effort. Instead, we make the reasonable cutoff choice $\lambda_0=1$ and find, within the accuracy of our calculation, that $C\approx 1.27$ and $\tilde{C}\approx 0.58$. 



Given our result for the anomalous dimensions $\eta$ and $\tilde{\eta}$,
we may relate these to the anomalous dimension
$\eta_\psi$ of the fermion field 
and the fermionic dynamic exponent, $z$.
Let us therefore recall that, at the quantum critical point,
the retarded single-particle Green function assumes for small frequencies
and momenta the following scaling form, 
\begin{equation}
  \label{eq:Gscale}
  G^{\alpha} (\omega + i0^+,\bm{k} )  \propto  
\frac{1}{\left[A_\omega\,\text{sgn}({\omega}) 
      |\omega|^{1/z} - \xi_{\bm{k}}^{\alpha} \right]^{1-\eta_\psi/2} },
\end{equation}
where $A_\omega$ is some dimensionful constant with positive imaginary part and real
part depending on the sign of $\omega$, see Eq.~(\ref{eq:GlowenergyMetlitskiSachdev}).
For $\omega = k_{\bot}=0$, this implies
 \begin{equation}
  G^{\alpha} (i0^+ , k_{\parallel} , 0)  \propto   k_{\parallel}^{ - 1 + \eta_{\psi} /2 } .
 \label{eq:Galpha1}
 \end{equation}
On the other hand, from the definition of  $\tilde{\eta}_{\Lambda}$ 
in Eq.~(\ref{eq:etaetadef}) we see that
$ \tilde{Z}_{\Lambda} \propto \Lambda^{\tilde{\eta} } \propto k_{\parallel}^{\tilde{\eta}/2}$,
where we have used the fact that $k_{\parallel}$ scales as $\Lambda^2$.
Hence, 
 \begin{equation}
  G^{\alpha} (i0^+ , k_{\parallel} , 0 )  \propto  \tilde{Z}_{\Lambda} / k_{\parallel}
 \propto
 k_{\parallel}^{ - 1 + \tilde{\eta}/2  } .
 \label{eq:Galpha2}
 \end{equation}
Comparing this with Eq.~(\ref{eq:Galpha1}), we conclude that
 \begin{equation}
  \eta_{\psi} = \tilde{\eta} .
 \end{equation}
Next, setting $\bd{k} =0$ in Eq.~(\ref{eq:Gscale}), we find
\begin{equation}
  \label{eq:Gscaleomega1}
  G^{\alpha} (\omega + i0^+ ,0)  \propto  
 \omega^{ - ( 1 - \eta_{\psi} / 2 ) / z }.
\end{equation}
On the other hand, 
from the definition of  ${\eta}_{\Lambda}$ 
in Eq.~(\ref{eq:etaetadef}) we infer that
$ {Z}_{\Lambda} \propto \Lambda^{{\eta} } \propto \omega^{  {\eta} / (2z) }$
(using the fact that  $\omega$ scales as $\Lambda^{2 z}$), 
so that
\begin{equation}
  \label{eq:Gscaleomega2}
  G^{\alpha} (0,\omega + i0^+ )  \propto  Z_{\Lambda} / \omega  \propto
 \omega^{ - ( 1- \eta / (2z) )  }.
\end{equation}
Comparing this with Eq.~(\ref{eq:Gscaleomega1}), we see that
 $1- \eta / (2z) = ( 1 - \eta_{\psi} / 2 ) / z$, or
  \begin{equation}
 z = 1 + \frac{\eta - \tilde{\eta}}{2} .
 \label{eq:zzb}
 \end{equation}
Writing Eq.~(\ref{eq:etafix}) as $\eta = z_b -2 + \delta \eta $ 
 with 
\begin{equation}
 \delta \eta  = 
\frac{( z_b-2)^2}{2} C ( \lambda_0 ) + {\cal{O}} ( (z_b -2 )^3 ) ,
 \label{eq:deltaetafix}
 \end{equation}
we see that Eq.~(\ref{eq:zzb}) can also be written as
 \begin{eqnarray}
z & =  & \frac{z_b}{2}  + \frac{\delta \eta - \tilde{\eta}}{2} 
 \nonumber
 \\
  & = &  \frac{z_b}{2}    +   \frac{ (z_b-2)^2}{4} 
 \Bigl[  {C} ( \lambda_0 ) - \tilde{C} ( \lambda_0 ) \Bigr]  +
  {\cal{O}} ( (z_b -2 )^3 ) .
 \nonumber
 \\
 & &
 \label{eq:zres}
 \end{eqnarray}
According to the above calculation, 
the scaling relation $z = z_b /2$ for the 
fermionic dynamic exponent acquires
a correction of order $ (z_b -2 )^2$, such that the effective theory appears to have two different time scales.
However, according to Metlitski and Sachdev,\cite{Metlitski10a} the strong correlations imply that $z = z_b /2$ should be satisfied exactly.
This discrepancy 
might be due to the fact that our truncation of the FRG flow equations 
introduces approximations which violate the general scaling theory.
In fact, this seems to be true also in the case of the loop expansion.
In the previous works of Metlitski and Sachdev\cite{Metlitski10a} and
Mross {\em et al.},\cite{Mross10}
the three-loop correction to the fermionic anomalous 
dimension $\eta_{\psi}$ is calculated via the momentum
dependence of the self-energy, and according to scaling theory it is argued that a 
corresponding correction should appear if one calculates $\eta_{\psi}$ 
via the frequency dependence of the self-energy. 
However, as pointed out in Ref.~\onlinecite{Chubukov10},
the correction in the frequency dependence of the self-energy 
(the term $\delta \eta$ in our notation)
appears only beyond the three-loop order.
The FRG is not based on an expansion in powers of loops, so that diagrams beyond
three loops are \mbox{included in our truncation.}


\section{Summary and conclusions}
\label{sec:conclusions}

In this work, we have used a functional renormalization group approach to
calculate the anomalous dimension $\eta_{\psi}$ of the fermion field
and the fermionic dynamic exponent $z$ 
of an effective low-energy field theory describing the
Ising-nematic quantum critical point in two-dimensional metals.
In the limit  $N ( z_b -2 ) \ll 1 $ (where $N$ is the number of fermion flavors and
$z_b$ is the bosonic dynamic exponent), we have been able to
explicitly calculate the  fermionic anomalous dimension $\eta_{\psi}$
of the system, with the result
 \begin{equation}
 {\eta}_{\psi}   \approx 0.3 \left(z_b-2\right)^2   + {\cal{O}} ( (z_b -2 )^3 )  .
 \label{eq:psires}
 \end{equation}
If we extrapolate this expression to the physically relevant case, $z_b =3$, 
we obtain $\eta_{\psi} \approx 0.3$,  which is 
larger than the estimate $\eta_{\psi} \approx 0.068$ given by
Metlitski and Sachdev,\cite{Metlitski10a}
but still 
smaller than
the estimate  $\eta_{\psi} \approx 0.6$ obtained from an extrapolation of the
corresponding
expression given by Mross {\it{et al.}}.\cite{Mross10} 
{
Given the 
different types of approximations, it is not surprising that different values for the exponent
$\eta_\psi$
are found. This fact shows that the recent calculations done so far are not yet fully under control
in the physically relevant case of $z_b =3$ and $N=2$.
It is however reassuring that our method also finds a finite value for $\eta_\psi$.
In particular, we note that in the previous calculations based on the 
field-theoretical renormalization group,  
the values of the exponent $\eta_\psi$ are obtained within a loop-expansion truncated at the
three-loop order, while in our FRG approach the truncation does not rely on
a loop-expansion, so that a certain class of diagrams beyond the three-loop level are effectively re-summed
to higher  order  within our truncation. 
On the other hand, 
at the lowest order, namely at the one-loop level in the field theoretical RG and
neglecting the vertex corrections in the FRG, the results obtained using the FRG 
and other renormalization group methods all
coincide with the well-known results obtained within the RPA. 
}
Finally, in contrast to previous works,\cite{Metlitski10a,Mross10} 
we have explicitly shown that both the frequency and the momentum dependence 
of the self-energy give rise to anomalous corrections to the one-loop result.
While our calculations in principle lead to a small correction term to 
the fermionic dynamic critical exponent, $z$, the scaling theory of 
Metlitski and Sachdev\cite{Metlitski10a} implies the exact identity
$z=z_b/2$.


The calculations presented in this work can be extended in several directions.
Because our FRG approach does not rely on the smallness of the parameter $N ( z_b -2)$,
with some numerical effort
it should be possible to extract 
the anomalous dimension $\eta_{\psi}$ for
general $N $ and $z_b -2$ from Eqs.~(\ref{eq:etaint1}) and (\ref{eq:etabarint1}).
In this case the vertices appearing in these expressions
cannot be calculated analytically but must be represented as one-dimensional
integrals, so that the evaluation of the anomalous dimensions in 
 Eqs.~(\ref{eq:etaint1}) and (\ref{eq:etabarint1}) requires 
rather complicated numerical integrations, which is beyond the scope of this work.
In principle our FRG flow equations also allow us to calculate the
entire frequency-dependence of the self-energies, but this 
seems to be numerically even more expensive.

In contrast to the strategy adopted in 
the Hertz-Millis approach,\cite{Hertz76,Millis93} 
in the present  problem it is not possible to integrate over the
fermionic degrees of freedom to obtain an effective bosonic theory with
regular vertices. We have therefore explicitly
retained both bosonic and fermionic degrees of freedom in
our FRG calculation. 
The FRG approach developed in this work should also be useful to discuss
other model systems  where gapless fermionic and bosonic excitations 
are strongly coupled.

\begin{figure}[bt]    
   \centering
  \includegraphics[width=0.45\textwidth]{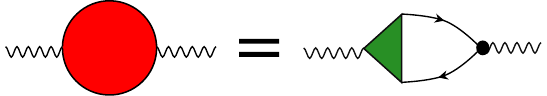}
  \caption{%
(Color online) 
Exact skeleton equation relating the bosonic self-energy
to the exact fermionic propagators and the three-legged vertex
with two fermion and one boson leg, see Eq.~(\ref{eq:skeletonPi}).
The black dot denotes the bare
vertex with two fermionic and one bosonic external leg
given in Eq.~(\ref{eq:Gamma0}).
}
    \label{fig:skeletonpolar}
  \end{figure}
\vspace{.5cm}

\section*{ACKNOWLEDGMENTS}
We would like to thank Max Metlitski for useful discussions.
This work was financially supported by the DFG via FOR 723.

\begin{appendix}

\section*{Appendix A: Skeleton equations}
\setcounter{equation}{0}
\renewcommand{\theequation}{A\arabic{equation}}
\label{sec:appendixskeleton}

In sections~\ref{sec:novertexcorrections} and \ref{sec:vertex}, we have combined
FRG flow equations with skeleton equations for the bosonic self-energy
and the bosonic three-legged vertex to obtain a closed system of equations.
In this appendix, we briefly describe the derivation of these skeleton equations.

Skeleton equations relating
vertex functions of different order
follow from the general Dyson-Schwinger equation 
for the generating functional for the connected Green function, 
which is a simple
consequence of the
invariance of the integration measure of the functional integral under infinitesimal shifts.
The derivation of the skeleton equation for the
bosonic self-energy of models  of the type considered in this work has
 been discussed in detail in Refs.~\onlinecite{Bartosch09a,Kopietz10}, so
let us here only quote the result,
\begin{align}
   & {\Pi} ( \bar{K} ) = 
 \int_K \sum_{\alpha, \sigma }
  { G}^{\alpha} ( K  ) G^{\alpha} ( K + \bar{K} ) 
\nonumber \\
& \qquad \qquad  {} \times \Gamma_0^{\alpha}
  \Gamma^{ \bar{\psi}^{\alpha} \psi^{\alpha} \phi  } ( K; K + \bar{K} ; -  \bar{K}  ).
  \label{eq:skeletonPi}
\end{align}
This exact identity is shown diagrammatically in Fig.~\ref{fig:skeletonpolar}. 

To derive the skeleton equation for the
three-boson vertex, 
let us start from the Dyson-Schwinger equation 
given in Eq.~(11.27a) of Ref.~\onlinecite{Kopietz10}.
After taking two successive derivatives 
with respect to $\phi_{{\bar K}_2}$ and $\phi_{{\bar K}_3}$, we obtain
\begin{equation}
  \label{eq:DSGamma3}
  \frac{\delta^3 \Gamma}{\delta \phi_{{\bar K}_1} \, \delta \phi_{{\bar K}_2} \, \delta \phi_{{\bar K}_3}} = \sum_{\alpha,\sigma} \Gamma_0^{\alpha} \int_{ {K}} 
  \frac{ \delta^{4} {\mathcal{G}}_c}{ 
      \delta \bar{\jmath}_{ K}^\alpha  \delta j_{ K  + \bar{K}_1 }^\alpha  \delta \phi_{ \bar{K}_2} \delta \phi_{ \bar{K}_3}} .
\end{equation}
Here, $ \Gamma [ \bar{\psi} , \psi , \phi ]$
is the generating functional of the one-line irreducible vertices, and
${\cal{G}}_c [ \bar{j} , j , J ]$ is the generating functional
of the connected Green functions, and is a functional of the
sources $\bar{j} , j$ and $J$ conjugate to the fields $\psi, \bar{\psi}$, and $\phi$. 
We now use Eq. (6.82) of Ref.~\onlinecite{Kopietz10} and set all external fields equal to zero. The desired skeleton equation can then be written as
\begin{widetext}
\begin{align}
  & \Gamma^{\phi\phi\phi}(\bar{K}_1, \bar{K}_2, - \bar{K}_1  - \bar{K}_2) = 
 N\sum_\alpha  \Gamma_0^{\alpha}       \left[
 \int_K G^\alpha(K) G^\alpha(K + \bar{K}_1) \Gamma^{\bar{\psi}^\alpha \psi^\alpha \phi \phi} (K ; K + \bar{K}_1; \bar{K}_2, -\bar{K}_1 - \bar{K}_2) \right. \nonumber \\
&\left.  {}  + \int_K G^\alpha(K) G^\alpha(K + \bar{K}_1) G^\alpha(K + \bar{K}_1 + \bar{K}_2)  \right. \nonumber \\
& \left. \qquad \times \Gamma^{\bar{\psi}^\alpha \psi^\alpha \phi} (K + \bar{K}_1 + \bar{K}_2; K + \bar{K}_1 ; \bar{K}_2) \Gamma^{\bar{\psi}^\alpha \psi^\alpha \phi} (K; K + \bar{K}_1 + \bar{K}_2; -\bar{K}_1 - \bar{K}_2) \right. \nonumber \\
 & \left. {} +  \int_K G^\alpha(K) G^\alpha(K + \bar{K}_1) G^\alpha(K - \bar{K}_2) \Gamma^{\bar{\psi}^\alpha \psi^\alpha \phi} (K - \bar{K}_2; K + \bar{K}_1 ; - \bar{K}_1 - \bar{K}_2) \Gamma^{\bar{\psi}^\alpha \psi^\alpha \phi} (K; K - \bar{K}_2; \bar{K}_2) \right] .
\label{eq:SkeletonThreeLegged}
\end{align}
A graphical representation of this equation is given in Fig.~\ref{fig:Skeleton3leg}.
\begin{figure}[tb]
    \centering
    \includegraphics[width=15.5cm]{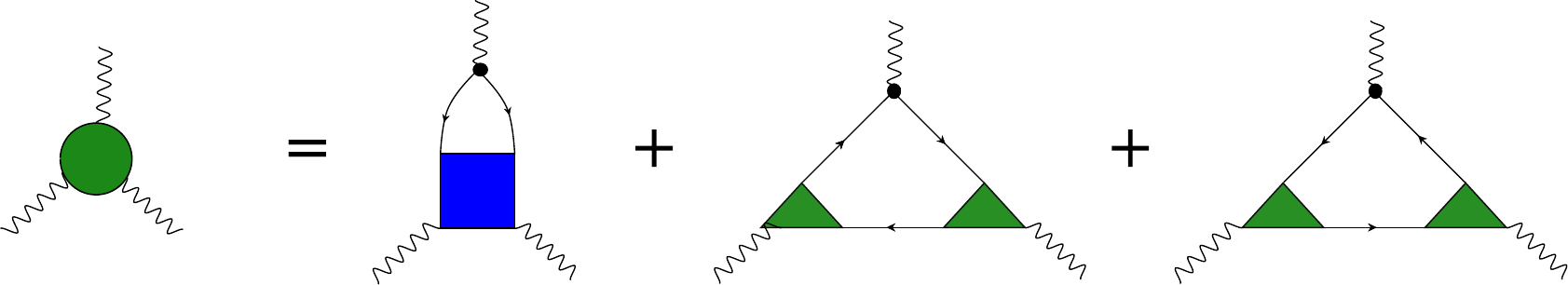}
    \caption{(Color online) Graph of the skeleton equation 
(\ref{eq:SkeletonThreeLegged}) for the three-legged bosonic vertex.}
 \label{fig:Skeleton3leg}
  \end{figure}
Let us check a known limit of this equation. After replacing the exact 
propagators $ G^\alpha(K)$ by  $ G_0^\alpha(K)$, 
the exact vertices $\Gamma^{\bar{\psi}^\alpha \psi^\alpha \phi} (K_1; K_2; \bar{K})$ 
by the bare vertices $\Gamma_0^{\alpha}$, and 
neglecting
the contribution involving  
$\Gamma^{\bar{\psi}^\alpha \psi^\alpha \phi \phi} (K_1; K_2; \bar{K}_1, \bar{K}_2)$,
we obtain from Eq.~(\ref{eq:SkeletonThreeLegged})
\begin{align}
    & \Gamma_0^{\phi\phi\phi}(\bar{K}_1, \bar{K}_2, - \bar{K}_1  - \bar{K}_2) = N\sum_\alpha \left(\Gamma_0^\alpha \right)^3 \left[
   \int_K G_0^\alpha(K) G_0^\alpha(K + \bar{K}_1) G_0^\alpha(K + \bar{K}_1 + \bar{K}_2)  \right. \nonumber \\
 & \qquad \qquad \qquad \qquad \qquad \qquad  \qquad \qquad \left. {} +  \int_K G_0^\alpha(K) G_0^\alpha(K + \bar{K}_1) G_0^\alpha(K - \bar{K}_2)  \right] ,
\label{eq:SkeletonThreeLegged0}
\end{align}
\end{widetext}
which is invariant under arbitrary permutations of $\bar{K}_1$, $\bar{K}_2$, and $\bar{K}_3 = - \bar{K}_1  - \bar{K}_2$.
In fact, in this approximation, the bosonic three-legged vertex can be identified
with the symmetrized closed fermion loop with three external legs and bare propagators
given in Eq.~(\ref{eq:initial}).
As concerns Eq.~(\ref{eq:SkeletonThreeLegged}), the symmetry under permutations of its arguments is less obvious. However, as the left-hand side of Eq.~(\ref{eq:DSGamma3}) is symmetric under permutations, and because all manipulations are exact, Eq.~(\ref{eq:SkeletonThreeLegged})  indeed fulfills this symmetry.

To obtain the approximation for the bosonic three-loop used
in Sec.~\ref{sec:vertex}, we adopt the same approximation strategy as in the
skeleton approximation used for the bosonic self-energy:
replacing the vertices $\Gamma^{\bar{\psi}^\alpha \psi^\alpha \phi} (K_1; K_2; \bar{K})$ and $\Gamma^{\bar{\psi}^\alpha \psi^\alpha \phi \phi} (K_1; K_2; \bar{K}_1, \bar{K}_2)$ by the 
bare vertices $\Gamma_0^\alpha$ and $0$, but retaining dressed
propagators $G^\alpha(K)$, we arrive at
Eqs.~(\ref{eq:Gamma3phires}) and (\ref{eq:L3resmod}).

\section*{Appendix B: The symmetrized three-loop}
\setcounter{equation}{0}
\renewcommand{\theequation}{B\arabic{equation}}
\label{sec:threeloop}

In the momentum-transfer cutoff scheme, all irreducible vertices involving only
bosonic external legs are finite at the initial scale and can be identified by the
symmetrized closed fermion loops with bare fermionic propagators.
In particular, the initial condition for the vertex with three external boson legs is
\begin{equation}
 \Gamma^{\phi \phi \phi }_0 ( \bar{K}_1 , \bar{K}_2 , \bar{K}_3 )  
=  2!N \sum_{\alpha}   (\Gamma_0^{\alpha} )^3 L^{\alpha}_{3} ( - \bar{K}_1, - \bar{K}_2 , - \bar{K}_3 ),
 \label{eq:initial}
 \end{equation} 
where the symmetrized three-loop 
$L^{ \alpha}_{3} (  \bar{K}_1,  \bar{K}_2 ,  \bar{K}_3 )  $
can be expressed in terms of the non-symmetrized three-loop
 $\bar{L}^\alpha_3 ( \bar{K}_1 , \bar{K}_2, \bar{K}_3 )$, defined by
 \begin{equation}
 \bar{L}^\alpha_3 ( \bar{K}_1 , \bar{K}_2, \bar{K}_3 ) = \int_K G_0^{\alpha} ( K - \bar{K}_1 )
 G_0^{\alpha} ( K - \bar{K}_2 ) G_0^{\alpha} ( K - \bar{K}_3 ) 
 \label{eq:L3}
 \end{equation}
as follows,
 \begin{eqnarray}
 & &  L^{ \alpha}_{3} (  \bar{K}_1,  \bar{K}_2 ,  \bar{K}_3 )  
 \nonumber
 \\
& = &
 \frac{1}{3!} \Bigl[
 \bar{L}^{\alpha}_3 (  \bar{K}_1 , \bar{K}_1 + \bar{K}_2 ,0 ) 
 +   \bar{L}^{\alpha}_3 (  \bar{K}_2 , \bar{K}_1 + \bar{K}_2 ,0 )
 \nonumber
 \\
 & & \hspace{3mm} + \bar{L}^{\alpha}_{3} (  \bar{K}_2 , \bar{K}_2 + \bar{K}_3,0  ) 
 +   \bar{L}^{\alpha}_{3}  (  \bar{K}_3 , \bar{K}_2 + \bar{K}_3 ,0  )
 \nonumber
 \\
 & & \hspace{3mm}   + \bar{L}^{\alpha}_{3}  (  \bar{K}_3 , \bar{K}_3 + \bar{K}_1, 0  ) 
 +     \bar{L}^{\alpha}_{3}   ( \bar{K}_1 , \bar{K}_3 + \bar{K}_1 ,0 )
 \Bigr]. \hspace{9mm}
 \end{eqnarray}
Actually, taking into account energy-momentum conservation,
we may set $\bar{K}_3 = - \bar{K}_1 - \bar{K}_2$, so that
we need
 \begin{eqnarray}
 & &  L^{ \alpha}_{3} (  \bar{K}_1,  \bar{K}_2 ,  - \bar{K}_1 - \bar{K}_2 )  
 \nonumber
 \\
& = &
 \frac{1}{3!} \Bigl[
 \bar{L}^{\alpha}_3 (  \bar{K}_1 , \bar{K}_1 + \bar{K}_2 ,0 ) 
 +   \bar{L}^{\alpha}_3 (  - \bar{K}_1 , - \bar{K}_1 - \bar{K}_2 ,0 )
 \nonumber
 \\
 & & \hspace{3mm} + \bar{L}^{\alpha}_{3} (  \bar{K}_2 ,  \bar{K}_1 + \bar{K}_2 ,0  ) 
 +   \bar{L}^{\alpha}_{3}  (  - \bar{K}_2,  - \bar{K}_1 - \bar{K}_2  ,0  )
 \nonumber
 \\
 & & \hspace{3mm}   + \bar{L}^{\alpha}_{3}  (   \bar{K}_1, - \bar{K}_2 , 0  ) 
 +     \bar{L}^{\alpha}_{3}   ( - \bar{K}_1 ,  \bar{K}_2 ,0 )
 \Bigr] .
 \label{eq:Lalphasym}
 \end{eqnarray}
For our model, the  three-loops can be calculated analytically 
using the method outlined in the appendix of
Ref.~\onlinecite{Pirooznia08}.
Consider first the non-symmetrized three-loop
defined in Eq.~(\ref{eq:L3}).
To perform the loop integration,
we decompose the integrand in partial fractions, 
then carry out the $k_{\parallel}$ integration by means of the method of
residues, and finally perform the $\ktau$-integration.
Using the notation
$\bar{K}_i = ( i \bar{\omega}_i , \bar{\bd{k}}_i )$,  
the result can be written as
 \begin{eqnarray}
& & \bar{L}^\alpha_3 ( \bar{K}_1 , \bar{K}_2, \bar{K}_3 )  =  
 \frac{1}{2 \pi}
 \int \frac{d k_{\bot}}{2 \pi}  \sum_{ i=1}^{3}   i \bar{\omega}_i 
\prod_{ \stackrel{j=1}{ j \neq i} }^3
 \frac{1}{ \Omega^{\alpha}_{ ij} ( k_{\bot} )}
 \nonumber
 \\
 &  & = \frac{1}{2 \pi}  \int \frac{d k_{\bot}}{2 \pi} 
\Biggl[
 \frac{ i \bar{\omega}_1}{ \Omega_{12}^{\alpha} ( k_{\bot} ) 
 \Omega_{13}^{\alpha} ( k_{\bot} ) }
+ \frac{ i \bar{\omega}_2}{ \Omega_{23}^{\alpha} ( k_{\bot} ) 
 \Omega_{21}^{\alpha} ( k_{\bot} ) }
 \nonumber
 \\
 & &   \hspace{20mm} 
 +
\frac{ i \bar{\omega}_3}{ \Omega_{31}^{\alpha} ( k_{\bot} ) 
 \Omega_{32}^{\alpha} ( k_{\bot} ) }
 \Biggr],
 \label{eq:Lbar1}
 \end{eqnarray}
where we have defined
 \begin{eqnarray}
& & \Omega^{\alpha}_{ij} ( {k}_{\bot} )  =  i  \bar{\omega}_i - i \bar{\omega}_j 
 + \xi^{\alpha}_{ \bd{k} - \bar{\bd{k}}_i }
 -  \xi^{\alpha}_{ \bd{k} - \bar{\bd{k}}_j }
 \nonumber
 \\
 &  &  =
i \omega_{ij}  -
\alpha ( \bar{k}_{\parallel i} - \bar{k}_{\parallel j} )
 + ( \bar{k}_{\bot i}^2 - \bar{k}_{\bot j}^2 ) 
- 2 q_{ij} k_{\bot} 
 \nonumber
 \\
& &  = - 2 q_{ij} [ k_{\bot} - k_{ij}],
 \end{eqnarray}
with
 \begin{subequations}
 \begin{eqnarray}
  \omega_{ ij} & = & \bar{\omega}_{ i} - \bar{\omega}_{ j} \; \; ,\; \; 
  q_{ ij}  =  \bar{k}_{\bot i} - \bar{k}_{\bot j},
 \\
 k_{ij} & = & \frac{ \bar{k}_{\bot i} + \bar{k}_{\bot j}}{2} + \frac{   i \omega_{ij} 
- \alpha ( \bar{k}_{\parallel i} - \bar{k}_{\parallel j} )   }{2 q_{ij}}.
\end{eqnarray}
\end{subequations}
The remaining $k_{\bot}$-integration
in Eq.~(\ref{eq:Lbar1}) can now be done using the residue theorem and we finally obtain
  \begin{eqnarray}
 \bar{L}^\alpha_3 ( \bar{K}_1 , \bar{K}_2, \bar{K}_3 ) &  = &  
 - \frac{1}{8 \pi} \Biggl[
 \nonumber
 \\
 & &
 \frac{ \bar{\omega}_1}{q_{12} q_{13}} \frac{ \Theta ( {\rm Im} (k_{12}) ) - 
 \Theta ( {\rm Im} (k_{13}) ) }{ k_{12} - k_{13} }
 \nonumber
 \\
 & + & 
 \frac{ \bar{\omega}_2}{q_{23} q_{21}} \frac{ \Theta ( {\rm Im} (k_{23}) ) - 
 \Theta ( {\rm Im} (k_{21}) ) }{ k_{23} - k_{21} }
 \nonumber
 \\
& + & \frac{ \bar{\omega}_3}{q_{31} q_{32}} \frac{ \Theta ( {\rm Im} (k_{31}) ) - 
 \Theta ( {\rm Im} (k_{32}) ) }{ k_{31} - k_{32} }
 \Biggr]. \hspace{7mm}
 \end{eqnarray}
Substituting this expression into Eq.~(\ref{eq:Lalphasym}) 
and defining
 \begin{eqnarray}
 x_i & = & \frac{ \barktaui}{ \bar{k}_{\bot i}} \; \; , 
 \; \; s_i = \frac{ \bar{k}_{\bot i}}{ \bar{k}_{\bot 1} +   \bar{k}_{\bot 2} },
 \end{eqnarray}
we obtain for the symmetrized three-loop,
 \begin{widetext}
 \begin{eqnarray}
  L^{ \alpha}_{3} (  \bar{K}_1,  \bar{K}_2 ,  - \bar{K}_1 - \bar{K}_2 )  
 & = &    \frac{1}{ 3! 2 \pi } 
 \frac{1}{ 
 s_1 s_2 \left[ 
  \frac{ \bar{k}_{\parallel 1}}{ \bar{k}_{\bot 1}} 
 -    \frac{ \bar{k}_{\parallel 2}}{ \bar{k}_{\bot 2}}  -  i \alpha ( x_1 - x_2 )  \right]^2  
-    \bar{k}_{\bot 1}  \bar{k}_{\bot 2}  
 } \Biggl[
 \bigl( s_1 | x_1 | +  s_2 | x_2 | \bigr)
 \Theta ( - x_1 x_2 )
 \nonumber
 \\
&  & \hspace{-30mm}
 + \bigl( s_1 | x_1 | - | s_1 x_1 + s_2 x_2 | \bigr)
 \Theta ( - x_1 (  s_1 x_1 + s_2 x_2 ) )
 + \bigl( s_2 | x_2 | - | s_1 x_1 + s_2 x_2 | \bigr)
 \Theta ( - x_2 (  s_1 x_1 + s_2 x_2 ) )
 \Biggr].
\end{eqnarray}
Alternatively, this expression can be written as
\begin{eqnarray}
  L^{ \alpha}_{3} (  \bar{K}_1,  \bar{K}_2 ,  - \bar{K}_1 - \bar{K}_2 )  
 & = &    \frac{1}{ 3! 2 \pi } 
 \frac{1}{ s_1 s_2
 \left[  \frac{ \bar{k}_{\parallel 1}}{ \bar{k}_{\bot 1}} 
 -    \frac{ \bar{k}_{\parallel 2}}{ \bar{k}_{\bot 2}} -  i \alpha ( x_1 - x_2 )  \right]^2  
 -  \bar{k}_{\bot 1}  \bar{k}_{\bot 2}  } 
\Biggl[
 \bigl( s_1   x_1  -  s_2 x_2  \bigr)
 [ \Theta (  x_1 ) - \Theta ( x_2 ) ]
 \nonumber
 \\
&  & \hspace{-20mm}
 + \bigl( 2 s_1 x_1  + s_2 x_2 \bigr)
 [ \Theta ( x_1) - \Theta (   s_1 x_1 + s_2 x_2 ) ]
 + \bigl( 2 s_2 x_2  + s_1 x_1 \bigr)
 [ \Theta ( x_2) - \Theta (   s_1 x_1 + s_2 x_2 ) ]
 \Biggr]
 \nonumber
 \\
 & = &  \frac{1}{4 \pi}
 \frac{ s_1 x_1 \Theta ( x_1 ) + s_2 x_2 \Theta ( x_2 ) - ( s_1 x_1 + s_2 x_2 ) \Theta
( s_1 x_1 + s_2 x_2 )  }{  s_1 s_2
 \left[  \frac{ \bar{k}_{\parallel 1}}{ \bar{k}_{\bot 1}} 
 -    \frac{ \bar{k}_{\parallel 2}}{ \bar{k}_{\bot 2}} -  i \alpha ( x_1 - x_2 )  \right]^2  
 -  \bar{k}_{\bot 1}  \bar{k}_{\bot 2}  }
\nonumber
 \\
 & = & 
\frac{1}{4 \pi} 
 \left( \frac{1}{ \bar{k}_{\bot 1}} +    \frac{1}{ \bar{k}_{\bot 2}} \right)
 \frac{ \barktauone \Theta \left( \frac{ \barktauone}{\bar{k}_{\bot 1}} \right) + 
 \barktautwo \Theta \left(   \frac{ \barktautwo}{\bar{k}_{\bot 2}}   \right) - 
 ( \barktauone + \barktautwo ) 
 \Theta \left( \frac{\barktauone + \barktautwo}{
 \bar{k}_{\bot 1} + \bar{k}_{\bot 2} } \right)
       }{ 
 \left[  \frac{ \bar{k}_{\parallel 1}}{ \bar{k}_{\bot 1}} 
 -    \frac{ \bar{k}_{\parallel 2}}{ \bar{k}_{\bot 2}} -  
 i \alpha (  \frac{ \barktauone}{ \bar{k}_{\bot 1}} -    
 \frac{ \barktautwo}{ \bar{k}_{\bot 2}}   )  \right]^2  
 -  ( \bar{k}_{\bot 1}  +  \bar{k}_{\bot 2}  )^2 } .
 \label{eq:L3res}
\end{eqnarray}
\end{widetext}
For a different effective model for the nematic quantum critical point
involving a quadratic energy dispersion (and hence a compact Fermi surface)
the scaling properties of fermion loops have recently been analyzed 
by Thier and Metzner \cite{Thier11}, who found that also in this case
the fermion loops exhibit a singular dependence on momenta and frequencies.
However, to obtain consistent scaling properties
of the effective interactions between bosonic fluctuations described by the fermion
loops, one should use one-loop renormalized fermion propagators
$G^{\alpha} ( K ) = \frac{Z}{ i \ktau - Z \xi^{\alpha}_{\bd{k}}}$
in the loop integrations, which
can be formally justified from the skeleton equation for the 
irreducible vertex with three external bosonic legs, as discussed in Appendix A. 
The corresponding expression for the renormalized three-loop
can be obtained from Eq.~(\ref{eq:L3res})  by
replacing all external frequencies by
$ \barktau \rightarrow \barktau / Z$ and multiplying 
the loop by an overall factor of $Z$. The result is given in Eq.~(\ref{eq:L3resmod}).

\end{appendix}


\begin{thebibliography}{47}
\expandafter\ifx\csname natexlab\endcsname\relax\def\natexlab#1{#1}\fi
\expandafter\ifx\csname bibnamefont\endcsname\relax
  \def\bibnamefont#1{#1}\fi
\expandafter\ifx\csname bibfnamefont\endcsname\relax
  \def\bibfnamefont#1{#1}\fi
\expandafter\ifx\csname citenamefont\endcsname\relax
  \def\citenamefont#1{#1}\fi
\expandafter\ifx\csname url\endcsname\relax
  \def\url#1{\texttt{#1}}\fi
\expandafter\ifx\csname urlprefix\endcsname\relax\def\urlprefix{URL }\fi
\providecommand{\bibinfo}[2]{#2}
\providecommand{\eprint}[2][]{\url{#2}}

\bibitem[{\citenamefont{Ando et~al.}(2002)\citenamefont{Ando, Segawa, Komiya,
  and Lavrov}}]{Ando02}
\bibinfo{author}{\bibfnamefont{Y.}~\bibnamefont{Ando}},
  \bibinfo{author}{\bibfnamefont{K.}~\bibnamefont{Segawa}},
  \bibinfo{author}{\bibfnamefont{S.}~\bibnamefont{Komiya}}, \bibnamefont{and}
  \bibinfo{author}{\bibfnamefont{A.~N.} \bibnamefont{Lavrov}},
  \bibinfo{journal}{Phys. Rev. Lett.} \textbf{\bibinfo{volume}{88}},
  \bibinfo{pages}{137005} (\bibinfo{year}{2002}).

\bibitem[{\citenamefont{Borzi et~al.}(2007)\citenamefont{Borzi, Grigera,
  Farrell, Perry, Lister, Lee, Tennant, Maeno, and Mackenzie}}]{Borzi07}
\bibinfo{author}{\bibfnamefont{R.~A.} \bibnamefont{Borzi}},
  \bibinfo{author}{\bibfnamefont{S.~A.} \bibnamefont{Grigera}},
  \bibinfo{author}{\bibfnamefont{J.}~\bibnamefont{Farrell}},
  \bibinfo{author}{\bibfnamefont{R.~S.} \bibnamefont{Perry}},
  \bibinfo{author}{\bibfnamefont{S.~J.~S.} \bibnamefont{Lister}},
  \bibinfo{author}{\bibfnamefont{S.~L.} \bibnamefont{Lee}},
  \bibinfo{author}{\bibfnamefont{D.~A.} \bibnamefont{Tennant}},
  \bibinfo{author}{\bibfnamefont{Y.}~\bibnamefont{Maeno}}, \bibnamefont{and}
  \bibinfo{author}{\bibfnamefont{A.~P.} \bibnamefont{Mackenzie}},
  \bibinfo{journal}{Science} \textbf{\bibinfo{volume}{315}},
  \bibinfo{pages}{214} (\bibinfo{year}{2007}).

\bibitem[{\citenamefont{Kohsaka et~al.}(2007)\citenamefont{Kohsaka, Taylor,
  Fujita, Schmidt, Lupien, Hanaguri, Azuma, Takano, Eisaki, Takagi
  et~al.}}]{Kohsaka07}
\bibinfo{author}{\bibfnamefont{Y.}~\bibnamefont{Kohsaka}},
  \bibinfo{author}{\bibfnamefont{C.}~\bibnamefont{Taylor}},
  \bibinfo{author}{\bibfnamefont{K.}~\bibnamefont{Fujita}},
  \bibinfo{author}{\bibfnamefont{A.}~\bibnamefont{Schmidt}},
  \bibinfo{author}{\bibfnamefont{C.}~\bibnamefont{Lupien}},
  \bibinfo{author}{\bibfnamefont{T.}~\bibnamefont{Hanaguri}},
  \bibinfo{author}{\bibfnamefont{M.}~\bibnamefont{Azuma}},
  \bibinfo{author}{\bibfnamefont{M.}~\bibnamefont{Takano}},
  \bibinfo{author}{\bibfnamefont{H.}~\bibnamefont{Eisaki}},
  \bibinfo{author}{\bibfnamefont{H.}~\bibnamefont{Takagi}},
  \bibinfo{author}{\bibfnamefont{S.}~\bibnamefont{Uchida}}, \bibnamefont{and}
  \bibinfo{author}{\bibfnamefont{J.~C.} \bibnamefont{Davis}},
  \bibinfo{journal}{Science} \textbf{\bibinfo{volume}{315}},
  \bibinfo{pages}{1380} (\bibinfo{year}{2007}).

\bibitem[{\citenamefont{Hinkov et~al.}(2008)\citenamefont{Hinkov, Haug,
  Fauqu\'{e}, Bourges, Sidis, Ivanov, Bernhard, Lin, and Keimer}}]{Hinkov08}
\bibinfo{author}{\bibfnamefont{V.}~\bibnamefont{Hinkov}},
  \bibinfo{author}{\bibfnamefont{D.}~\bibnamefont{Haug}},
  \bibinfo{author}{\bibfnamefont{B.}~\bibnamefont{Fauqu\'{e}}},
  \bibinfo{author}{\bibfnamefont{P.}~\bibnamefont{Bourges}},
  \bibinfo{author}{\bibfnamefont{Y.}~\bibnamefont{Sidis}},
  \bibinfo{author}{\bibfnamefont{A.}~\bibnamefont{Ivanov}},
  \bibinfo{author}{\bibfnamefont{C.}~\bibnamefont{Bernhard}},
  \bibinfo{author}{\bibfnamefont{C.~T.} \bibnamefont{Lin}}, \bibnamefont{and}
  \bibinfo{author}{\bibfnamefont{B.}~\bibnamefont{Keimer}},
  \bibinfo{journal}{Science} \textbf{\bibinfo{volume}{319}},
  \bibinfo{pages}{597} (\bibinfo{year}{2008}).

\bibitem[{\citenamefont{Daou et~al.}(2010)\citenamefont{Daou, Chang, LeBoeuf,
  Cyr-Choini{\`e}re, Lalibert{\'e}, Doiron-Leyraud, Ramshaw, Liang, Bonn, Hardy, Taillefer}}]{Daou10}
\bibinfo{author}{\bibfnamefont{R.}~\bibnamefont{Daou}},
  \bibinfo{author}{\bibfnamefont{J.}~\bibnamefont{Chang}},
  \bibinfo{author}{\bibfnamefont{D.}~\bibnamefont{LeBoeuf}},
  \bibinfo{author}{\bibfnamefont{O.}~\bibnamefont{Cyr-Choini{\`e}re}},
  \bibinfo{author}{\bibfnamefont{F.}~\bibnamefont{Lalibert{\'e}}},
  \bibinfo{author}{\bibfnamefont{N.}~\bibnamefont{Doiron-Leyraud}},
  \bibinfo{author}{\bibfnamefont{B.}~\bibnamefont{Ramshaw}},
  \bibinfo{author}{\bibfnamefont{R.}~\bibnamefont{Liang}},
  \bibinfo{author}{\bibfnamefont{D.}~\bibnamefont{Bonn}},
  \bibinfo{author}{\bibfnamefont{W.}~\bibnamefont{Hardy}}, \bibnamefont{and}
  \bibinfo{author}{\bibfnamefont{L.}~\bibnamefont{Taillefer}},
  \bibinfo{journal}{Nature}
  \textbf{\bibinfo{volume}{463}}, \bibinfo{pages}{519} (\bibinfo{year}{2010}).

\bibitem[{\citenamefont{Nandi et~al.}(2010)\citenamefont{Nandi, Kim, Kreyssig,
  Fernandes, Pratt, Thaler, Ni, Bud'ko, Canfield, Schmalian, McQueeney, Goldman}}]{Nandi10}
\bibinfo{author}{\bibfnamefont{S.}~\bibnamefont{Nandi}},
  \bibinfo{author}{\bibfnamefont{M.~G.} \bibnamefont{Kim}},
  \bibinfo{author}{\bibfnamefont{A.}~\bibnamefont{Kreyssig}},
  \bibinfo{author}{\bibfnamefont{R.~M.} \bibnamefont{Fernandes}},
  \bibinfo{author}{\bibfnamefont{D.~K.} \bibnamefont{Pratt}},
  \bibinfo{author}{\bibfnamefont{A.}~\bibnamefont{Thaler}},
  \bibinfo{author}{\bibfnamefont{N.}~\bibnamefont{Ni}},
  \bibinfo{author}{\bibfnamefont{S.~L.} \bibnamefont{Bud'ko}},
  \bibinfo{author}{\bibfnamefont{P.~C.} \bibnamefont{Canfield}},
  \bibinfo{author}{\bibfnamefont{J.}~\bibnamefont{Schmalian}},
  \bibinfo{author}{\bibfnamefont{R.~J.} \bibnamefont{McQueeney}},
  \bibnamefont{and} \bibinfo{author}{\bibfnamefont{A.~I.}
  \bibnamefont{Goldman}}, \bibinfo{journal}{Phys. Rev. Lett.}
  \textbf{\bibinfo{volume}{104}}, \bibinfo{pages}{057006}
  (\bibinfo{year}{2010}).

\bibitem[{\citenamefont{Fradkin et~al.}(2010)\citenamefont{Fradkin, Kivelson,
  Lawler, Eisenstein, and Mackenzie}}]{Fradkin10}
\bibinfo{author}{\bibfnamefont{E.}~\bibnamefont{Fradkin}},
  \bibinfo{author}{\bibfnamefont{S.~A.} \bibnamefont{Kivelson}},
  \bibinfo{author}{\bibfnamefont{M.~J.} \bibnamefont{Lawler}},
  \bibinfo{author}{\bibfnamefont{J.~P.} \bibnamefont{Eisenstein}},
  \bibnamefont{and} \bibinfo{author}{\bibfnamefont{A.~P.}
  \bibnamefont{Mackenzie}}, \bibinfo{journal}{Annu. Rev. Cond. Mat. Phys.}
  \textbf{\bibinfo{volume}{1}}, \bibinfo{pages}{153} (\bibinfo{year}{2010}).

\bibitem[{\citenamefont{Pomeranchuk}(1958)}]{Pomeranchuk58}
\bibinfo{author}{\bibfnamefont{I.}~\bibnamefont{Pomeranchuk}},
  \bibinfo{journal}{Soviet. Phys. JETP} \textbf{\bibinfo{volume}{8}},
  \bibinfo{pages}{361} (\bibinfo{year}{1958}).

\bibitem[{\citenamefont{Halboth and Metzner}(2000)}]{Halboth00}
\bibinfo{author}{\bibfnamefont{C.~J.} \bibnamefont{Halboth}} \bibnamefont{and}
  \bibinfo{author}{\bibfnamefont{W.}~\bibnamefont{Metzner}},
  \bibinfo{journal}{Phys. Rev. Lett.} \textbf{\bibinfo{volume}{85}},
  \bibinfo{pages}{5162} (\bibinfo{year}{2000}).

\bibitem[{\citenamefont{Hertz}(1976)}]{Hertz76}
\bibinfo{author}{\bibfnamefont{J.~A.} \bibnamefont{Hertz}},
  \bibinfo{journal}{Phys. Rev. B} \textbf{\bibinfo{volume}{14}},
  \bibinfo{pages}{1165} (\bibinfo{year}{1976}).

\bibitem[{\citenamefont{Millis}(1993)}]{Millis93}
\bibinfo{author}{\bibfnamefont{A.~J.} \bibnamefont{Millis}},
  \bibinfo{journal}{Phys. Rev. B} \textbf{\bibinfo{volume}{48}},
  \bibinfo{pages}{7183} (\bibinfo{year}{1993}).

\bibitem[{\citenamefont{Belitz et~al.}(2005)\citenamefont{Belitz, Kirkpatrick,
  and Vojta}}]{Belitz05}
\bibinfo{author}{\bibfnamefont{D.}~\bibnamefont{Belitz}},
  \bibinfo{author}{\bibfnamefont{T.~R.} \bibnamefont{Kirkpatrick}},
  \bibnamefont{and} \bibinfo{author}{\bibfnamefont{T.}~\bibnamefont{Vojta}},
  \bibinfo{journal}{Rev. Mod. Phys.} \textbf{\bibinfo{volume}{77}},
  \bibinfo{pages}{579} (\bibinfo{year}{2005}).

\bibitem[{\citenamefont{L\"ohneysen et~al.}(2007)\citenamefont{L\"ohneysen,
  Rosch, Vojta, and W\"olfle}}]{Loehneysen07}
\bibinfo{author}{\bibfnamefont{H.~v.} \bibnamefont{L\"ohneysen}},
  \bibinfo{author}{\bibfnamefont{A.}~\bibnamefont{Rosch}},
  \bibinfo{author}{\bibfnamefont{M.}~\bibnamefont{Vojta}}, \bibnamefont{and}
  \bibinfo{author}{\bibfnamefont{P.}~\bibnamefont{W\"olfle}},
  \bibinfo{journal}{Rev. Mod. Phys.} \textbf{\bibinfo{volume}{79}},
  \bibinfo{pages}{1015} (\bibinfo{year}{2007}).

\bibitem[{\citenamefont{Vojta}(2003)}]{Vojta03}
\bibinfo{author}{\bibfnamefont{M.}~\bibnamefont{Vojta}}, \bibinfo{journal}{Rep.
  Prog. Phys.} \textbf{\bibinfo{volume}{66}}, \bibinfo{pages}{2069}
  (\bibinfo{year}{2003}).

\bibitem[{\citenamefont{Holstein et~al.}(1973)\citenamefont{Holstein, Norton,
  and Pincus}}]{Holstein73}
\bibinfo{author}{\bibfnamefont{T.}~\bibnamefont{Holstein}},
  \bibinfo{author}{\bibfnamefont{R.~E.} \bibnamefont{Norton}},
  \bibnamefont{and} \bibinfo{author}{\bibfnamefont{P.}~\bibnamefont{Pincus}},
  \bibinfo{journal}{Phys. Rev. B} \textbf{\bibinfo{volume}{8}},
  \bibinfo{pages}{2649} (\bibinfo{year}{1973}).

\bibitem[{\citenamefont{Reizer}(1989)}]{Reizer89}
\bibinfo{author}{\bibfnamefont{M.~Y.} \bibnamefont{Reizer}},
  \bibinfo{journal}{Phys. Rev. B} \textbf{\bibinfo{volume}{40}},
  \bibinfo{pages}{11571} (\bibinfo{year}{1989}).

\bibitem[{\citenamefont{Lee and Nagaosa}(1992)}]{Lee92}
\bibinfo{author}{\bibfnamefont{P.~A.} \bibnamefont{Lee}} \bibnamefont{and}
  \bibinfo{author}{\bibfnamefont{N.}~\bibnamefont{Nagaosa}},
  \bibinfo{journal}{Phys. Rev. B} \textbf{\bibinfo{volume}{46}},
  \bibinfo{pages}{5621} (\bibinfo{year}{1992}).

\bibitem[{\citenamefont{Halperin et~al.}(1993)\citenamefont{Halperin, Lee, and
  Read}}]{Halperin93}
\bibinfo{author}{\bibfnamefont{B.~I.} \bibnamefont{Halperin}},
  \bibinfo{author}{\bibfnamefont{P.~A.} \bibnamefont{Lee}}, \bibnamefont{and}
  \bibinfo{author}{\bibfnamefont{N.}~\bibnamefont{Read}},
  \bibinfo{journal}{Phys. Rev. B} \textbf{\bibinfo{volume}{47}},
  \bibinfo{pages}{7312} (\bibinfo{year}{1993}).

\bibitem[{\citenamefont{Motrunich}(2005)}]{Motrunich05}
\bibinfo{author}{\bibfnamefont{O.~I.} \bibnamefont{Motrunich}},
  \bibinfo{journal}{Phys. Rev. B} \textbf{\bibinfo{volume}{72}},
  \bibinfo{pages}{045105} (\bibinfo{year}{2005}).

\bibitem[{\citenamefont{Lee and Lee}(2005)}]{SSLee05}
\bibinfo{author}{\bibfnamefont{S.-S.} \bibnamefont{Lee}} \bibnamefont{and}
  \bibinfo{author}{\bibfnamefont{P.~A.} \bibnamefont{Lee}},
  \bibinfo{journal}{Phys. Rev. Lett.} \textbf{\bibinfo{volume}{95}},
  \bibinfo{pages}{036403} (\bibinfo{year}{2005}).

\bibitem[{\citenamefont{Rech et~al.}(2006)\citenamefont{Rech, P\'epin, and
  Chubukov}}]{Rech06}
\bibinfo{author}{\bibfnamefont{J.}~\bibnamefont{Rech}},
  \bibinfo{author}{\bibfnamefont{C.}~\bibnamefont{P\'epin}}, \bibnamefont{and}
  \bibinfo{author}{\bibfnamefont{A.~V.} \bibnamefont{Chubukov}},
  \bibinfo{journal}{Phys. Rev. B} \textbf{\bibinfo{volume}{74}},
  \bibinfo{pages}{195126} (\bibinfo{year}{2006}).

\bibitem[{\citenamefont{Gonz\'ales et~al.}(1994)\citenamefont{Gonz\'ales, Guinea, and
  Vozmediano}}]{Gonzales94}
\bibinfo{author}{\bibfnamefont{J.}~\bibnamefont{Gonz\'ales}},
  \bibinfo{author}{\bibfnamefont{F.}~\bibnamefont{Guinea}}, \bibnamefont{and}
  \bibinfo{author}{\bibfnamefont{M.~A.~H.} \bibnamefont{Vozmediano}},
  \bibinfo{journal}{Nucl. Phys. B} \textbf{\bibinfo{volume}{424}},
  \bibinfo{pages}{595} (\bibinfo{year}{1994}).

\bibitem[{\citenamefont{Giuliani et~al.}(2010)\citenamefont{Giuliani, Mastropietro, and
  Porta}}]{Giuliani10}
\bibinfo{author}{\bibfnamefont{A.}~\bibnamefont{Giuliani}},
  \bibinfo{author}{\bibfnamefont{V.}~\bibnamefont{Mastropietro}}, \bibnamefont{and}
  \bibinfo{author}{\bibfnamefont{M.} \bibnamefont{Porta}},
  \bibinfo{journal}{Phys. Rev. B} \textbf{\bibinfo{volume}{82}},
  \bibinfo{pages}{121418(R)} (\bibinfo{year}{2010}).

\bibitem[{\citenamefont{Altshuler et~al.}(1994)\citenamefont{Altshuler, Ioffe,
  and Millis}}]{Altshuler94}
\bibinfo{author}{\bibfnamefont{B.~L.} \bibnamefont{Altshuler}},
  \bibinfo{author}{\bibfnamefont{L.~B.} \bibnamefont{Ioffe}}, \bibnamefont{and}
  \bibinfo{author}{\bibfnamefont{A.~J.} \bibnamefont{Millis}},
  \bibinfo{journal}{Phys. Rev. B} \textbf{\bibinfo{volume}{50}},
  \bibinfo{pages}{14048} (\bibinfo{year}{1994}).

\bibitem[{\citenamefont{Polchinski}(1994)}]{Polchinski94}
\bibinfo{author}{\bibfnamefont{J.}~\bibnamefont{Polchinski}},
  \bibinfo{journal}{Nucl. Phys. B} \textbf{\bibinfo{volume}{422}},
  \bibinfo{pages}{617 } (\bibinfo{year}{1994}).

\bibitem[{\citenamefont{Sedrakyan and Chubukov}(2011)}]{Sedrakyan09}
\bibinfo{author}{\bibfnamefont{T.~A.} \bibnamefont{Sedrakyan}} \bibnamefont{and}
  \bibinfo{author}{\bibfnamefont{A.~V.}~\bibnamefont{Chubukov}},
  \bibinfo{journal}{Phys. Rev. B} \textbf{\bibinfo{volume}{79}},
  \bibinfo{pages}{115129} (\bibinfo{year}{2009}).

\bibitem[{\citenamefont{Lee}(2008)}]{SSLee08}
\bibinfo{author}{\bibfnamefont{S.-S.} \bibnamefont{Lee}},
  \bibinfo{journal}{Phys. Rev. B} \textbf{\bibinfo{volume}{78}},
  \bibinfo{pages}{085129} (\bibinfo{year}{2008}).

\bibitem[{\citenamefont{Lee}(2009)}]{SSLee09}
\bibinfo{author}{\bibfnamefont{S.-S.} \bibnamefont{Lee}},
  \bibinfo{journal}{Phys. Rev. B} \textbf{\bibinfo{volume}{80}},
  \bibinfo{pages}{165102} (\bibinfo{year}{2009}).

\bibitem[{\citenamefont{Metlitski and
  Sachdev}(2010{\natexlab{a}})}]{Metlitski10a}
\bibinfo{author}{\bibfnamefont{M.~A.} \bibnamefont{Metlitski}}
  \bibnamefont{and} \bibinfo{author}{\bibfnamefont{S.}~\bibnamefont{Sachdev}},
  \bibinfo{journal}{Phys. Rev. B} \textbf{\bibinfo{volume}{82}},
  \bibinfo{pages}{075127} (\bibinfo{year}{2010}{\natexlab{a}}).

\bibitem[{\citenamefont{Metlitski and
  Sachdev}(2010{\natexlab{b}})}]{Metlitski10b}
\bibinfo{author}{\bibfnamefont{M.~A.} \bibnamefont{Metlitski}}
  \bibnamefont{and} \bibinfo{author}{\bibfnamefont{S.}~\bibnamefont{Sachdev}},
  \bibinfo{journal}{Phys. Rev. B} \textbf{\bibinfo{volume}{82}},
  \bibinfo{pages}{075128} (\bibinfo{year}{2010}{\natexlab{b}}).

\bibitem[{\citenamefont{Chubukov}(2010)}]{Chubukov10}
\bibinfo{author}{\bibfnamefont{A.~V.} \bibnamefont{Chubukov}},
  \bibinfo{journal}{Physics} \textbf{\bibinfo{volume}{3}}, \bibinfo{pages}{70}
  (\bibinfo{year}{2010}).

\bibitem[{\citenamefont{Mross et~al.}(2010)\citenamefont{Mross, McGreevy, Liu,
  and Senthil}}]{Mross10}
\bibinfo{author}{\bibfnamefont{D.~F.} \bibnamefont{Mross}},
  \bibinfo{author}{\bibfnamefont{J.}~\bibnamefont{McGreevy}},
  \bibinfo{author}{\bibfnamefont{H.}~\bibnamefont{Liu}}, \bibnamefont{and}
  \bibinfo{author}{\bibfnamefont{T.}~\bibnamefont{Senthil}},
  \bibinfo{journal}{Phys. Rev. B} \textbf{\bibinfo{volume}{82}},
  \bibinfo{pages}{045121} (\bibinfo{year}{2010}).

\bibitem[{\citenamefont{Nayak and Wilczek}(1994{\natexlab{a}})}]{Nayak94a}
\bibinfo{author}{\bibfnamefont{C.}~\bibnamefont{Nayak}} \bibnamefont{and}
  \bibinfo{author}{\bibfnamefont{F.}~\bibnamefont{Wilczek}},
  \bibinfo{journal}{Nucl. Phys. B} \textbf{\bibinfo{volume}{417}},
  \bibinfo{pages}{359 } (\bibinfo{year}{1994}{\natexlab{a}}).

\bibitem[{\citenamefont{Nayak and Wilczek}(1994{\natexlab{b}})}]{Nayak94b}
\bibinfo{author}{\bibfnamefont{C.}~\bibnamefont{Nayak}} \bibnamefont{and}
  \bibinfo{author}{\bibfnamefont{F.}~\bibnamefont{Wilczek}},
  \bibinfo{journal}{Nucl. Phys. B} \textbf{\bibinfo{volume}{430}},
  \bibinfo{pages}{534 } (\bibinfo{year}{1994}{\natexlab{b}}).

\bibitem[{\citenamefont{Wetterich}(1993)}]{Wetterich93}
\bibinfo{author}{\bibfnamefont{C.}~\bibnamefont{Wetterich}},
  \bibinfo{journal}{Phys. Lett. B} \textbf{\bibinfo{volume}{301}},
  \bibinfo{pages}{90} (\bibinfo{year}{1993}).

\bibitem[{\citenamefont{Berges et~al.}(2002)\citenamefont{Berges, Tetradis, and
  Wetterich}}]{Berges02}
\bibinfo{author}{\bibfnamefont{J.}~\bibnamefont{Berges}},
  \bibinfo{author}{\bibfnamefont{N.}~\bibnamefont{Tetradis}}, \bibnamefont{and}
  \bibinfo{author}{\bibfnamefont{C.}~\bibnamefont{Wetterich}},
  \bibinfo{journal}{Phys. Rep.} \textbf{\bibinfo{volume}{363}},
  \bibinfo{pages}{223} (\bibinfo{year}{2002}).

\bibitem[{\citenamefont{Kopietz et~al.}(2010)\citenamefont{Kopietz, Bartosch,
  and Sch\"{u}tz}}]{Kopietz10}
\bibinfo{author}{\bibfnamefont{P.}~\bibnamefont{Kopietz}},
  \bibinfo{author}{\bibfnamefont{L.}~\bibnamefont{Bartosch}}, \bibnamefont{and}
  \bibinfo{author}{\bibfnamefont{F.}~\bibnamefont{Sch\"{u}tz}},
  \emph{\bibinfo{title}{{Introduction to the Functional Renormalization
  Group}}} (\bibinfo{publisher}{Springer, Berlin}, \bibinfo{year}{2010}).

\bibitem[{\citenamefont{Metzner et~al.}(2012)\citenamefont{Metzner, Salmhofer,
  Honerkamp, Meden, and Sch\"onhammer}}]{Metzner12}
\bibinfo{author}{\bibfnamefont{W.}~\bibnamefont{Metzner}},
  \bibinfo{author}{\bibfnamefont{M.}~\bibnamefont{Salmhofer}},
  \bibinfo{author}{\bibfnamefont{C.}~\bibnamefont{Honerkamp}},
  \bibinfo{author}{\bibfnamefont{V.}~\bibnamefont{Meden}}, \bibnamefont{and}
  \bibinfo{author}{\bibfnamefont{K.}~\bibnamefont{Sch\"onhammer}},
  \bibinfo{journal}{Rev. Mod. Phys.} \textbf{\bibinfo{volume}{84}},
  \bibinfo{pages}{299} (\bibinfo{year}{2012}).

\bibitem[{\citenamefont{Baier et~al.}(2004)\citenamefont{Baier, Bick, and
  Wetterich}}]{Baier04}
\bibinfo{author}{\bibfnamefont{T.}~\bibnamefont{Baier}},
  \bibinfo{author}{\bibfnamefont{E.}~\bibnamefont{Bick}}, \bibnamefont{and}
  \bibinfo{author}{\bibfnamefont{C.}~\bibnamefont{Wetterich}},
  \bibinfo{journal}{Phys. Rev. B} \textbf{\bibinfo{volume}{70}},
  \bibinfo{pages}{125111} (\bibinfo{year}{2004}).

\bibitem[{\citenamefont{Baier et~al.}(2005)\citenamefont{Baier, Bick, and
  Wetterich}}]{Baier05}
\bibinfo{author}{\bibfnamefont{T.}~\bibnamefont{Baier}},
  \bibinfo{author}{\bibfnamefont{E.}~\bibnamefont{Bick}}, \bibnamefont{and}
  \bibinfo{author}{\bibfnamefont{C.}~\bibnamefont{Wetterich}},
  \bibinfo{journal}{Phys. Lett. B} \textbf{\bibinfo{volume}{605}},
  \bibinfo{pages}{144} (\bibinfo{year}{2005}).

\bibitem[{\citenamefont{Sch{\"u}tz et~al.}(2005)\citenamefont{Sch{\"u}tz,
  Bartosch, and Kopietz}}]{Schuetz05}
\bibinfo{author}{\bibfnamefont{F.}~\bibnamefont{Sch{\"u}tz}},
  \bibinfo{author}{\bibfnamefont{L.}~\bibnamefont{Bartosch}}, \bibnamefont{and}
  \bibinfo{author}{\bibfnamefont{P.}~\bibnamefont{Kopietz}},
  \bibinfo{journal}{Phys. Rev. B} \textbf{\bibinfo{volume}{72}},
  \bibinfo{pages}{035107} (\bibinfo{year}{2005}).

\bibitem[{\citenamefont{Wetterich}(2007)}]{Wetterich07}
\bibinfo{author}{\bibfnamefont{C.}~\bibnamefont{Wetterich}},
  \bibinfo{journal}{Phys. Rev. B} \textbf{\bibinfo{volume}{75}},
  \bibinfo{pages}{085102} (\bibinfo{year}{2007}).

\bibitem[{\citenamefont{Strack et~al.}(2008)\citenamefont{Strack, Gersch, and
  Metzner}}]{Strack08}
\bibinfo{author}{\bibfnamefont{P.}~\bibnamefont{Strack}},
  \bibinfo{author}{\bibfnamefont{R.}~\bibnamefont{Gersch}}, \bibnamefont{and}
  \bibinfo{author}{\bibfnamefont{W.}~\bibnamefont{Metzner}},
  \bibinfo{journal}{Phys. Rev. B} \textbf{\bibinfo{volume}{78}},
  \bibinfo{pages}{014522} (\bibinfo{year}{2008}).

\bibitem[{\citenamefont{Bartosch
  et~al.}(2009{\natexlab{a}})\citenamefont{Bartosch, Freire, Ramos~Cardenas,
  and Kopietz}}]{Bartosch09a}
\bibinfo{author}{\bibfnamefont{L.}~\bibnamefont{Bartosch}},
  \bibinfo{author}{\bibfnamefont{H.}~\bibnamefont{Freire}},
  \bibinfo{author}{\bibfnamefont{J.~J.} \bibnamefont{Ramos~Cardenas}},
  \bibnamefont{and} \bibinfo{author}{\bibfnamefont{P.}~\bibnamefont{Kopietz}},
  \bibinfo{journal}{J. Phys.: Condens. Matter} \textbf{\bibinfo{volume}{21}},
  \bibinfo{pages}{305602} (\bibinfo{year}{2009}{\natexlab{a}}).

\bibitem[{\citenamefont{Bartosch
  et~al.}(2009{\natexlab{b}})\citenamefont{Bartosch, Kopietz, and
  Ferraz}}]{Bartosch09b}
\bibinfo{author}{\bibfnamefont{L.}~\bibnamefont{Bartosch}},
  \bibinfo{author}{\bibfnamefont{P.}~\bibnamefont{Kopietz}}, \bibnamefont{and}
  \bibinfo{author}{\bibfnamefont{A.}~\bibnamefont{Ferraz}},
  \bibinfo{journal}{Phys. Rev. B} \textbf{\bibinfo{volume}{80}},
  \bibinfo{pages}{104514} (\bibinfo{year}{2009}{\natexlab{b}}).

\bibitem[{\citenamefont{Sachdev}(2011)}]{SachdevBook2nd}
\bibinfo{author}{\bibfnamefont{S.}~\bibnamefont{Sachdev}},
  \emph{\bibinfo{title}{{Quantum Phase Transitions}}}
  (\bibinfo{publisher}{Cambridge University Press, Cambridge},
  \bibinfo{year}{2011}), \bibinfo{edition}{2nd} ed.

\bibitem[{\citenamefont{Yamamoto and Si}(2010)}]{Yamamoto10}
\bibinfo{author}{\bibfnamefont{S.~J.} \bibnamefont{Yamamoto}} \bibnamefont{and}
  \bibinfo{author}{\bibfnamefont{Q.}~\bibnamefont{Si}},
  \bibinfo{journal}{Phys. Rev. B} \textbf{\bibinfo{volume}{81}},
  \bibinfo{pages}{205106} (\bibinfo{year}{2010}).

\bibitem[{\citenamefont{Isidori et~al.}(2010)\citenamefont{Isidori, Roosen,
  Bartosch, Hofstetter, and Kopietz}}]{Isidori10}
\bibinfo{author}{\bibfnamefont{A.}~\bibnamefont{Isidori}},
  \bibinfo{author}{\bibfnamefont{D.}~\bibnamefont{Roosen}},
  \bibinfo{author}{\bibfnamefont{L.}~\bibnamefont{Bartosch}},
  \bibinfo{author}{\bibfnamefont{W.}~\bibnamefont{Hofstetter}},
  \bibnamefont{and} \bibinfo{author}{\bibfnamefont{P.}~\bibnamefont{Kopietz}},
  \bibinfo{journal}{Phys. Rev. B} \textbf{\bibinfo{volume}{81}},
  \bibinfo{pages}{235120} (\bibinfo{year}{2010}).

\bibitem[{\citenamefont{Pirooznia et~al.}(2008)\citenamefont{Pirooznia,
  Sch\"utz, and Kopietz}}]{Pirooznia08}
\bibinfo{author}{\bibfnamefont{P.}~\bibnamefont{Pirooznia}},
  \bibinfo{author}{\bibfnamefont{F.}~\bibnamefont{Sch\"utz}}, \bibnamefont{and}
  \bibinfo{author}{\bibfnamefont{P.}~\bibnamefont{Kopietz}},
  \bibinfo{journal}{Phys. Rev. B} \textbf{\bibinfo{volume}{78}},
  \bibinfo{pages}{075111} (\bibinfo{year}{2008}).

\bibitem[{\citenamefont{Thier and Metzner}(2011)}]{Thier11}
\bibinfo{author}{\bibfnamefont{S.~C.} \bibnamefont{Thier}} \bibnamefont{and}
  \bibinfo{author}{\bibfnamefont{W.}~\bibnamefont{Metzner}},
  \bibinfo{journal}{Phys. Rev. B} \textbf{\bibinfo{volume}{84}},
  \bibinfo{pages}{155133} (\bibinfo{year}{2011}).




\end{thebibliography}
\end{document}